\documentclass[apj]{emulateapj}

\submitted{Accepted by ApJ, 2010 February 9}

\begin{document}
\title{A Systematic Search for X-ray Cavities in the Hot Gas of Galaxy Groups}

\shorttitle{X-ray Cavities in Galaxy Groups}

\shortauthors{Dong, Rasmussen \& Mulchaey}

\author{Ruobing Dong,\altaffilmark{1} Jesper
Rasmussen,\altaffilmark{2,3} and John S.~Mulchaey\altaffilmark{2}}

\altaffiltext{1}{Department of Astrophysical Sciences, Princeton
University, Princeton, NJ 08544, rdong@princeton.edu}

\altaffiltext{2}{Observatories of Carnegie Institution of Washington,
813 Santa Barbara Street, Pasadena, CA 91101}

\altaffiltext{3}{Chandra Fellow}

\begin{abstract}

We have performed a systematic search for X-ray cavities in the hot
gas of 51~galaxy groups with {\em Chandra} archival data. The cavities
are identified based on two methods: subtracting an elliptical
$\beta$--model fitted to the X-ray surface brightness, and performing
unsharp masking. 13 groups in the sample ($\sim 25\%$) are identified
as clearly containing cavities, with another 13 systems showing
tentative evidence for such structures. We find tight correlations
between the radial and tangential radii of the cavities, and between
their size and projected distance from the group center, in
quantitative agreement with the case for more massive clusters. This
suggests that similar physical processes are responsible for cavity
evolution and disruption in systems covering a large range in total
mass. We see no clear association between the detection of cavities
and the current 1.4~GHz radio luminosity of the central brightest
group galaxy, but there is a clear tendency for systems with a cool
core to be more likely to harbor detectable cavities.  To test the
efficiency of the adopted cavity detection procedures, we employ a set
of mock images designed to mimic typical {\em Chandra} data of our
sample, and find that the model-fitting approach is generally more
reliable than unsharp masking for recovering cavity properties.
Finally, we find that the detectability of cavities is strongly
influenced by a few factors, particularly the signal-to-noise ratio of
the data, and that the real fraction of X-ray groups with prominent
cavities could be substantially larger than the 25--50\% suggested by
our analysis.

\end{abstract}


\section{INTRODUCTION}

Feedback and heating from active galactic nuclei (AGN) is considered a
prime candidate for solving the ``cooling flow'' problem \citep{fab94}
in the hot gas of galaxy clusters \citep{bir04,dun06,raf06,mcn07},
groups, and giant elliptical galaxies \citep{jon02,mac06}, although
the details of this process are not yet fully understood.  Recent
observations using {\em Chandra} and {\em XMM-Newton} have produced a
large increase in the detection of X-ray surface brightness
depressions (``cavities'' or ``bubbles'') in many of these systems,
interpreted as buoyantly rising bubbles created by AGN
outbursts. Studies of such cavities in individual clusters, including
Hydra~A \citep{mcn00}, Perseus \citep{fab00}, A2052 \citep{bla01},
A2199 \citep{joh02}, and Centaurus \citep{san02}, indicate that the
outburst energy required to inflate these cavities would be sufficient
to balance cooling \citep{bir04,raf06}, explain the lack of gas
cooling below $T\approx 2$~keV in cluster cores \citep{pet01,kaa04},
and reproduce the bright end of the galaxy luminosity function
\citep{ben03,bow06,cro06,sij06}.

Detailed studies of larger samples of X-ray bright elliptical galaxies
and their surrounding cavities has further elucidated the possible
role of AGN feedback for the thermal and morphological properties of
the hot gas in these systems. \citet{bes05,bes06} combined observed
cavity properties and central radio powers of the systems in the
\citet{bir04} sample with the inferred fraction of radio-loud
ellipticals in the Sloan Digital Sky Survey, to show that the
time-averaged heating rate by radio AGN in massive ellipticals can
generally balance the cooling rate of the hot gas surrounding these
galaxies. Using a sample of X-ray luminous ellipticals with identified
cavities, \cite{all06} estimated AGN jet powers from cavity properties
and showed that these correlate tightly with the anticipated Bondi
accretion rate of the central supermassive black hole. Evidence that
radio AGN may be able to affect their gaseous surroundings also in
lower-mass ellipticals has been provided by \citet{die08a}, who
demonstrated that the amount of asymmetry in the hot gas morphology of
ellipticals correlates with the central radio AGN luminosity, even
down to the lowest detectable radio powers in relatively X-ray faint
galaxies.

These results all point to a close connection between central AGN
radio outbursts and the creation of X-ray cavities in the surrounding
gas. In terms of establishing the incidence and nature of such
cavities, most work so far has focused on studying cavity properties
and AGN interactions with the intracluster medium (ICM) within galaxy
clusters \citep{bir04, dun04, dun05, dun06, raf06, ste07, bir08,
raf08, die08}. Results suggest that roughly 2/3 of X-ray bright
cool-core clusters harbor detectable cavities \citep{dun06}, and that
these cavities appear to obey tight scaling relations between their
size and projected clustercentric distance, offering a means of
testing their nature and that of the inflating mechanism
\citep{die08}. Comparable work on giant elliptical galaxies that do
not represent central brightest cluster galaxies is so far limited
\citep{all06,mcn07}, but indicates a significantly smaller detectable
cavity fraction of $\sim 1/4$ \citep{mcn07}.  However, studies of AGN
heating and X-ray cavities within large samples of galaxy groups have
largely been absent. Although AGN outbursts in groups are assumed to
be smaller in scale and less energetic than those in clusters, they
may play a more prominent role in the evolution of the host structure
due to the shallower gravitational potential of groups. Estimating the
incidence and properties of X-ray cavities in groups is therefore an
important step toward understanding the role played by AGN outbursts
for the evolution of baryons on group scales.

In many X-ray bright clusters with high-quality data, X-ray cavities
are prominent and can be easily identified as X-ray surface brightness
depressions using visual inspection. This can be augmented by radio
data revealing ongoing AGN activity within the central cluster galaxy,
or the presence of radio lobes coincident with the cavities. In this
process, a number of factors could affect the detectability of X-ray
cavities however, including the position, orientation, and angular
extent of the cavities \citep{en02, die08, bru09}, along with
observational details such as the sensitivity of the data. These
issues are even more important in the group regime, where the lower
intrinsic X-ray luminosities can further impede cavity
identification. To search for X-ray cavities in groups, it is
therefore important to consider additional methods that can aid simple
visual inspection in identifying these structures.

In the present work, we select a sample of 51 galaxy groups from the
{\em Chandra} archive, in order to systematically search for and
characterize the presence of X-ray cavities in a large sample of
galaxy groups. In addition to visual inspection of the raw X-ray data,
we facilitate the detection process by employing two methods to first
characterize the large-scale group emission, {\em viz.}\ unsharp
masking and modeling of the surface brightness distribution. A good
handful of the groups in our sample have been previously reported to
harbor detectable cavities. Here we find that at least 13 of the
groups, and potentially as much as half of our group sample (26/51),
show evidence of X-ray cavities. We quantify the size, groupcentric
distance, and relationship with the central AGN radio luminosity for
all these structures, in addition to offering some considerations as
to the preferred method for identifying cavities in these relatively
X-ray faint systems.

The structure of this paper is as follows. In Section~\ref{sec:sample}
we present the group sample and outline the data reduction
process. Our approach to searching for cavities and establishing their
properties are described in detail in Section~\ref{sec:analysis}.
Section~\ref{sec:result} presents the results of our {\em Chandra}
analysis. To understand the efficiency of the two methods employed for
cavity detection, we generated a set of mock data sets, the analysis
of which is described in Section~\ref{sec:mock}. This is followed by a
discussion of our results in Section~\ref{sec:discuss} and conclusions
in Section~\ref{sec:conclude}. The cosmological parameters assumed in
this paper are $H_0=73$~km~s$^{-1}$~Mpc$^{-1}$, $\Omega_{m}=0.27$, and
$\Omega_{\Lambda}=0.73$. Unless otherwise stated, all uncertainties
are quoted at the 68\% confidence level.\\

\section{SAMPLE SELECTION AND DATA REDUCTION}\label{sec:sample}

The groups in this study were selected from the {\em Chandra} archival
samples of \citet{sun09} (43 groups) and \citet{ras07} (an additional
9 groups). To suppress the potential impact of instrumental artifacts
on our results, only groups for which the central bright regions are
covered by a single ACIS CCD were considered. This excluded one group
(A160) from the sample. Our final sample is listed in
Table~\ref{table:sample} and includes 51 groups, ranging in distance
from $\sim$ 20--550~Mpc. A few of the groups have been covered by
multiple {\em Chandra} observations. In those cases, we generally used
the longest observation unless this had a large offset between the
group center and the CCD aimpoint. We stress that the sample does not
contain individual giant elliptical galaxies, nor any galaxy clusters,
and that the two studies from which our sample is drawn were not
themselves designed for cavity studies. Hence, while the selection
criteria employed in those two studies favor fairly undisturbed X-ray
bright systems, the sample should not be inherently biased toward
systems with prominent cavities.

Standard data reduction and calibration was performed to all data sets
starting from level one event files. Chandra Interactive Analysis of
Observations software ({\sc ciao}) v.4.1.1 and Chandra Calibration
Database (CALDB) 4.1.1 were used in this work. New level one event
files were created with the ``acis$\_$process$\_$events'' task in {\sc
ciao}, including charge transfer inefficiency correction,
time-dependent gain adjustment, and screening for bad pixels using the
bad pixel map provided by the pipeline. For observations taken in Very
Faint mode, additional background screening was performed.
Grade/status filters were applied (excluding {\em ASCA} grades 1, 5,
and 7) along with Good Time Intervals filters to produce level~two
event files. Times of high background were eliminated based on
lightcurves extracted in regions away from the extended source center.
Images were then produced in the 0.3--2~keV band, chosen to optimize
the signal-to-noise ratio at the typical group temperatures of our
sample, using spatial bins of 1 or 2 pixels depending on source extent
and data quality.  Exposure maps were produced assuming a
monoenergetic distribution of source photons at the peak flux energy
(usually around 1~keV). The source image was normalized by the
exposure map, correcting for the effect of strongly variable exposure
near the detector edges.  We adopt a threshold of $1.5\%$ of the
maximum value of the exposure map, setting all pixels with exposure
below this value equal to zero.

\section{DATA ANALYSIS}\label{sec:analysis}

To search for small-scale X-ray structure and identify potential
cavities, we employ two methods: Modeling the surface brightness
distribution of the groups using a two-dimensional (elliptical)
$\beta$--model and performing an unsharp masking procedure.

\subsection{Elliptical $\beta$--Model Fitting}\label{sec:fitting}

With this method, we first aim to characterize the large-scale group
emission by means of an elliptical $\beta$--model \citep{cav76}. We
use the Sherpa package in {\sc ciao} to fit this model along with a
uniform background to the exposure-corrected 0.3--2~keV images of all
groups.  Bright point sources were identified visually and masked out
in all fitting, and only data from the central CCD were considered.
Free parameters of the model are $\beta$, $r_c$, ellipticity, position
angle, and the normalization, in addition to the local background level.

\begin{figure}
\begin{center}
\epsscale{1.17}
\plotone{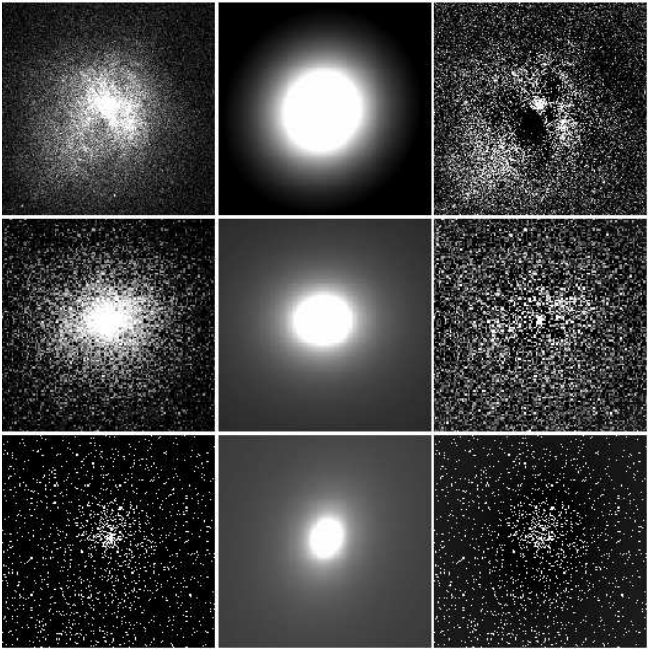}
\end{center}
\figcaption{Illustration of elliptical $\beta$--model fitting results
  for NGC\,5044 (top row), NGC\,3402 (middle row) and ESO\,552-020
  (bottom). From left to right: Input source image, fitted model, and
  residual image. \label{fit}}
\end{figure}

For each group, the best-fit model was subtracted from the input image
to produce a residual image. For a good fit, the residual image will
be flat almost everywhere (modulo Poisson fluctuations), with any
remaining structure revealing departures from the model, such as
cavities. The top panel of Figure~\ref{fit} shows an example of this,
displaying the results for NGC\,5044, the group with the largest
number of source counts and the most prominent cavities within our
sample. In this case, the two cavities are already visible in the
original image but become much more prominent in the residual image.
The middle panel of Figure~\ref{fit} shows the case of NGC\,3402, in
which cavities are not clearly detected. Here the fitted model
describes the group emission very well, and residual \lq\lq
structure\rq\rq \ can be largely ascribed to Poisson fluctuations.
Finally, images with low signal-to-noise ratio tend to have no
cavities detected, as illustrated in the bottom panel of
Figure~\ref{fit}. In this case, the relatively low number of counts
precludes any robust conclusion regarding the presence of cavities.

For some groups the model fitting fails to converge on sensible
parameters, returning, e.g., a best-fit core radius smaller than one
pixel or a model eccentricity higher than 0.99. The relevant systems
are 3C\,442A, A1139, A1238, A1692, A1992, A2092, A2462, NGC\,507,
RBS\,461, RXJ\,1206-0744, and UGC\,842. We ascribe the failure of the
fitting approach mainly to the fact that many of these data sets have
relatively low signal-to-noise ratio, coupled with the possibility
that some of those groups are simply not well described by an
elliptical $\beta$--model. We will return to these issues in
Section~\ref{sec:mock}.

\subsection{Unsharp Masking}\label{sec:masking}

\begin{figure*}
\begin{center}
\epsscale{1.0}
\plotone{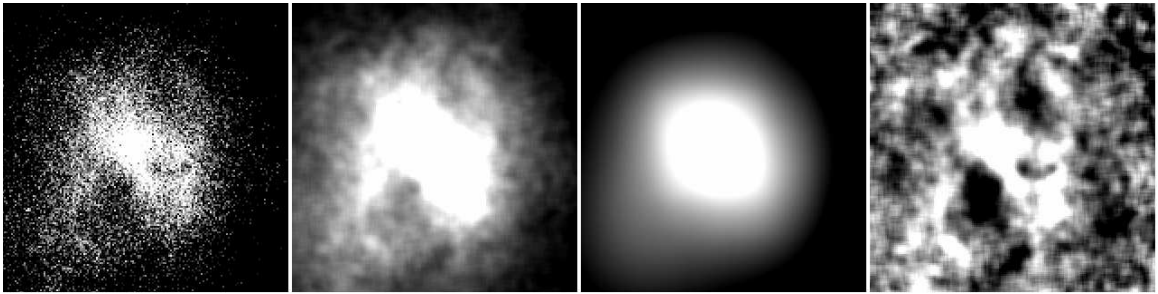}
\end{center}
\figcaption{Illustration of unsharp masking for NGC\,5044. From left
   to right: exposure-corrected image, small-scale smoothed image
   (Gaussian of $\sigma=3$~pixels), large-scale smoothed image
   ($\sigma =30$~pixels), and quotient image.
   \label{smooth}}
\end{figure*}

Another way to smooth the large-scale group emission and test for
structure on smaller scales is unsharp masking. This approach has been
successfully employed to uncover faint features at fine spatial detail
in deep cluster X-ray data (e.g., \citealt{fab06}). Here we perform
this procedure by smoothing the exposure-corrected images using the
``aconvolve'' task in {\sc ciao}. For each group, the data are first
smoothed using a wide Gaussian kernel that preserves the overall
morphology of the emission but erases small-scale structure. A
separate smoothing is performed on smaller scales to suppress (partly
noise--induced) pixel-to-pixel variations while preserving structure
on the likely scale of any cavities in the data. The latter image is
then divided by the former, with the resulting quotient image acting
as an analog of the residual image from the model fitting approach. In
principle, optimal choices for the characteristic smoothing scales
vary among the groups. In practice, however, we found that large and
small scales of 10--30 and 2--5~pixels, respectively, generally
produced the most visually compelling results, with the smaller values
generally preferred for more distant and compact systems.
Figure~\ref{smooth} illustrates this method applied to NGC\,5044. The
Figure shows the original image, the image smoothed by narrow and wide
Gaussians of $\sigma=3$ and 30~pixels, respectively, and the result of
dividing the former by the latter. The result can be directly compared
to the residual image from model fitting shown in the top row in
Figure~\ref{fit}.

\begin{figure}
\begin{center}
\epsscale{1.17} 
\plotone{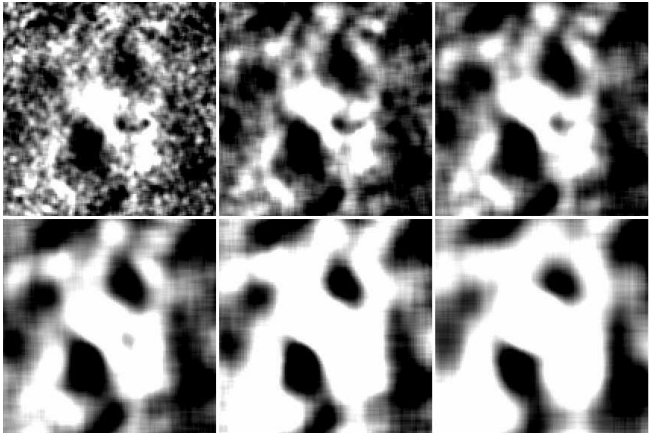}
\end{center}
\figcaption{The effect of choosing different widths of the small-scale
  Gaussian kernel in unsharp masking for NGC\,5044. From left to
  right, top to bottom: quotient images with $\sigma=2$, 4, 6, 8, 10,
  and 15 pixels (the large-scale kernel has $\sigma=30$~pixels).
  \label{smooth_scales}}
\end{figure}

To illustrate the impact of different choices for the small smoothing
scale, Figure~\ref{smooth_scales} shows the quotient image for
NGC\,5044 with six different small scales employed along with a fixed
large-scale kernel of $\sigma =30$~pixels.  All the quotient images
clearly reveal X-ray depressions at the positions of the cavities, but
it is easier to visually recognize the cavities as such in the images
produced with relatively narrow small-scale kernels: The cavities in
these images are more obvious and have a higher contrast with their
surroundings. On the other hand, when the small scale becomes larger
than 8~pixels (i.e.\ $>1/4$ of the larger scale), distinguishing the
cavities from other brightness depressions becomes more difficult as
the sharp contrast between the cavities and their immediate
surroundings is lost. However, we emphasize that NGC\,5044 is by far
the highest-flux system in our sample, and that the choice of small
smoothing scale is less straightforward for significantly fainter
systems. As such, a major drawback of this method compared with
surface brightness modeling is the need to make appropriate choices
for the smoothing scales.

\subsection{Cavity Identification and Characterization}\label{sec:detection}

Based on the images described in Sections~\ref{sec:fitting} and
\ref{sec:masking}, visual inspection for X-ray cavities was performed
for all groups. Cavities were mainly identified in the residual images
from model fitting, but the appearance of the unsmoothed
exposure-corrected images was also taken into consideration. In most
cases the impression gained from these two images agree. In the cases
for which the fitting approach did not converge on reasonable model
values (typically for sources of low signal-to-noise ratio), cavities
were identified visually based on the raw images, aided by the
quotient images from unsharp masking. For the groups judged to show
evidence of cavities, differences in signal-to-noise ratio and in the
contrast between the cavities and their surroundings led us to further
subdivide the sample, leading to the following classification scheme:

{\em Groups with certain cavities} (denoted the ``C''--sample in the
  following): Cavities are clearly detected upon visual inspection of
  the residual images, and their presence is at least indicative in
  the raw and unsharp masked images as well. These cavities always
  present a high contrast with their surroundings, including the
  presence of a bright rim.

{\em Groups with possible cavities} (``P''--sample), fulfilling one of
  two conditions: (1) Cavities are apparent in only one of the
  residual and raw images. (2) Both the residual image and the raw image
  show a hint, if not conclusive, of the presence of
  cavities. Cavities in this category always have a low contrast with
  their surroundings.

{\em Groups with no cavities} (``N''--sample): Neither raw nor
  residual images show any visually obvious evidence of cavities. This
  category also includes cases in which the surface brightness
  depression in the residual image forms a ring around the group
  center. This is likely an artifact of the fitting procedure, owing
  to the presence of a strong central and extended excess above the
  best-fit $\beta$--model.

We emphasize that cavity detection was done solely on the basis of
X-ray data, and that the presence or morphology of central radio
emission in a group was not considered for this purpose. Furthermore,
as indicated by comparison with other existing studies (see
Section~\ref{sec:result}), we have likely been conservative in our
classification. It is, therefore, conceivable that groups in the
P--sample may generally have a high probability of containing
cavities.

In addition to identifying cavities we also provide quantitative
estimates of their properties. In each case, the cavity center was
first defined as the centroid of the X-ray surface brightness
depression. In a few cases where the deepest depression point of the
cavity in the residual image from model fitting differed significantly
from the apparent geometrical center, we considered the cavity center
to be the midpoint between the two locations. The location of cavity
boundaries were identified as sharp drops in X-ray surface brightness,
equivalent to where the pixel counts in the residual images become
negative. On this basis, the cavities can be generally viewed as
ellipses, with a major axis $a$ in the tangential direction
(perpendicular to the line joining the cavity with the group center)
and a minor axis $b$ in the radial direction.  The cavity distance $D$
was defined as the distance between the cavity center and the group
center, with the latter defined as the location of the central peak of
the extended X-ray emission. Since the accuracy of these somewhat
subjective measurements generally depends in an unquantifiable manner
on factors such as the signal-to-noise ratio of the data, we do not
provide uncertainties on these results, and we stress that they should
all be considered approximate. We do note, however, that the inferred
properties of the cavities in the P-sample are generally more
uncertain than those of the C-- sample.

\begin{figure*}
\includegraphics[width=0.49\textwidth]{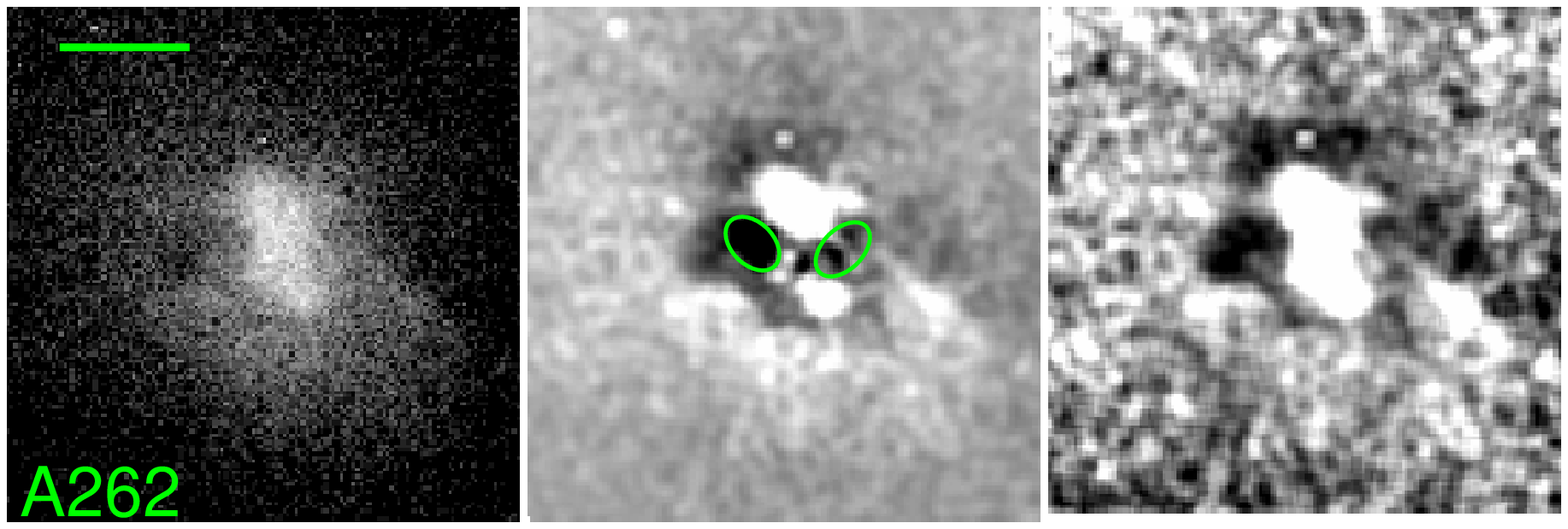}\hfill
\includegraphics[width=0.49\textwidth]{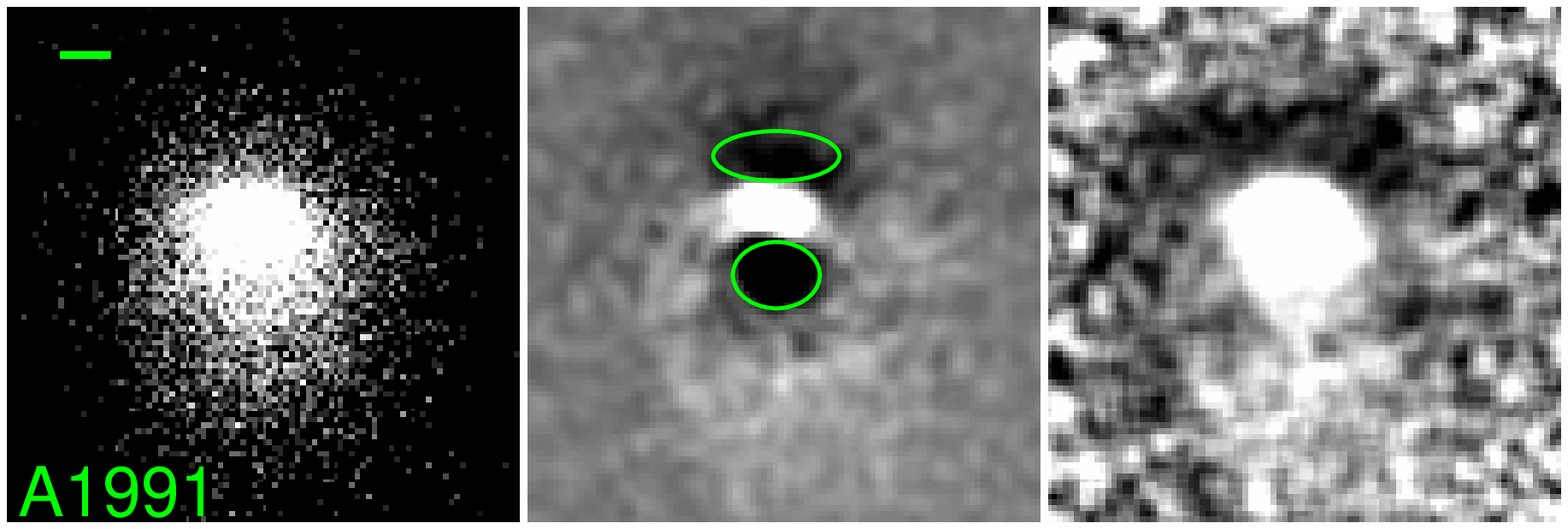}\\
\includegraphics[width=0.49\textwidth]{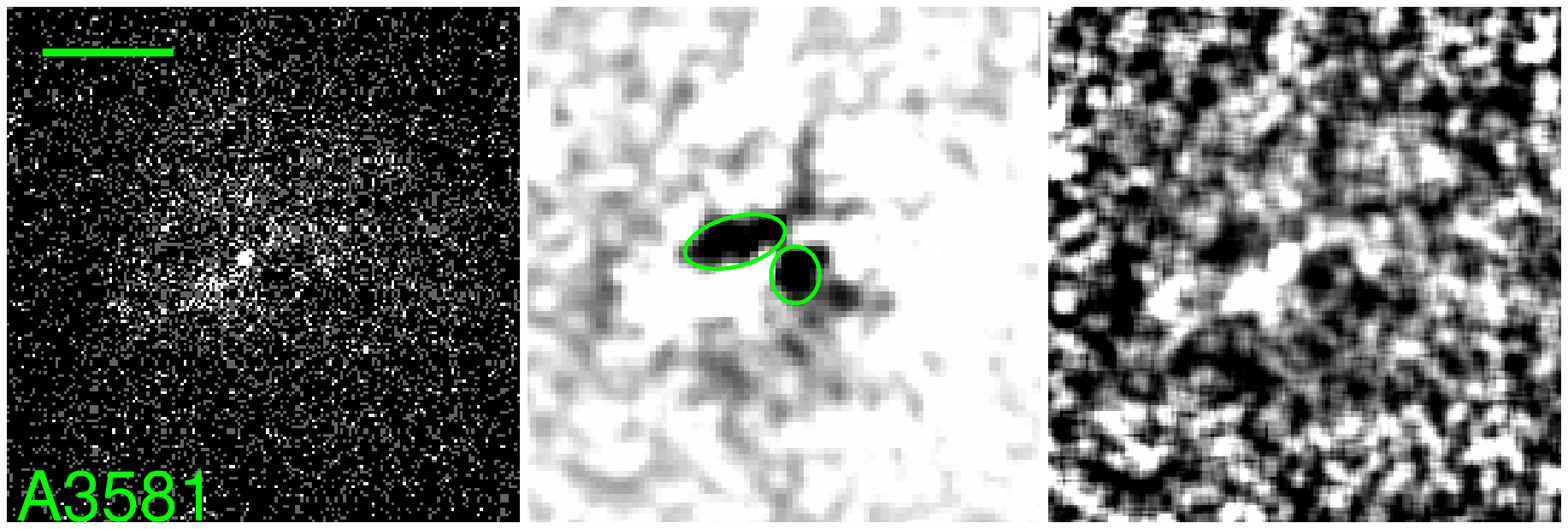}\hfill
\includegraphics[width=0.49\textwidth]{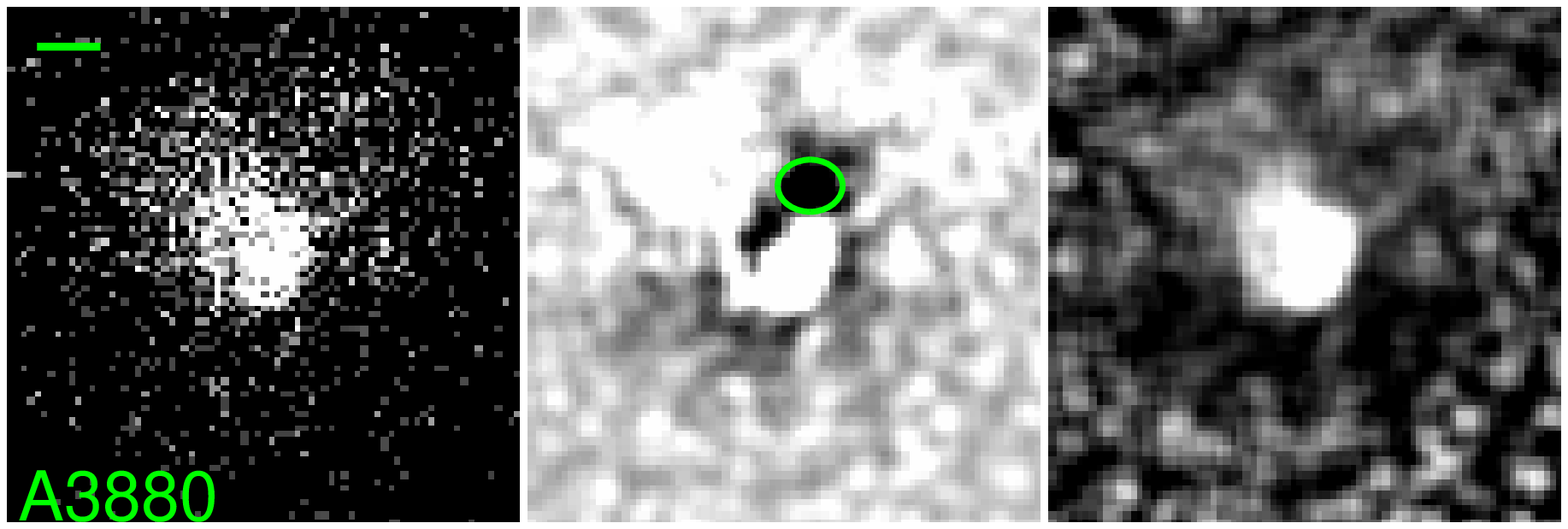}\\
\includegraphics[width=0.49\textwidth]{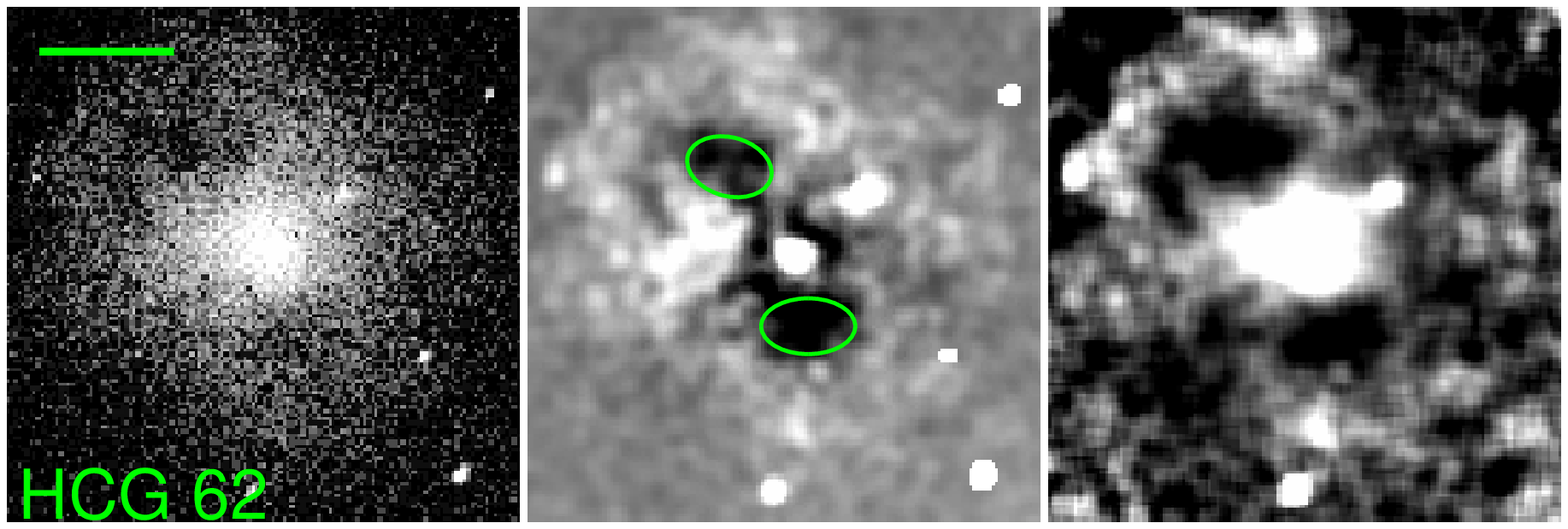}\hfill
\includegraphics[width=0.49\textwidth]{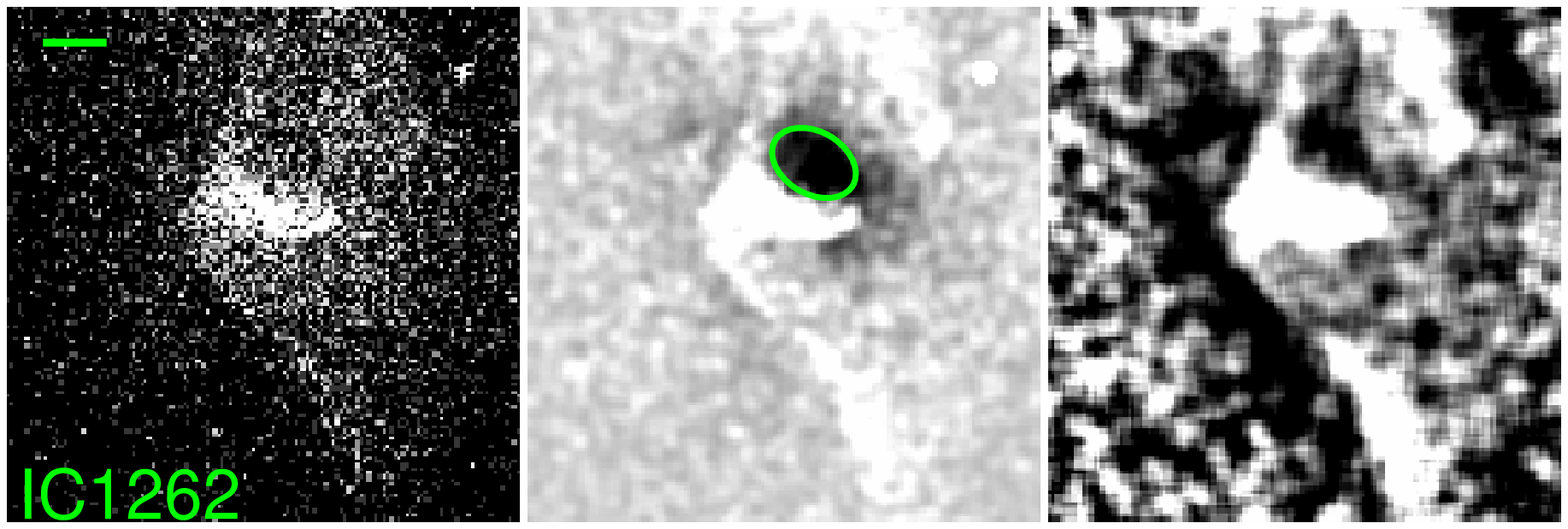}\\
\includegraphics[width=0.49\textwidth]{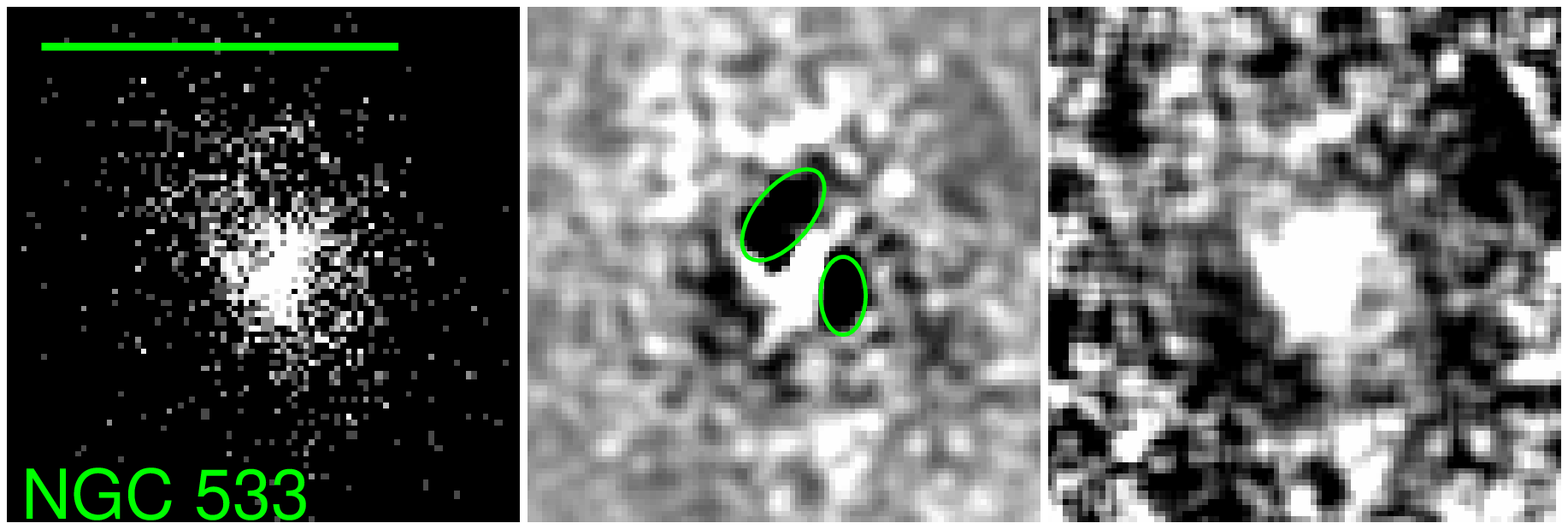}\hfill
\includegraphics[width=0.49\textwidth]{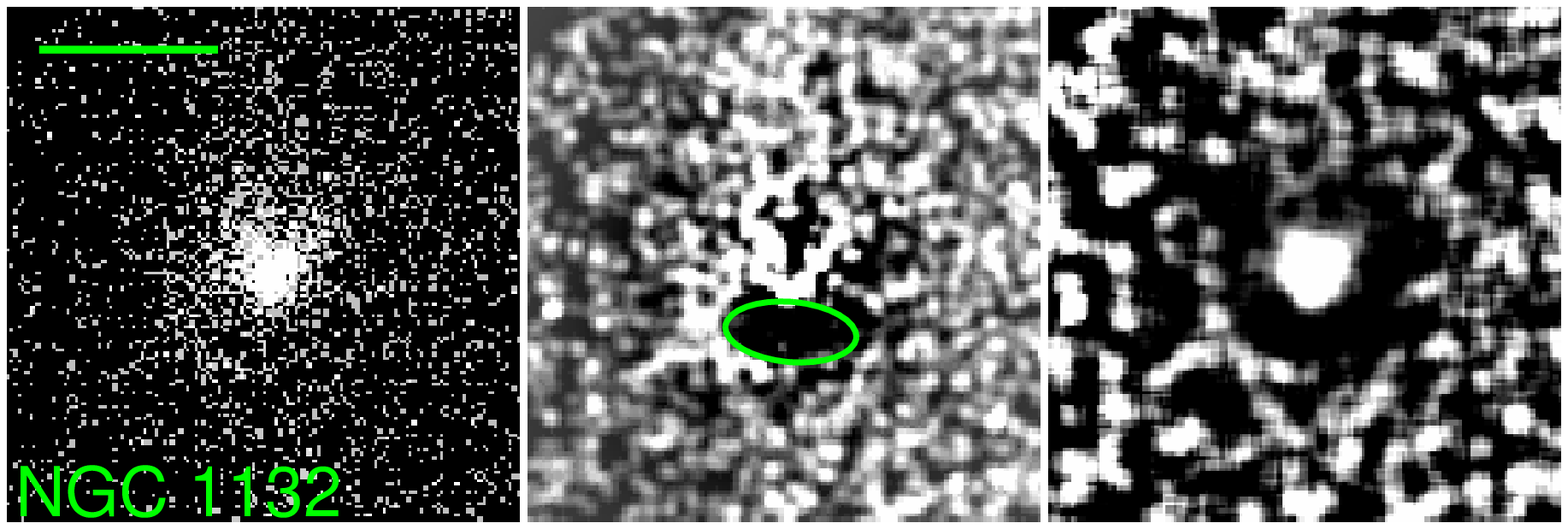}\\
\includegraphics[width=0.49\textwidth]{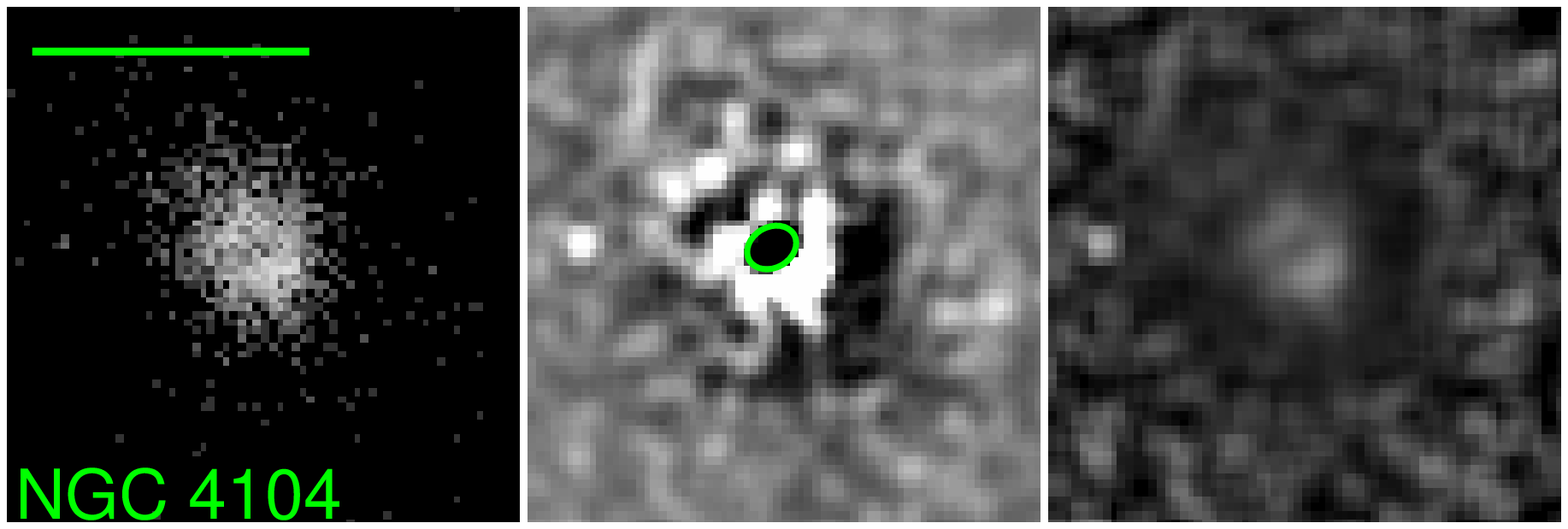}\hfill
\includegraphics[width=0.49\textwidth]{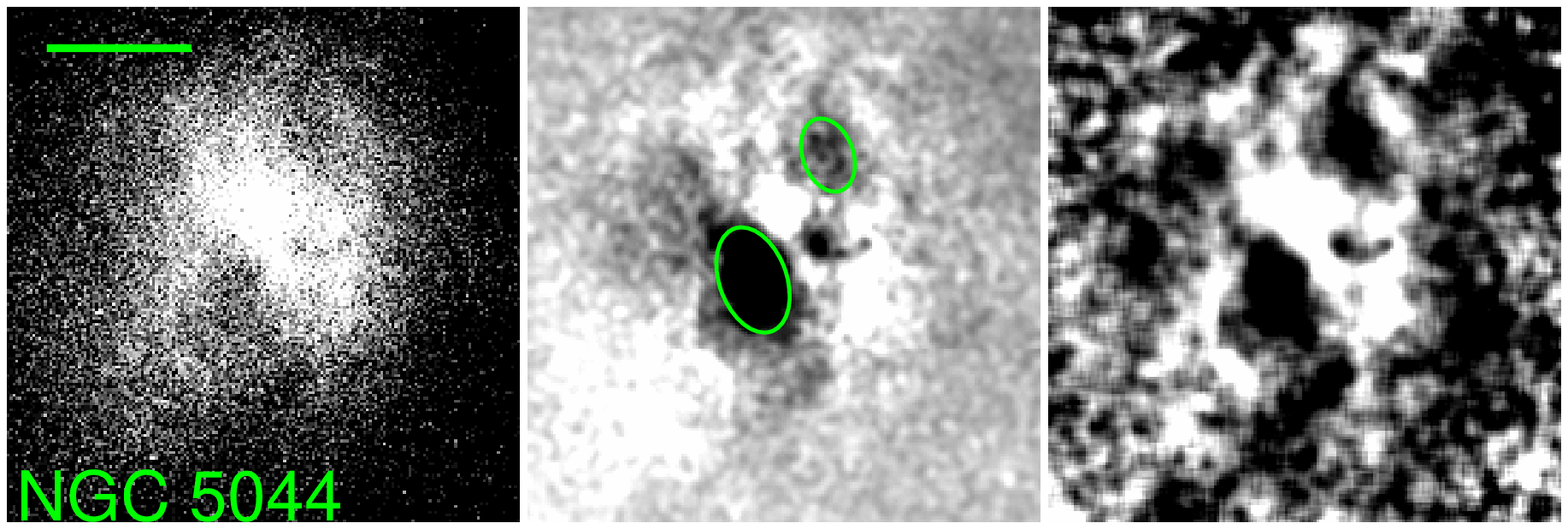}\\
\includegraphics[width=0.49\textwidth]{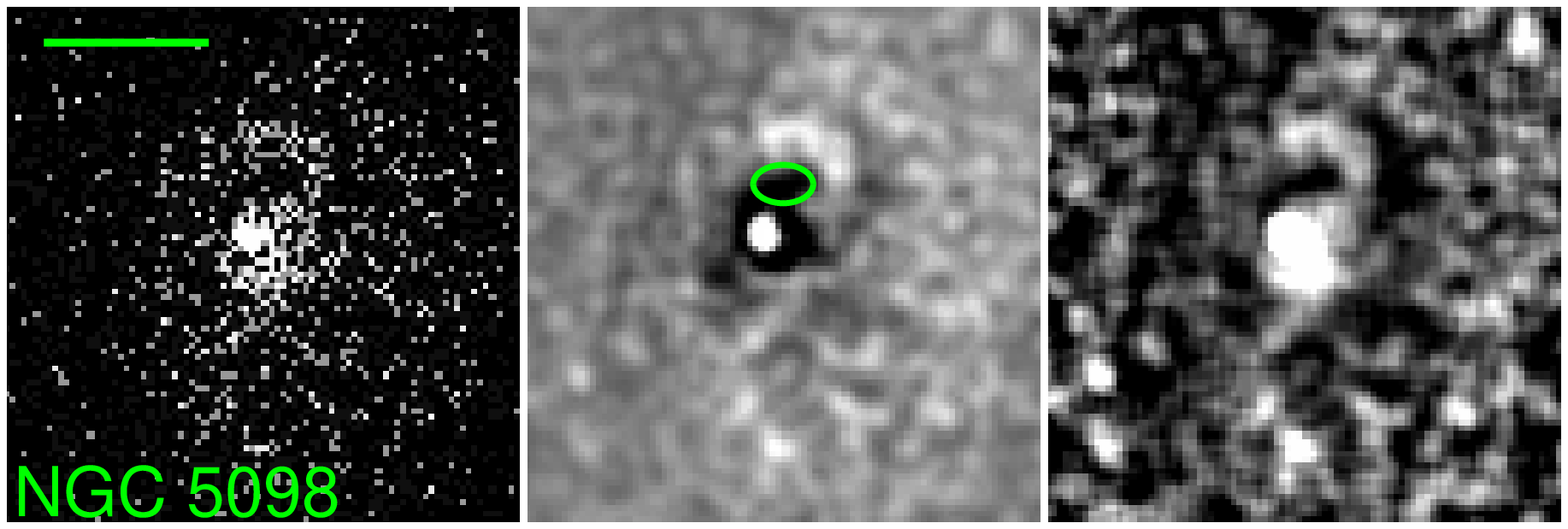}\hfill
\includegraphics[width=0.49\textwidth]{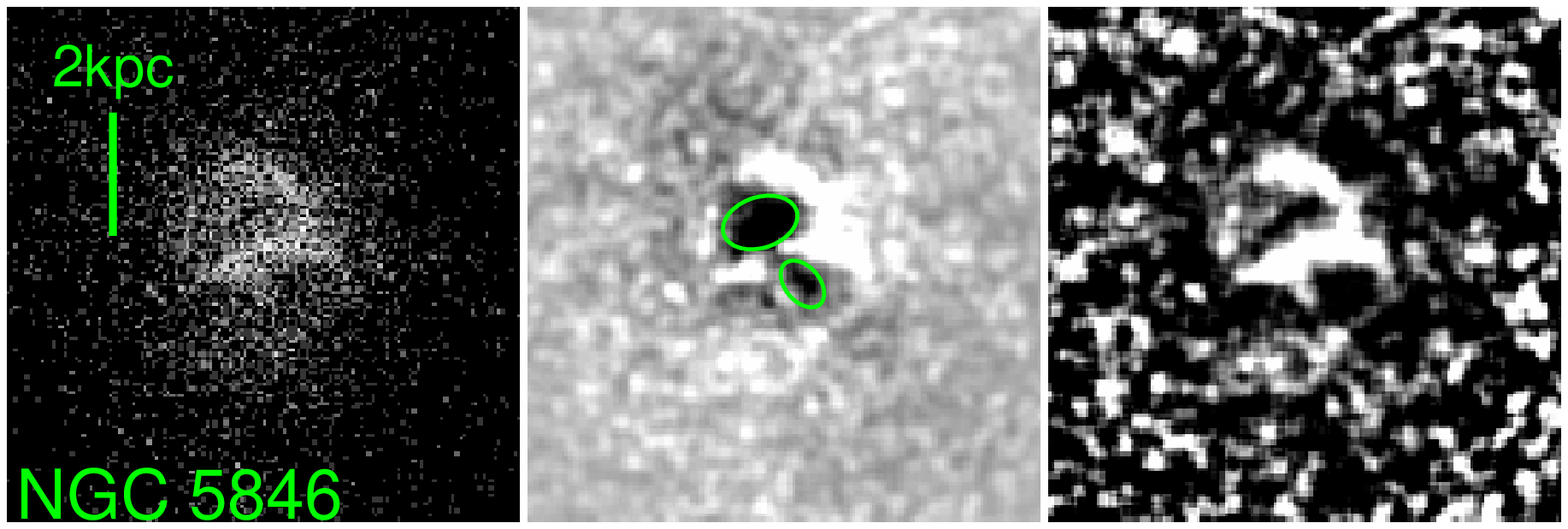}\\
\includegraphics[width=0.49\textwidth]{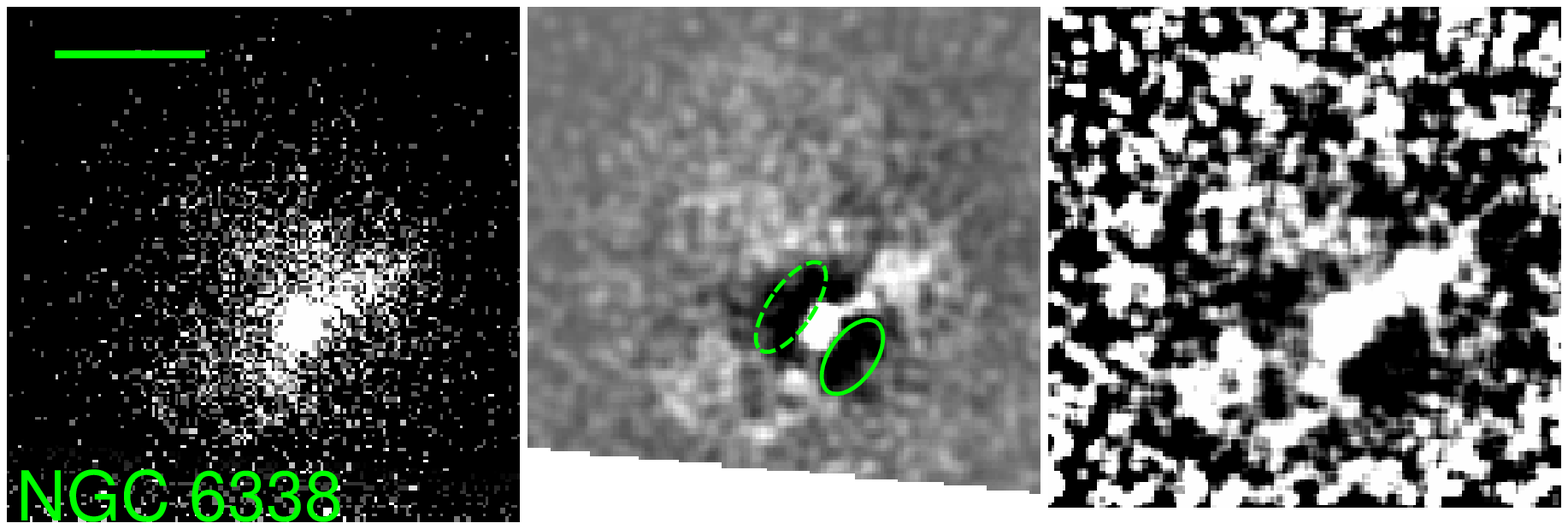}
\figcaption{The 13 groups in the C--sample with clearly identified
  cavities. For each group, left panel shows the exposure-corrected
  image, middle panel the residual image from model fitting smoothed
  by a Gaussian of $\sigma= 2$ or 3~pixels, and right panel the
  quotient image from unsharp masking. Ellipses overlaid on the
  residual images outline identified cavities. Horizontal bars mark a
  physical scale of 10~kpc. Where appropriate, vertical bars are used
  to indicate a different scale.
  \label{certain_cavity}}
\end{figure*}

\begin{figure*}
\includegraphics[width=0.49\textwidth]{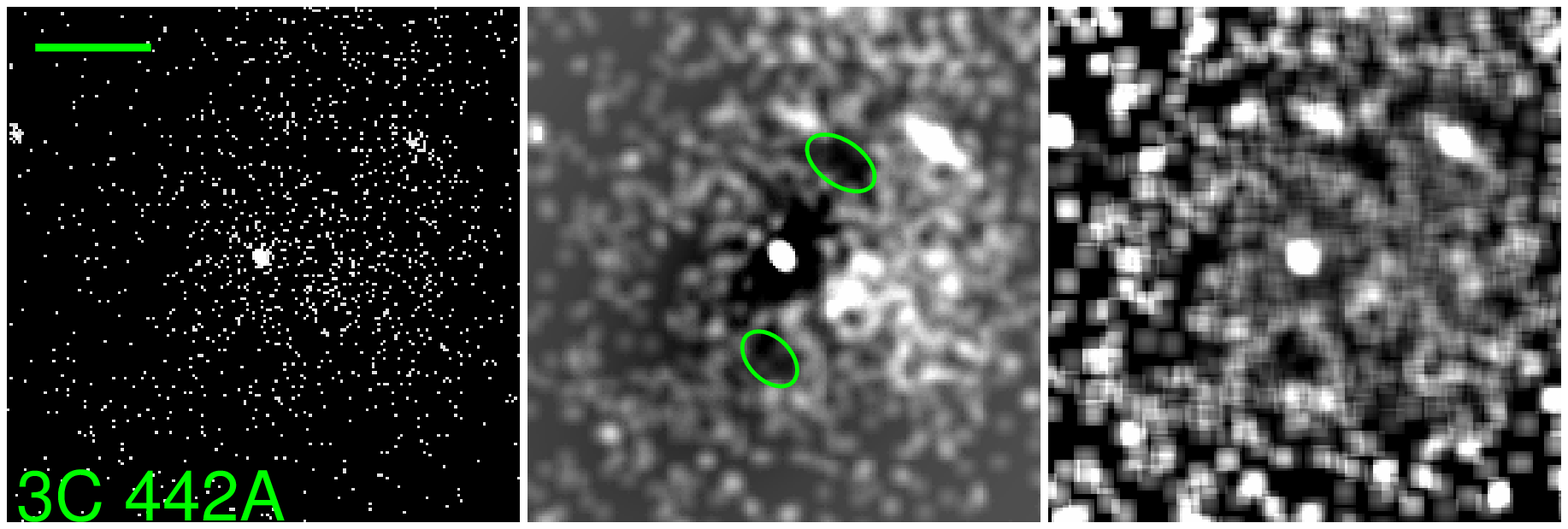}\hfill
\includegraphics[width=0.49\textwidth]{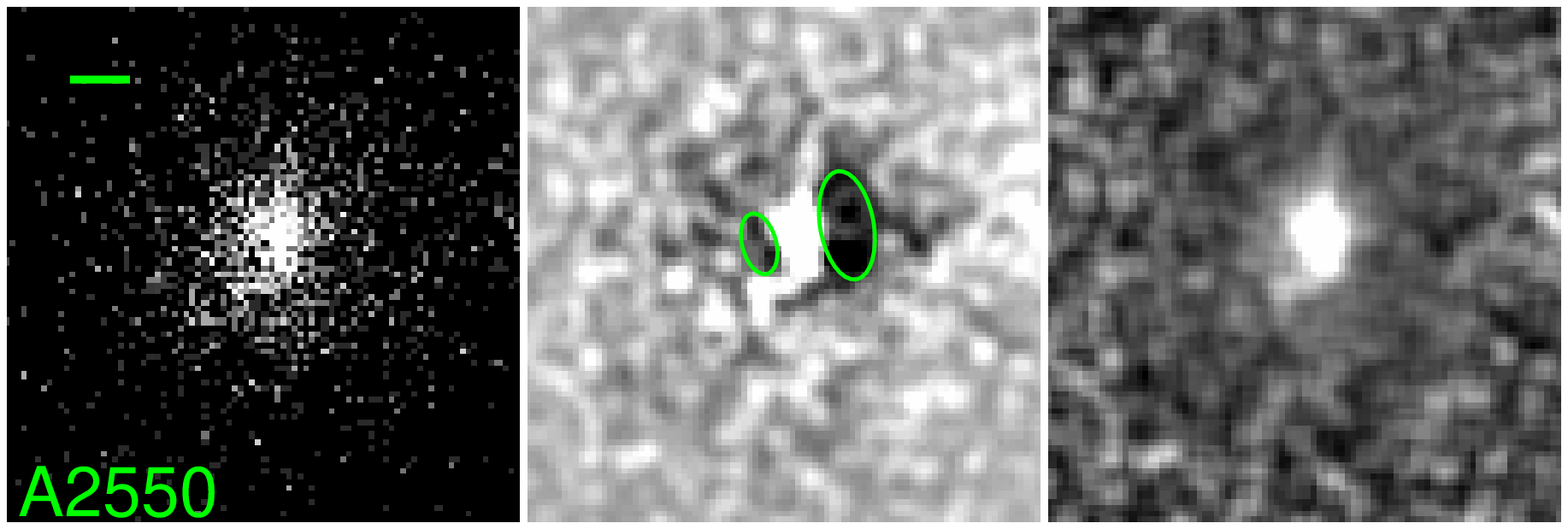}\\
\includegraphics[width=0.49\textwidth]{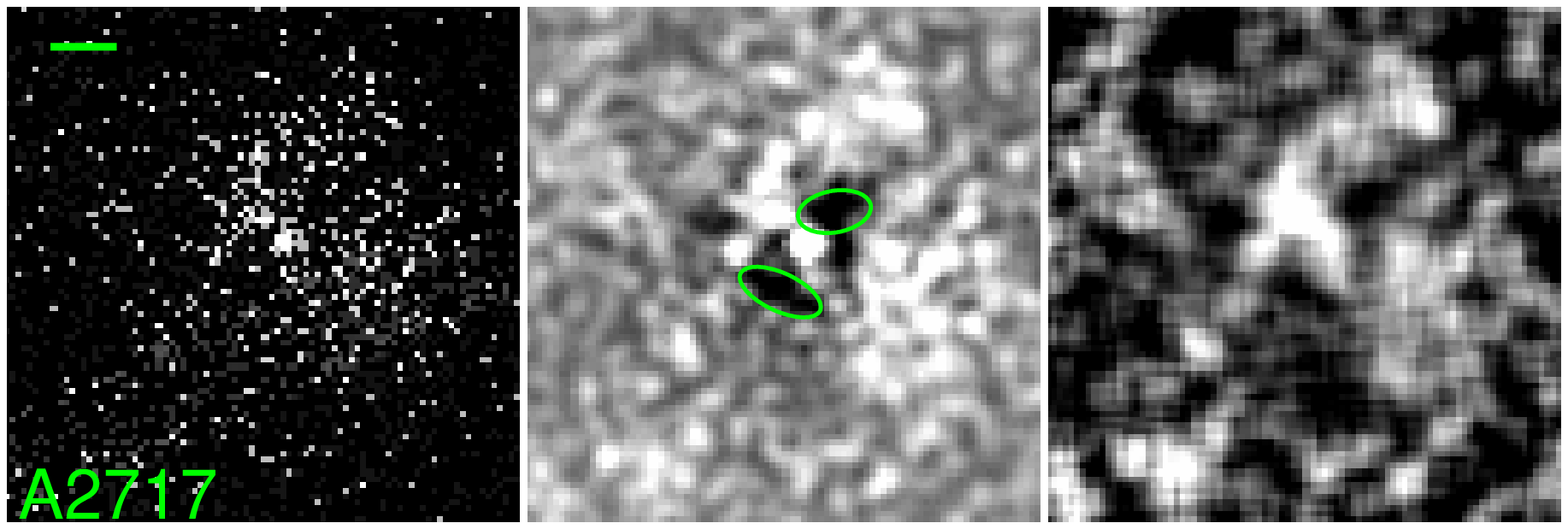}\hfill
\includegraphics[width=0.49\textwidth]{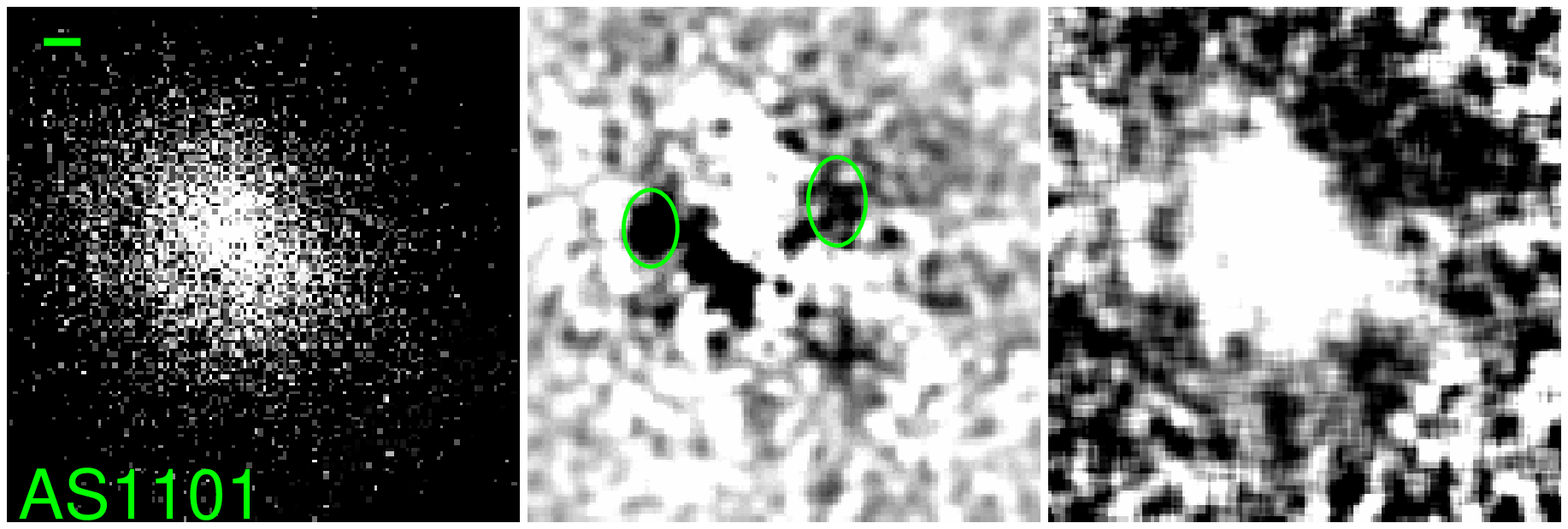}\\
\includegraphics[width=0.49\textwidth]{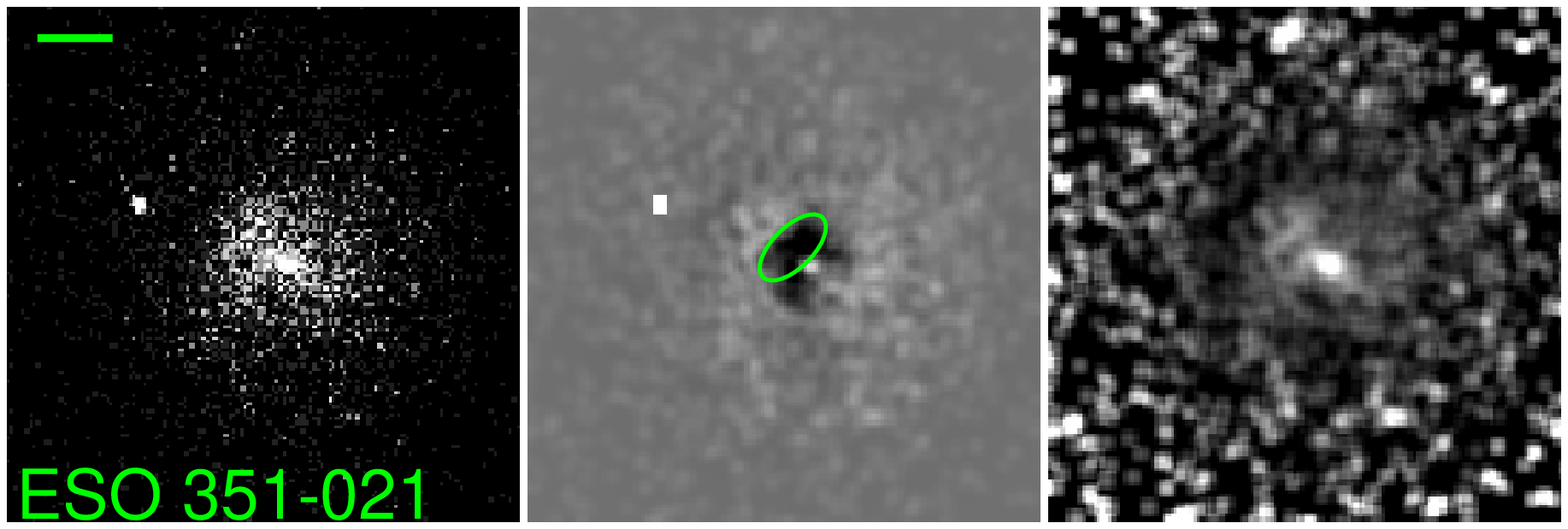}\hfill
\includegraphics[width=0.49\textwidth]{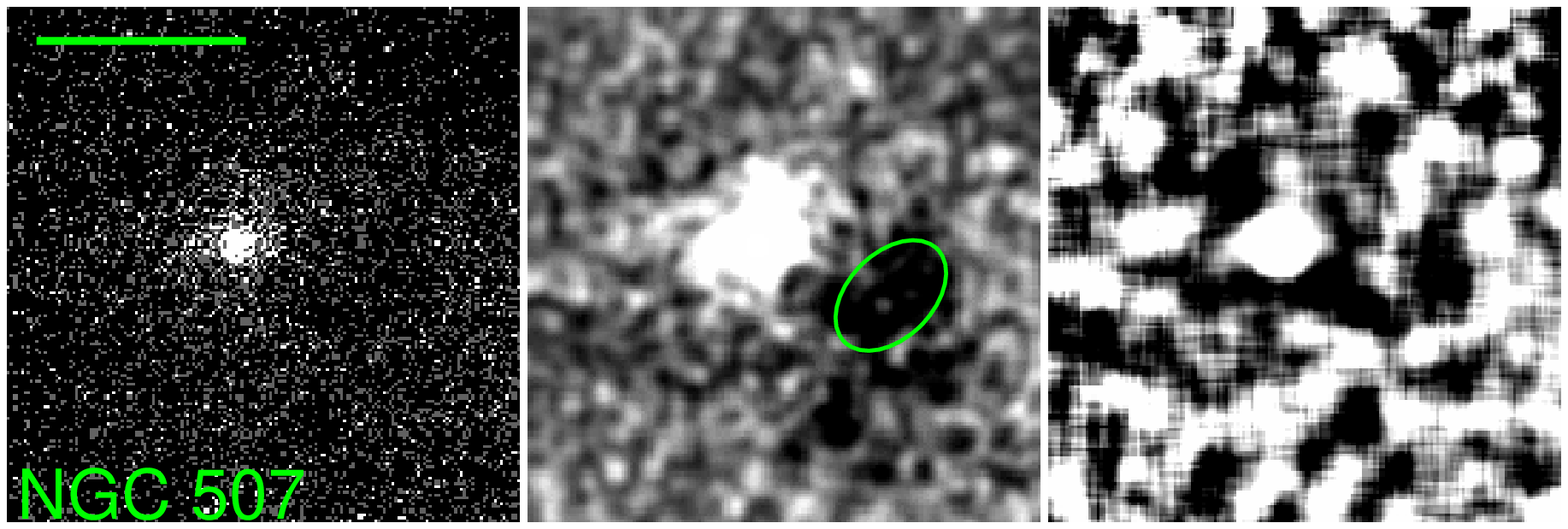}\\
\includegraphics[width=0.49\textwidth]{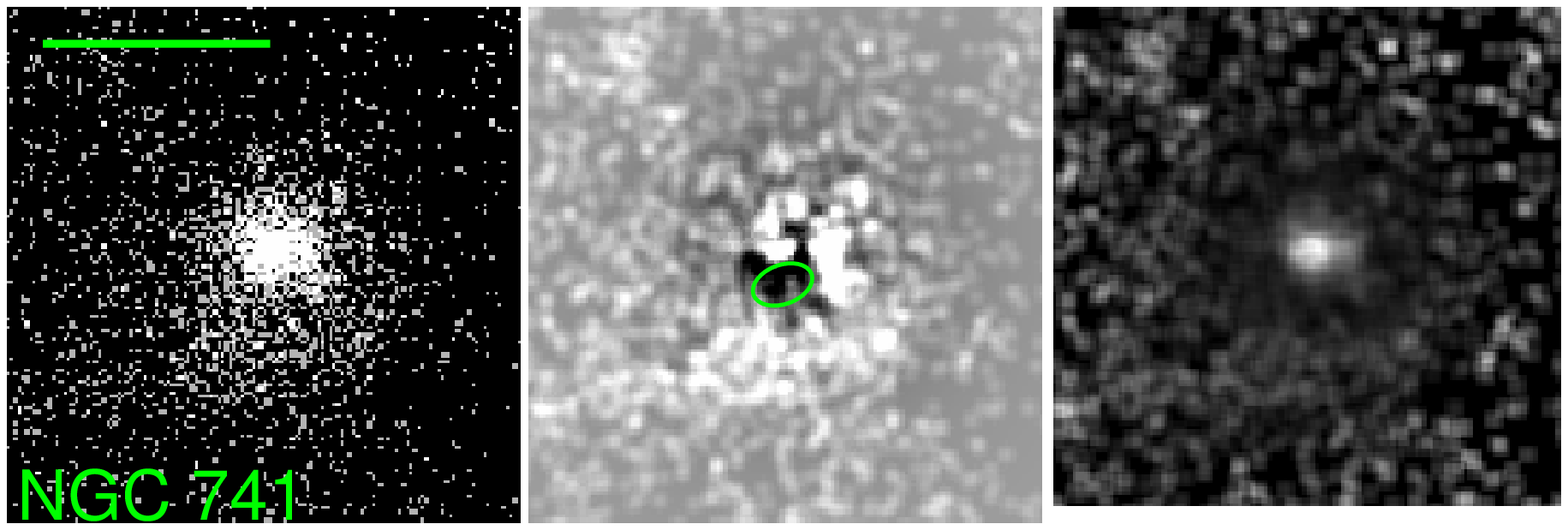}\hfill
\includegraphics[width=0.49\textwidth]{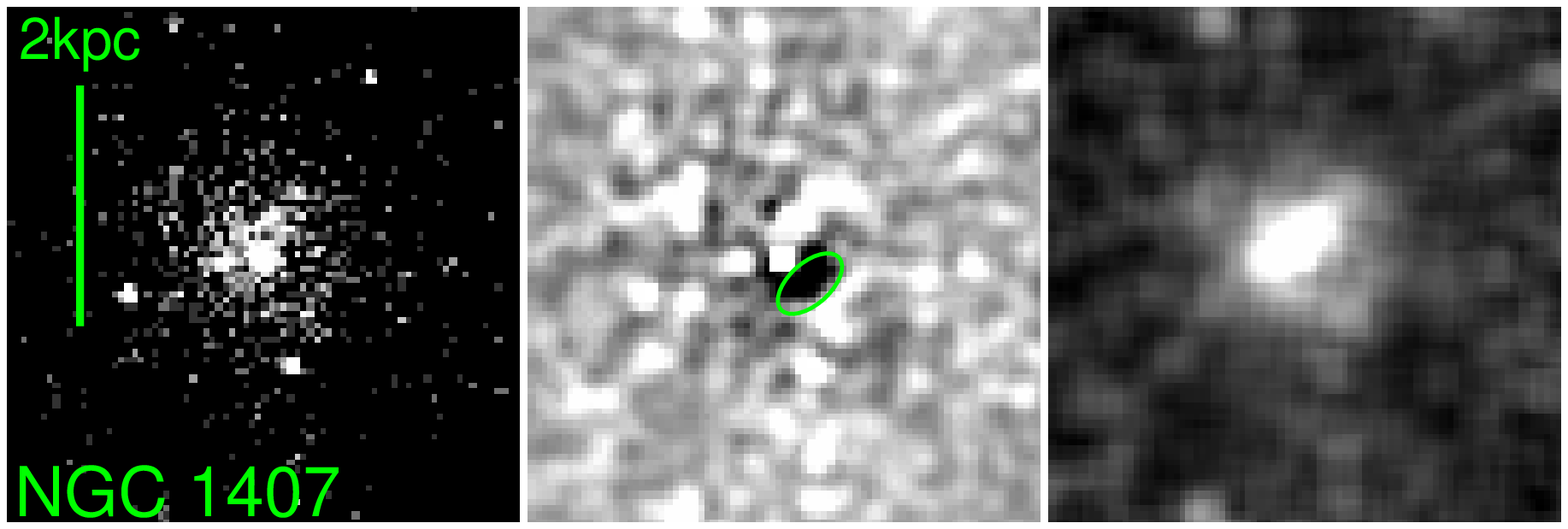}\\
\includegraphics[width=0.49\textwidth]{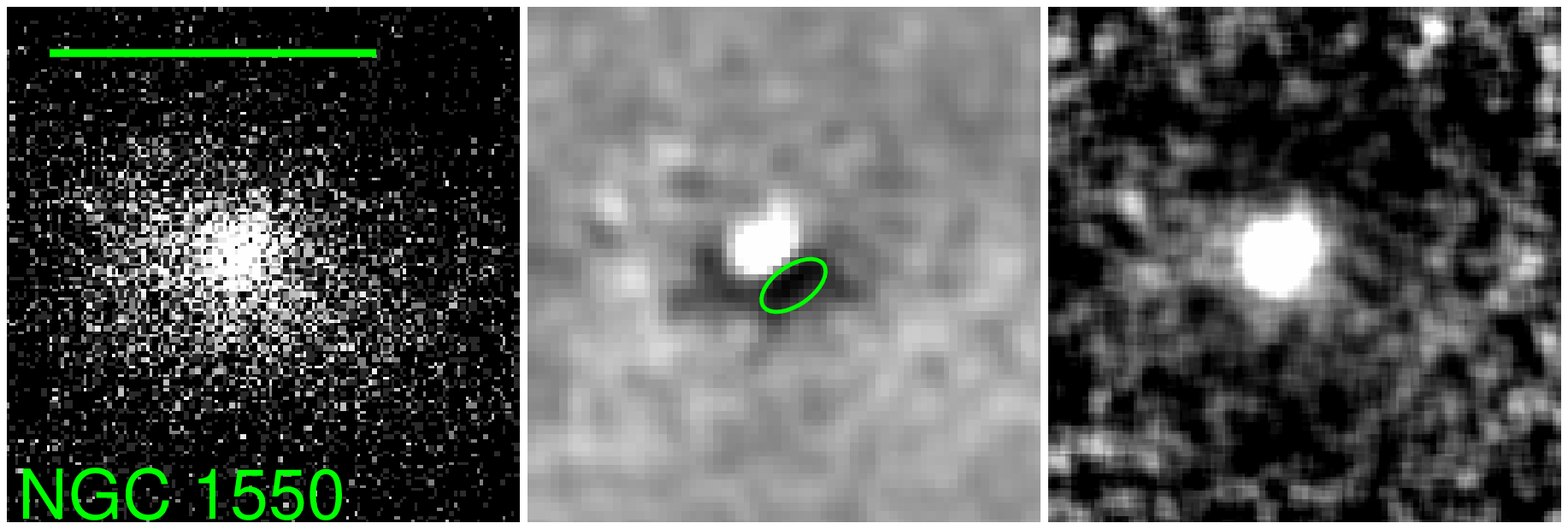}\hfill
\includegraphics[width=0.49\textwidth]{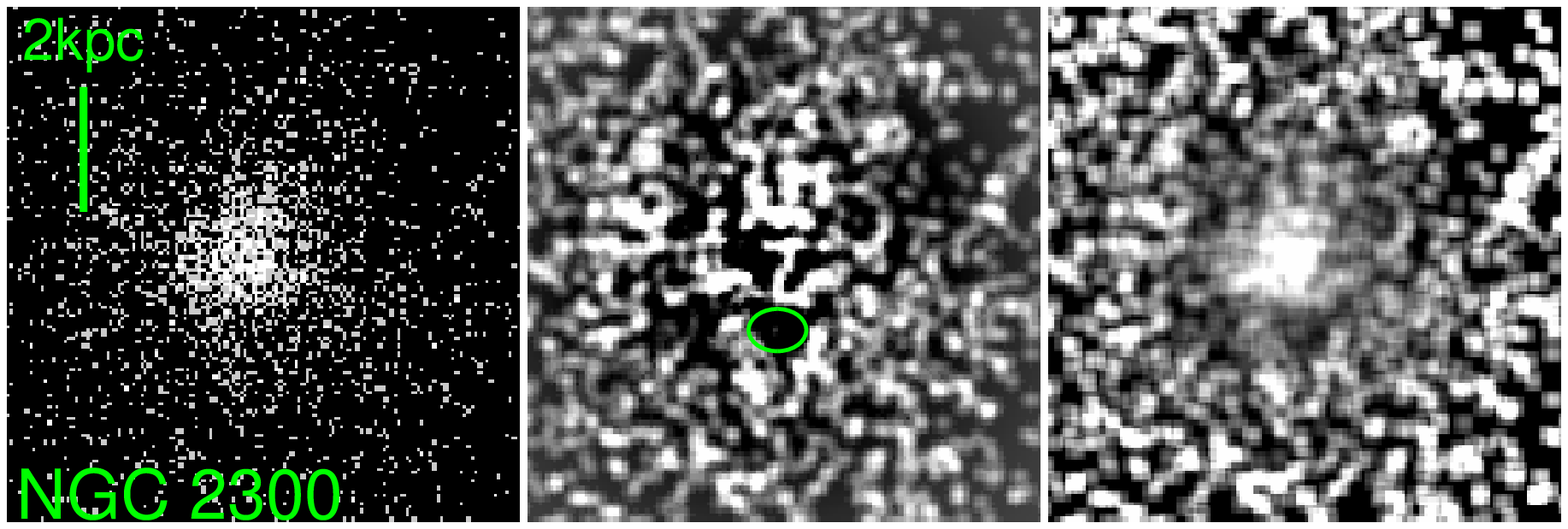}\\
\includegraphics[width=0.49\textwidth]{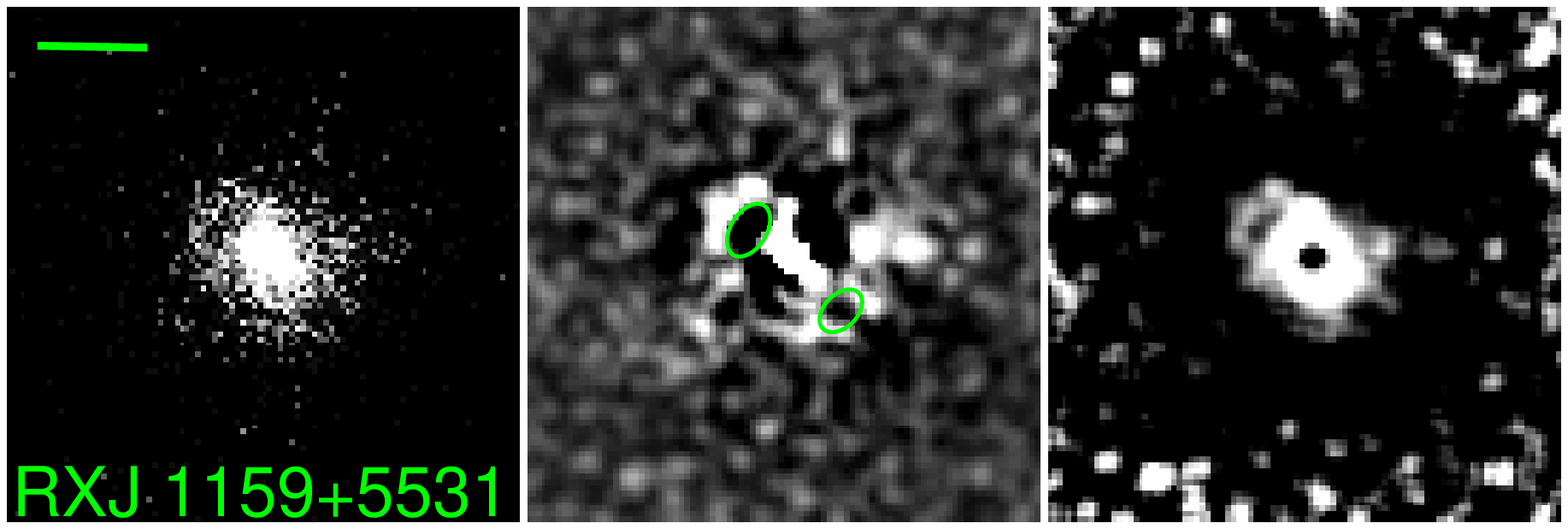}\hfill
\includegraphics[width=0.49\textwidth]{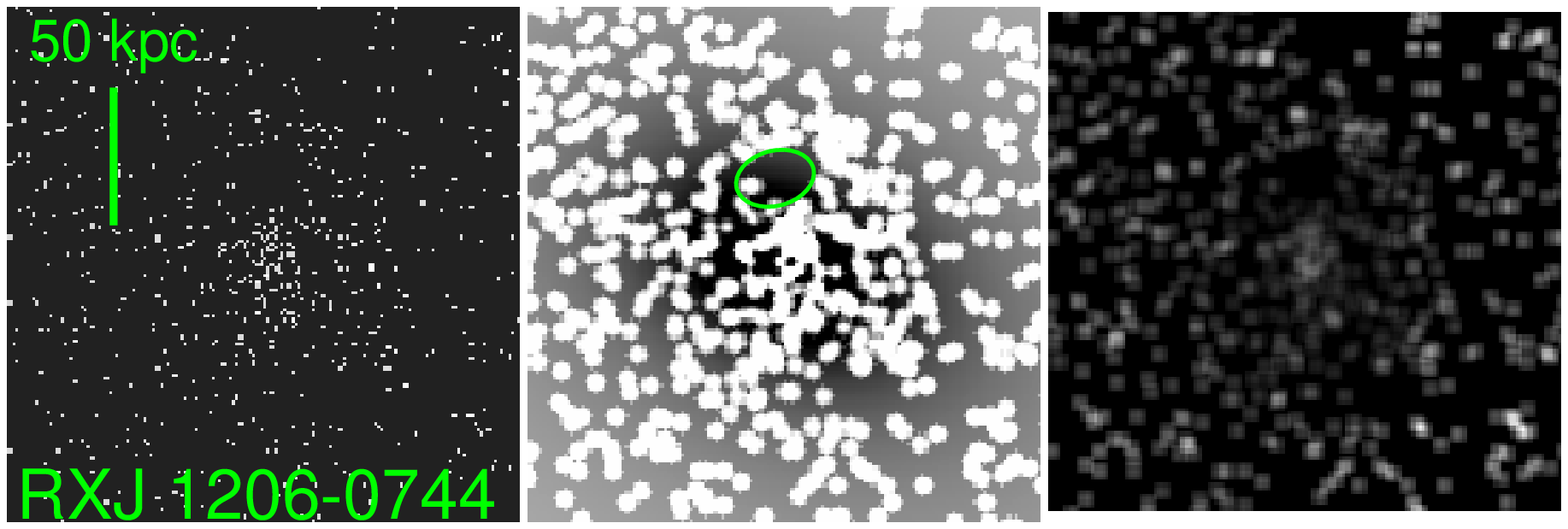}\\
\includegraphics[width=0.49\textwidth]{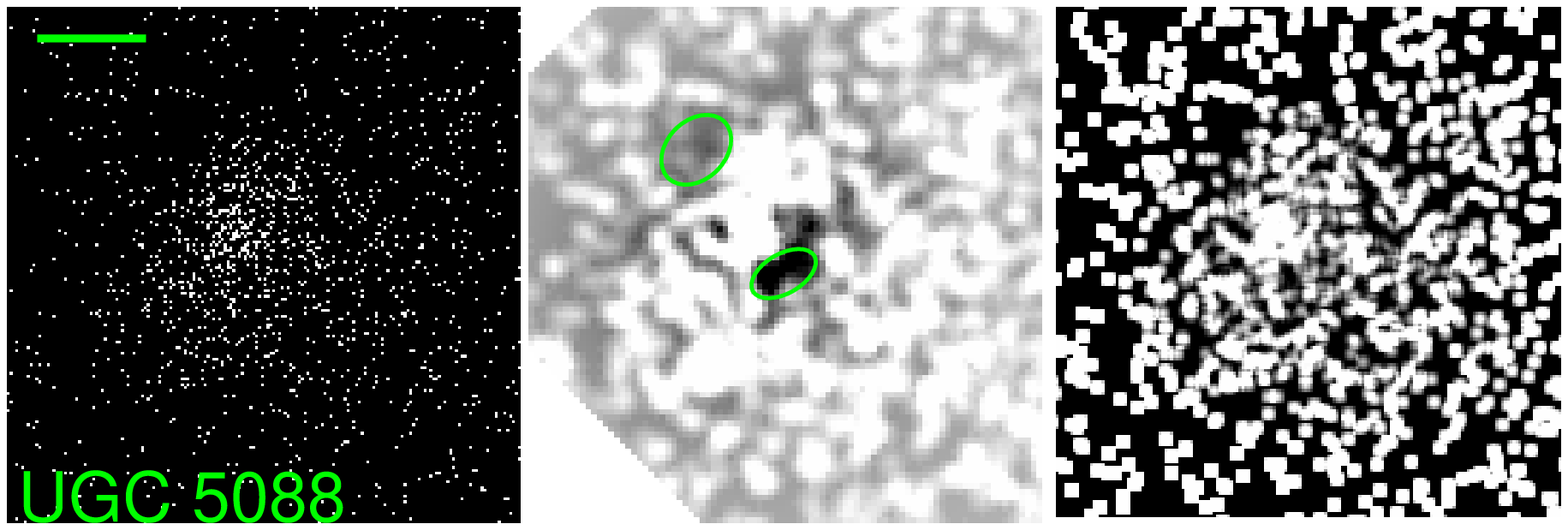}
\figcaption{As Figure~\ref{certain_cavity}, but for the 13 groups in
   the P--sample with tentative evidence of cavities.
   \label{possible_cavity}}
\end{figure*}

\begin{figure*}
\includegraphics[width=0.33\textwidth]{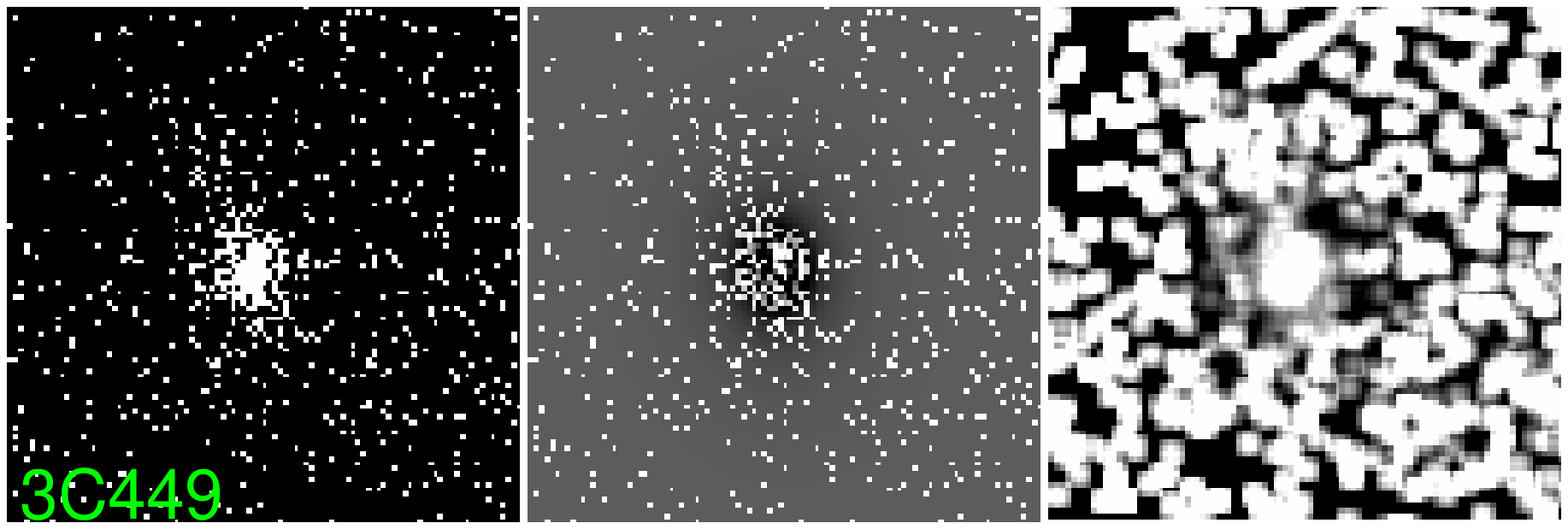}\hfill
\includegraphics[width=0.33\textwidth]{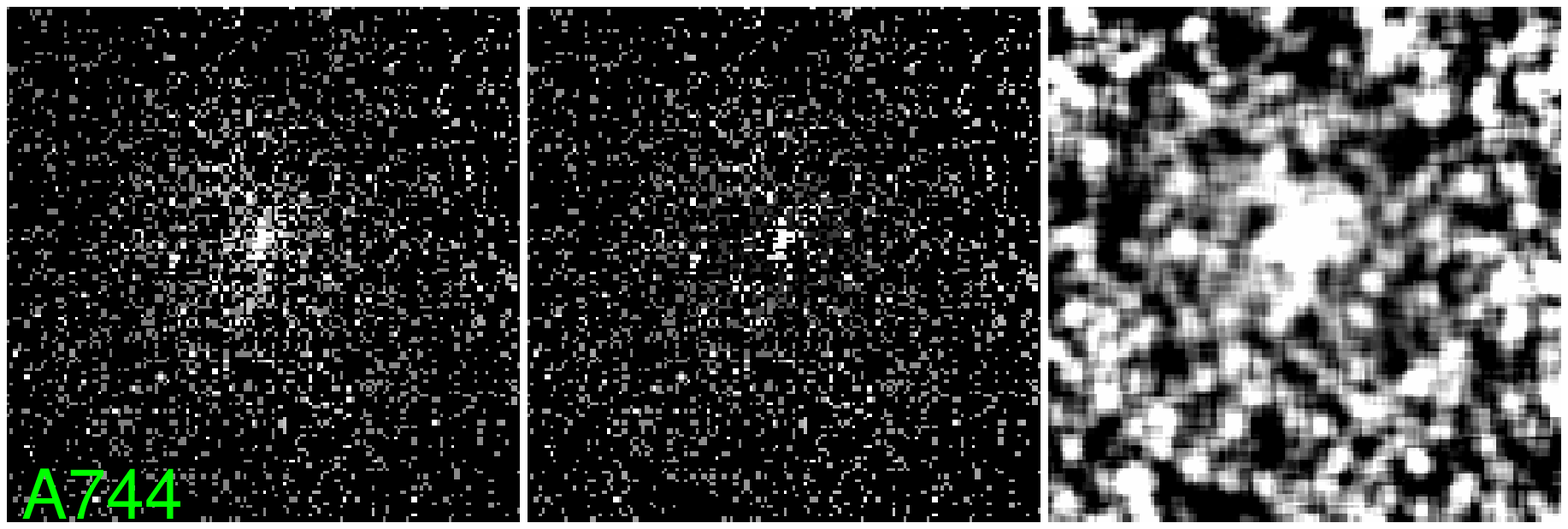}\hfill
\includegraphics[width=0.33\textwidth]{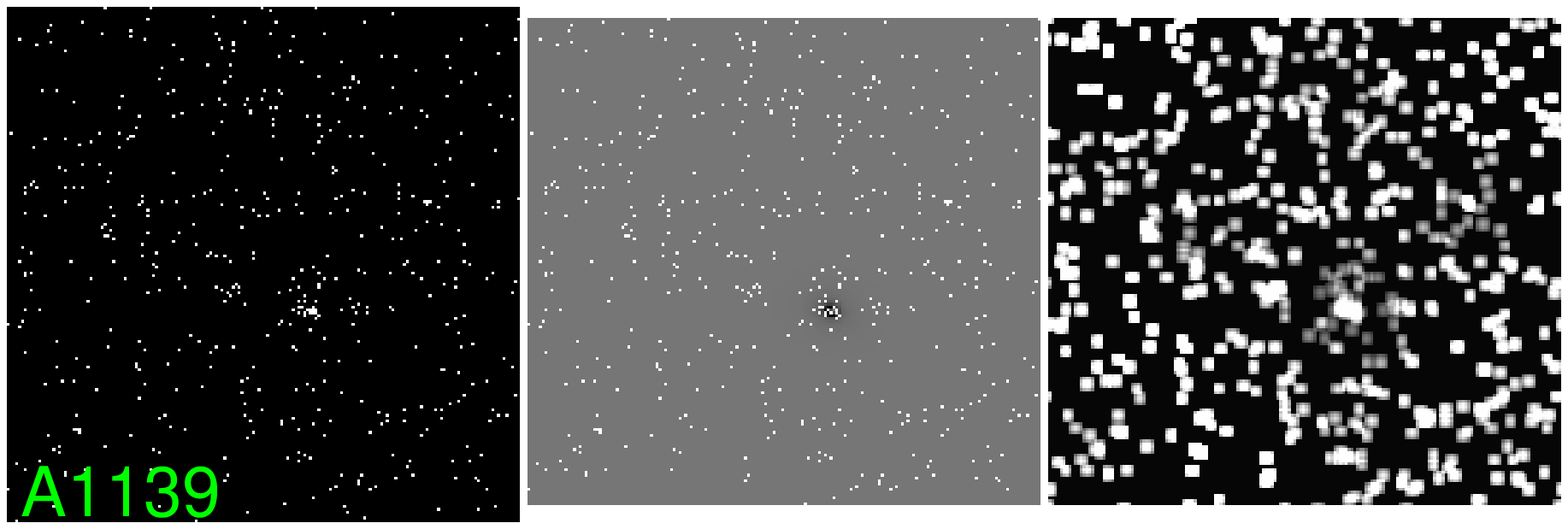}\\ 
\includegraphics[width=0.33\textwidth]{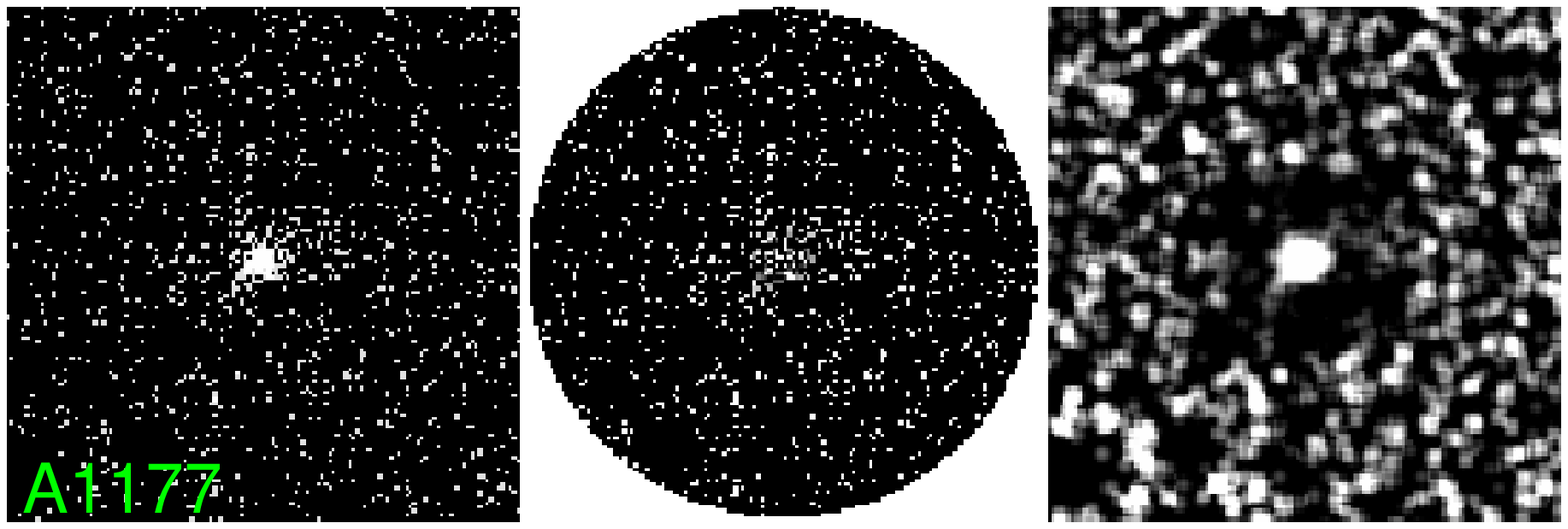}\hfill
\includegraphics[width=0.33\textwidth]{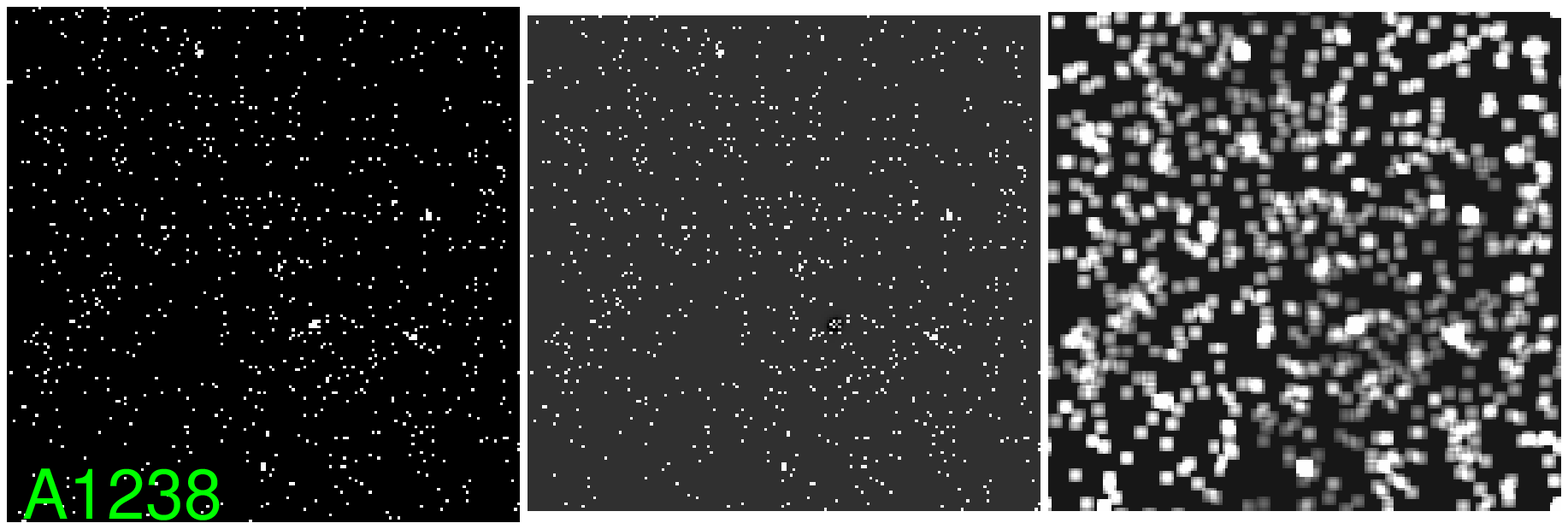}\hfill
\includegraphics[width=0.33\textwidth]{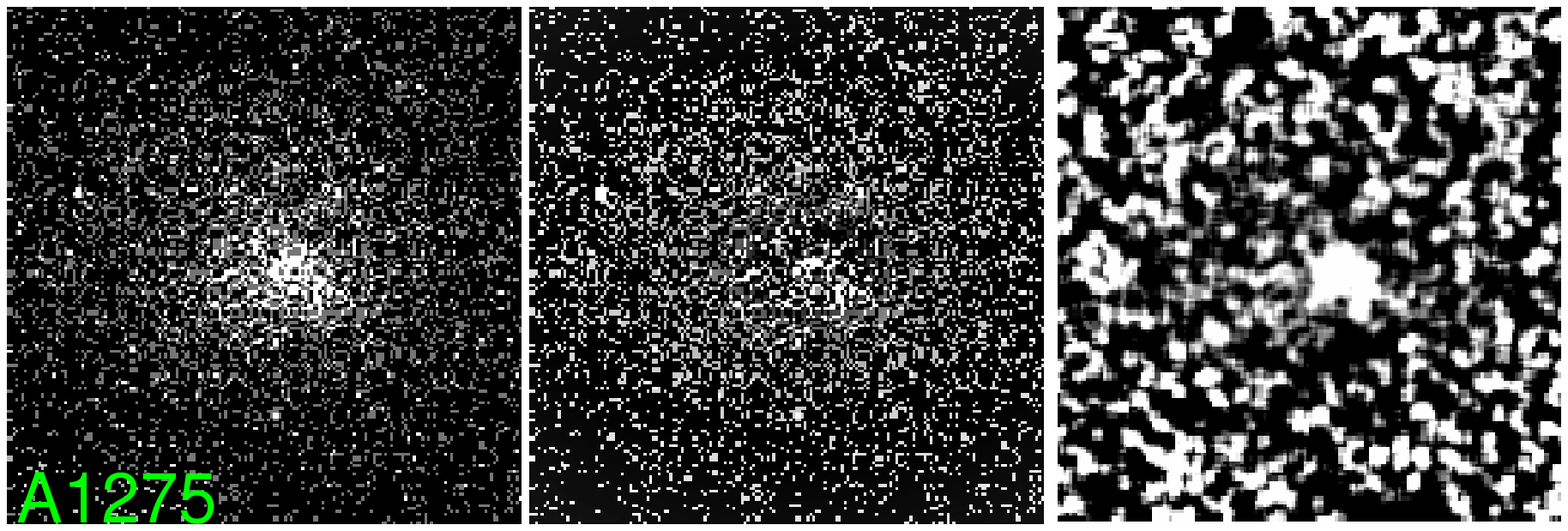}\\
\includegraphics[width=0.33\textwidth]{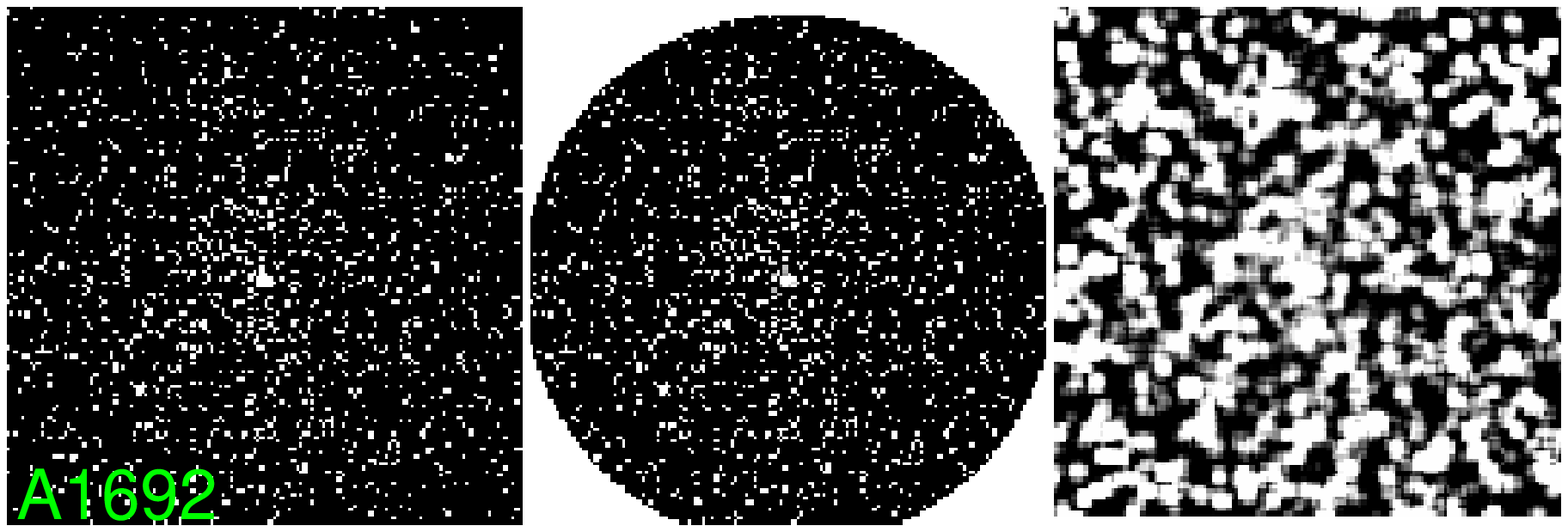}\hfill
\includegraphics[width=0.33\textwidth]{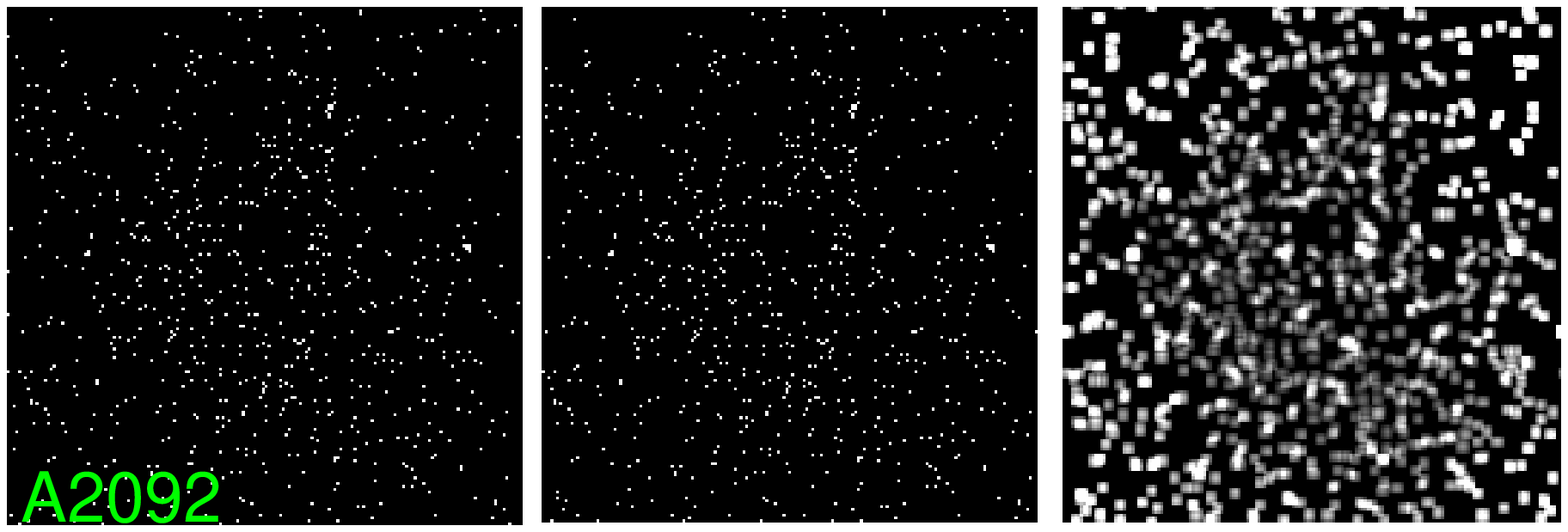}\hfill
\includegraphics[width=0.33\textwidth]{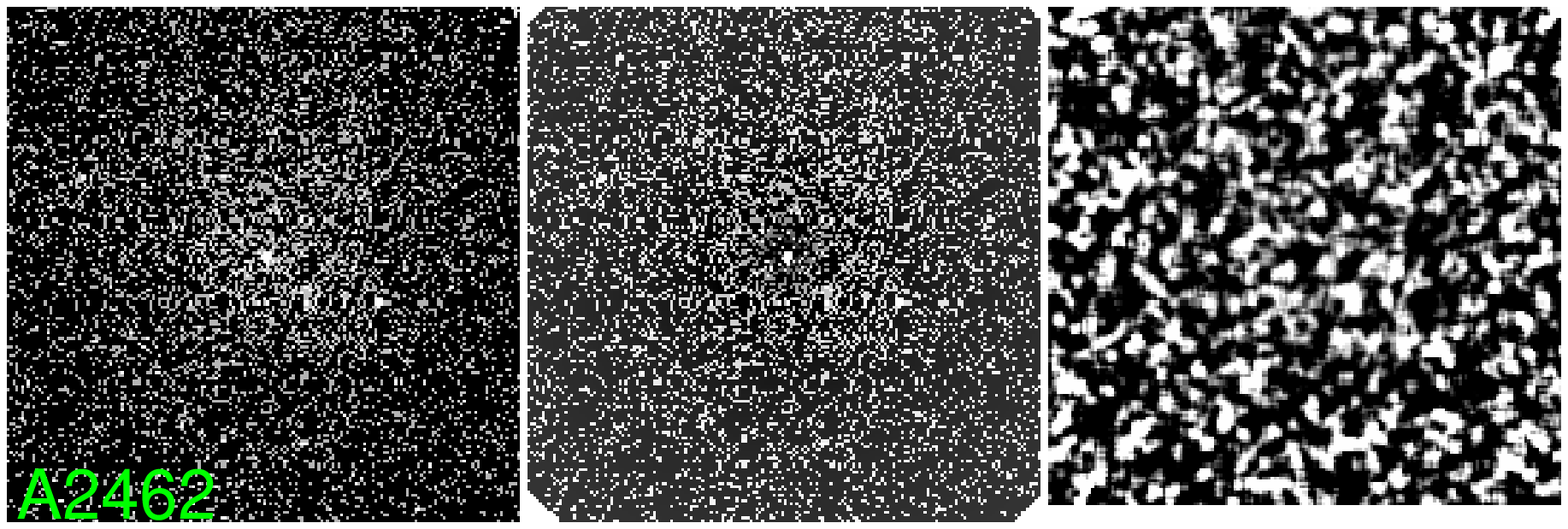}\\ 
\includegraphics[width=0.33\textwidth]{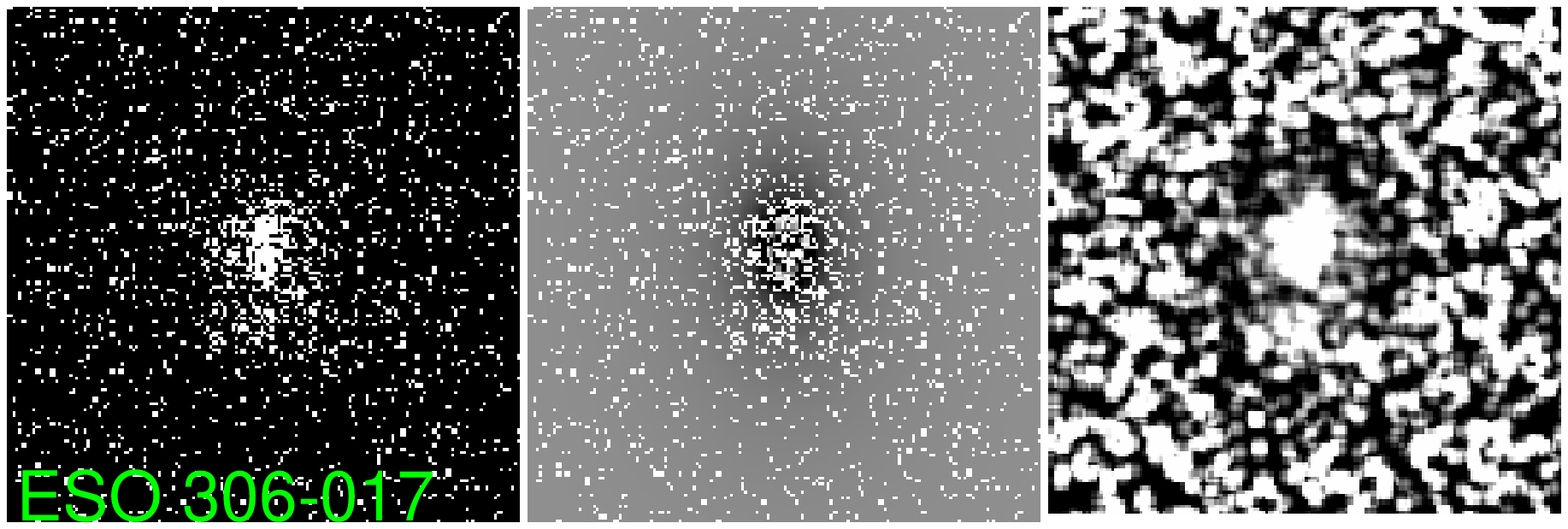}\hfill
\includegraphics[width=0.33\textwidth]{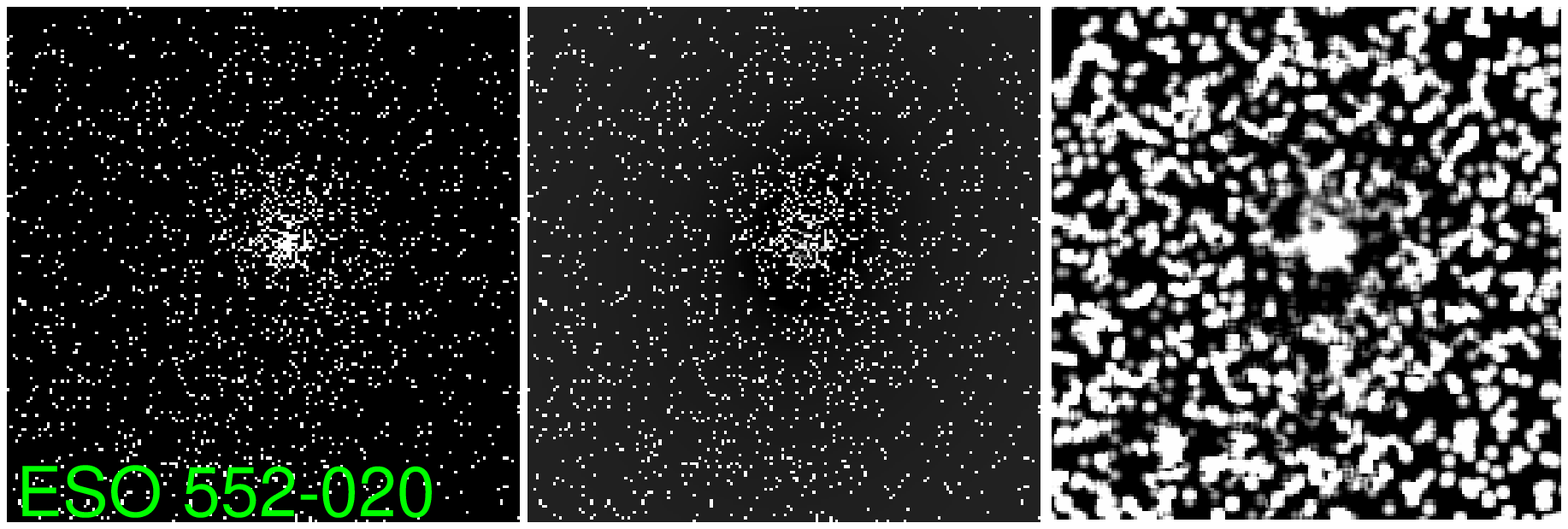}\hfill
\includegraphics[width=0.33\textwidth]{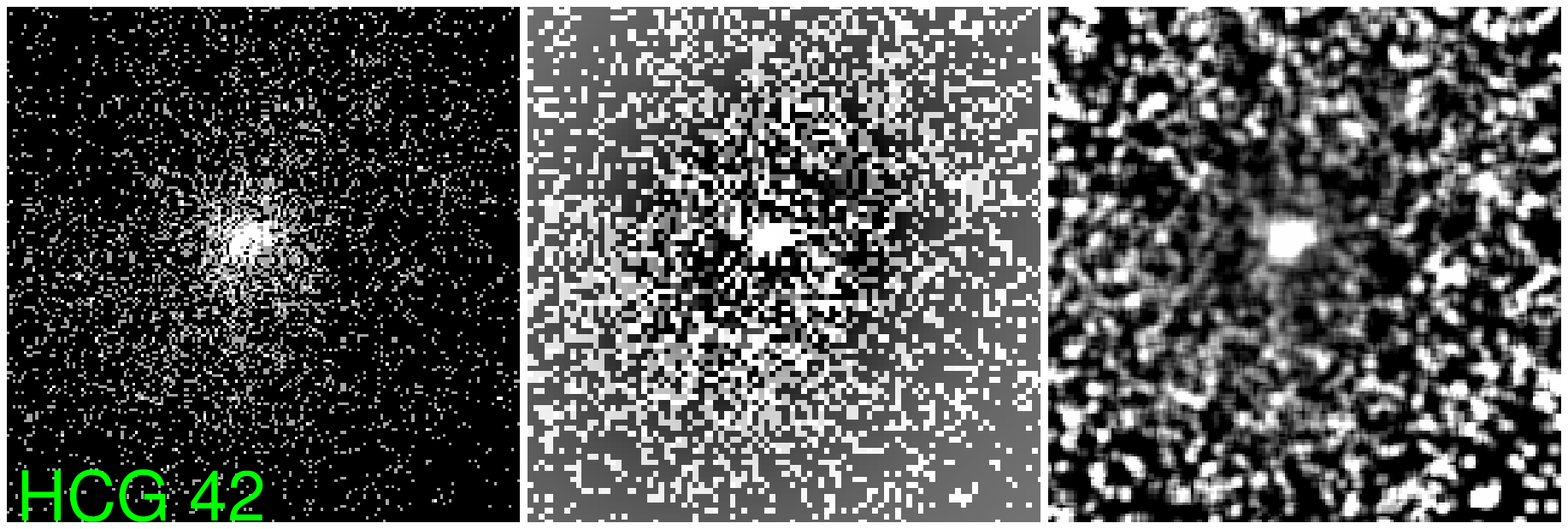}\\
\includegraphics[width=0.33\textwidth]{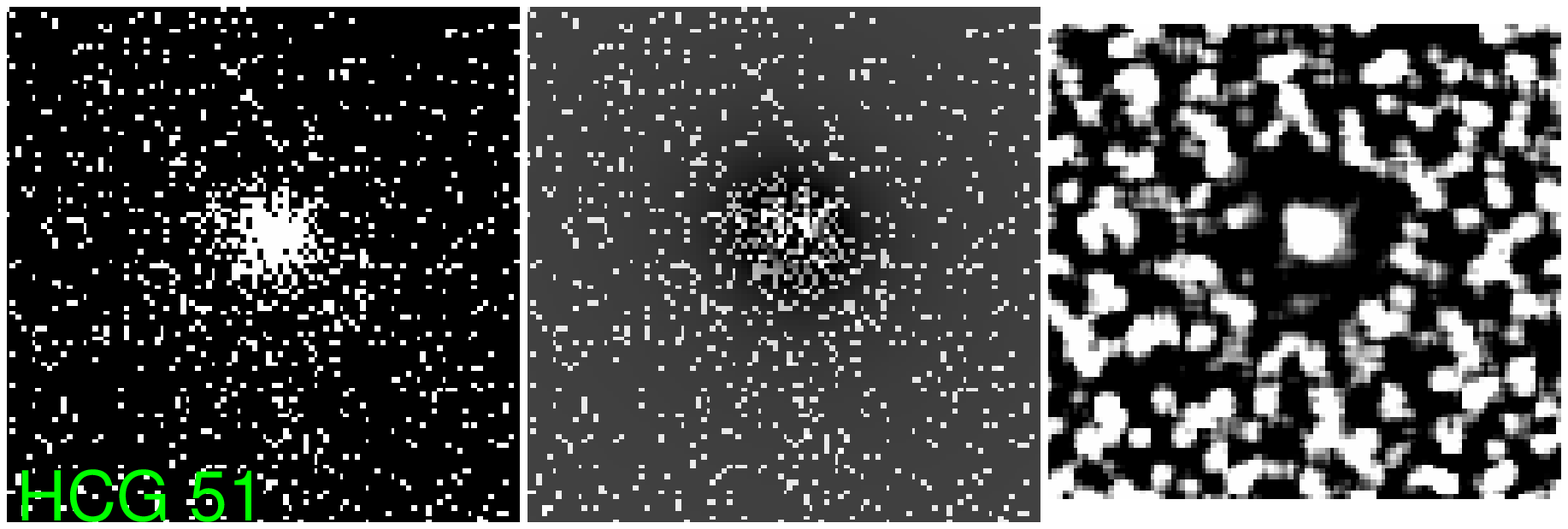}\hfill
\includegraphics[width=0.33\textwidth]{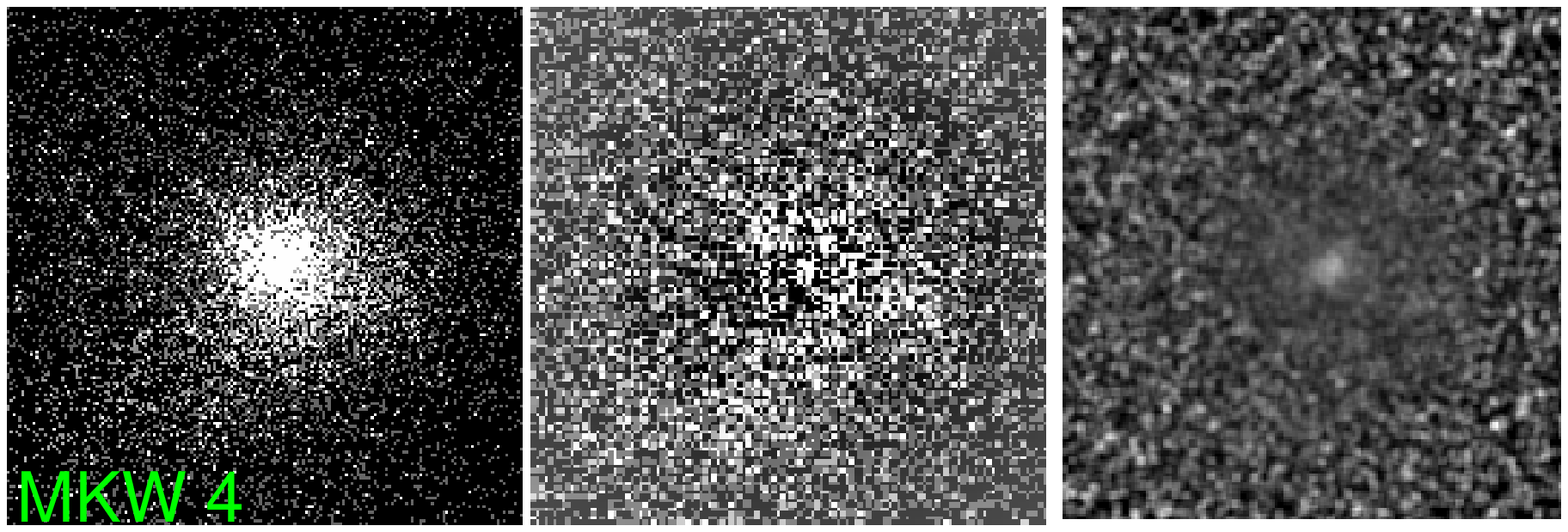}\hfill
\includegraphics[width=0.33\textwidth]{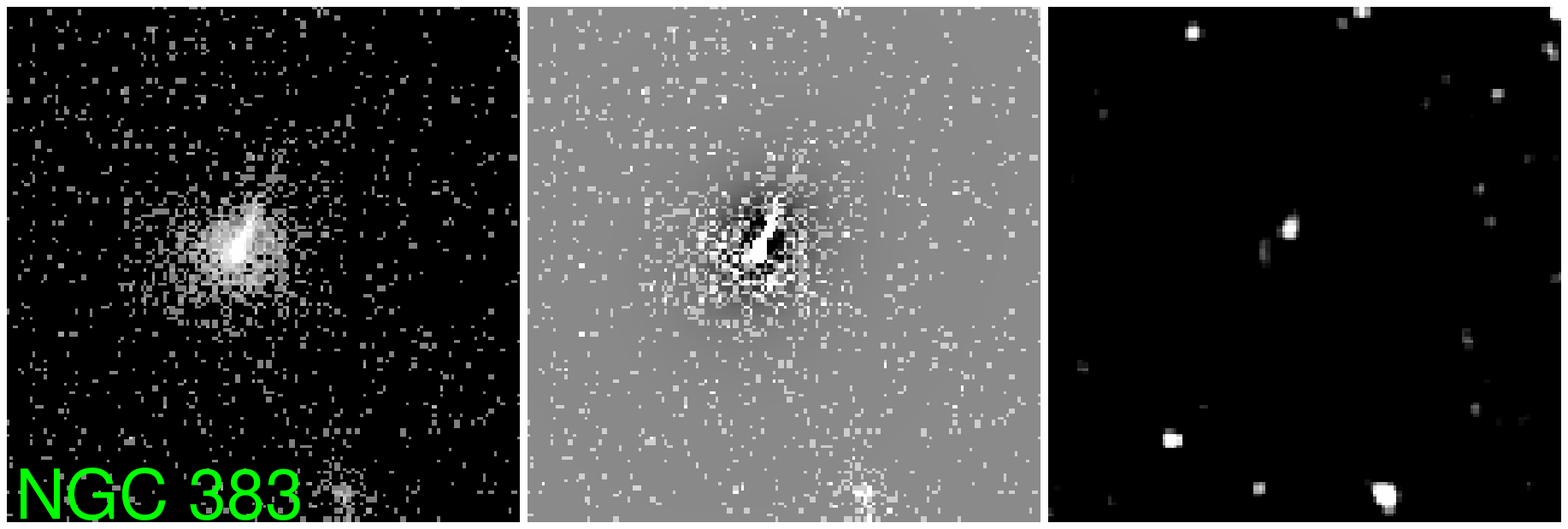}\\ 
\includegraphics[width=0.33\textwidth]{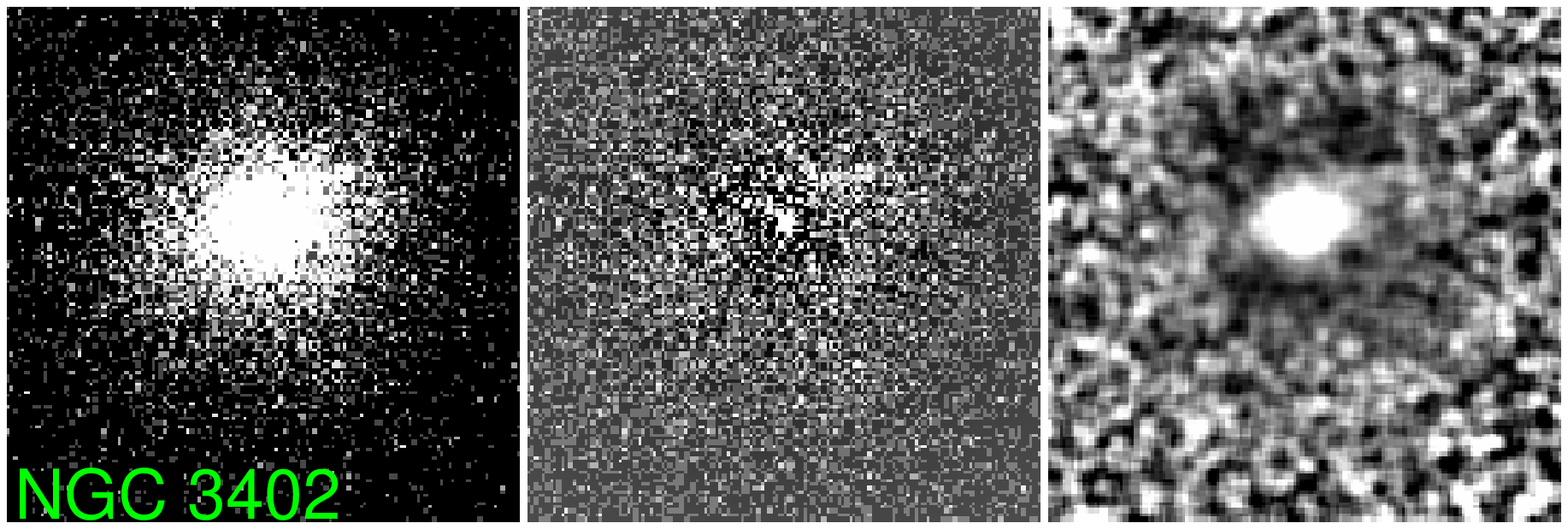}\hfill
\includegraphics[width=0.33\textwidth]{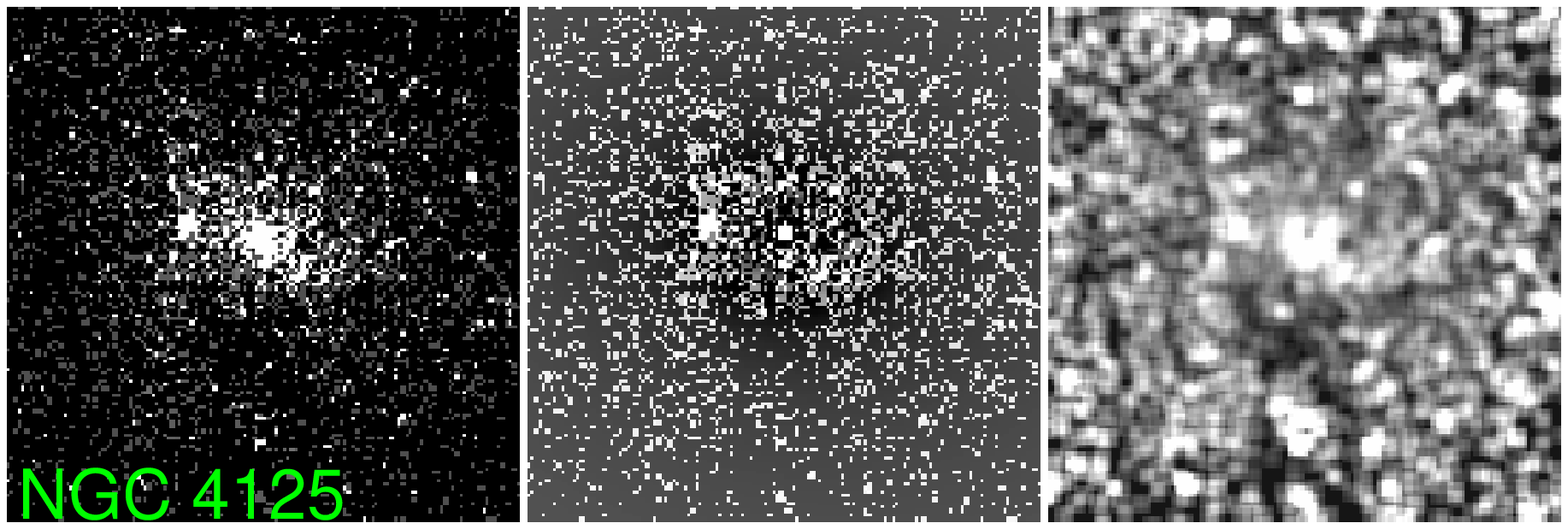}\hfill
\includegraphics[width=0.33\textwidth]{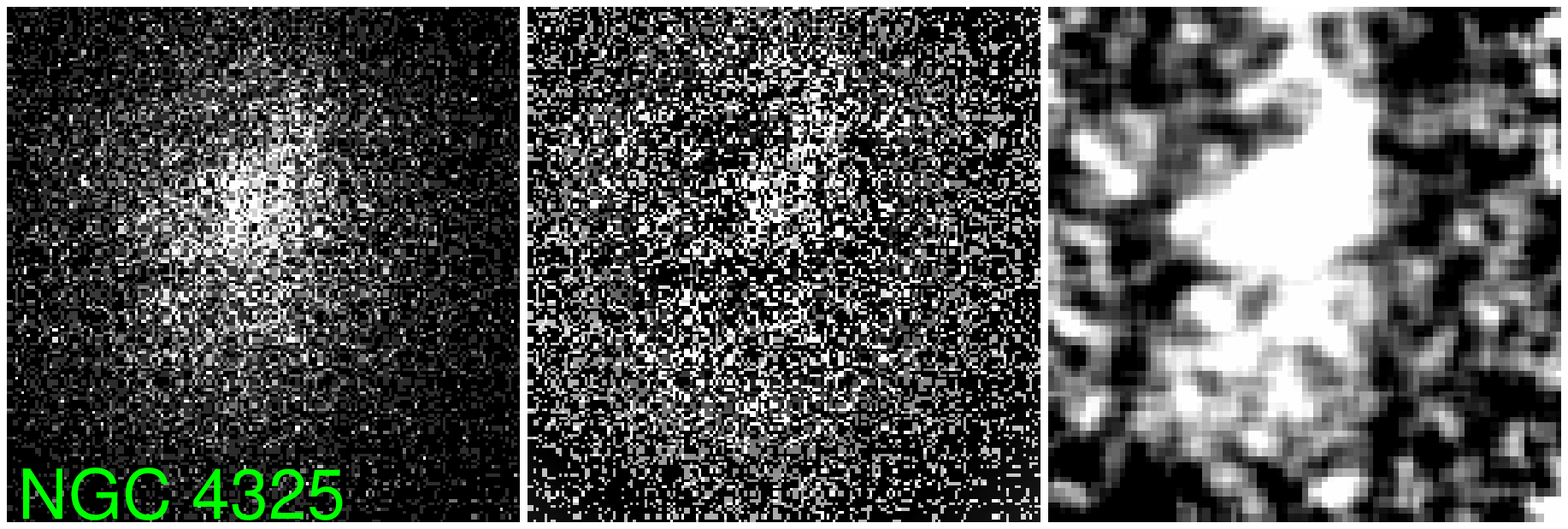}\\
\includegraphics[width=0.33\textwidth]{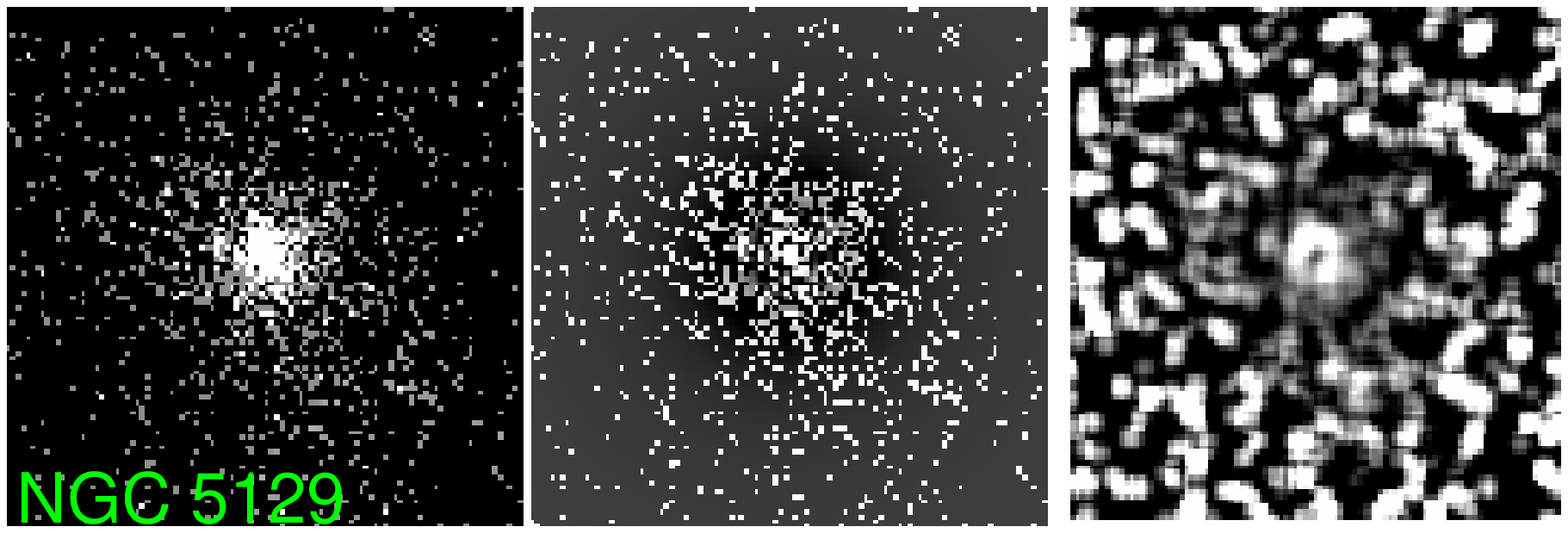}\hfill
\includegraphics[width=0.33\textwidth]{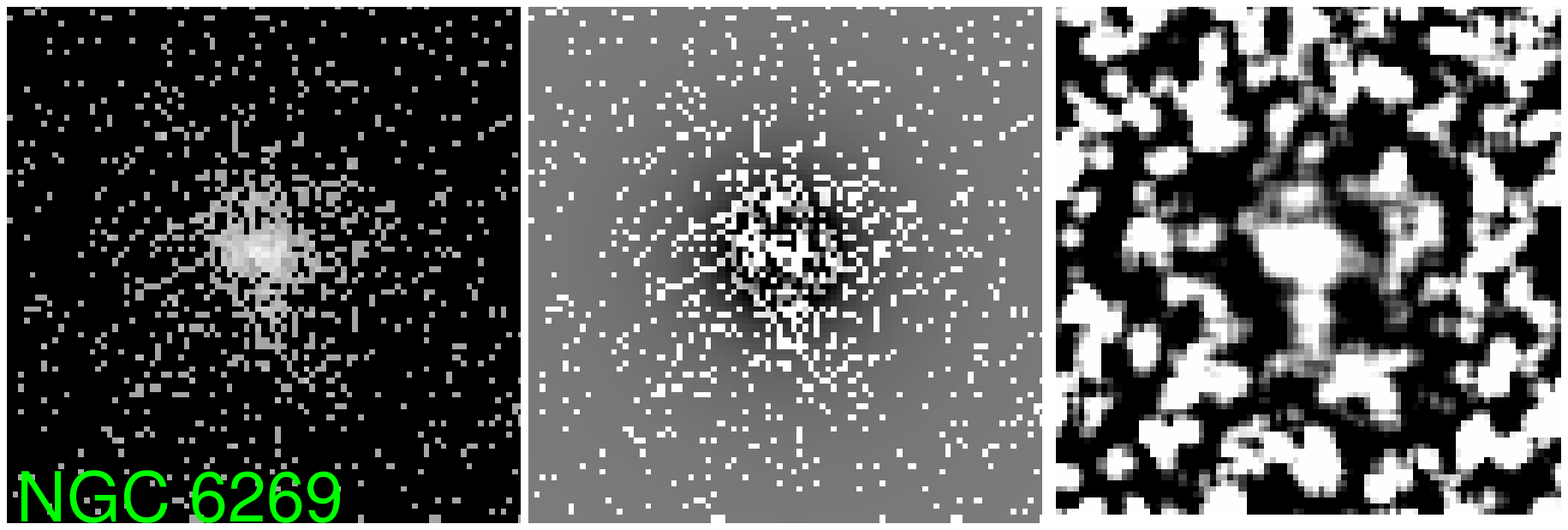}\hfill
\includegraphics[width=0.33\textwidth]{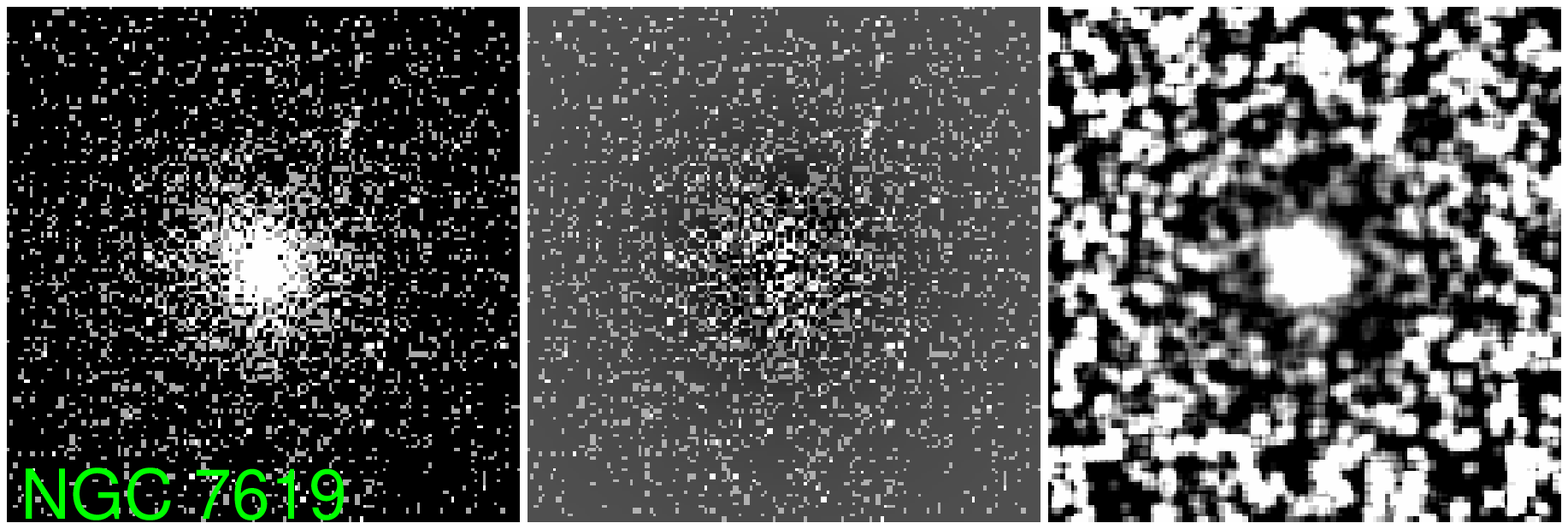}\\ 
\includegraphics[width=0.33\textwidth]{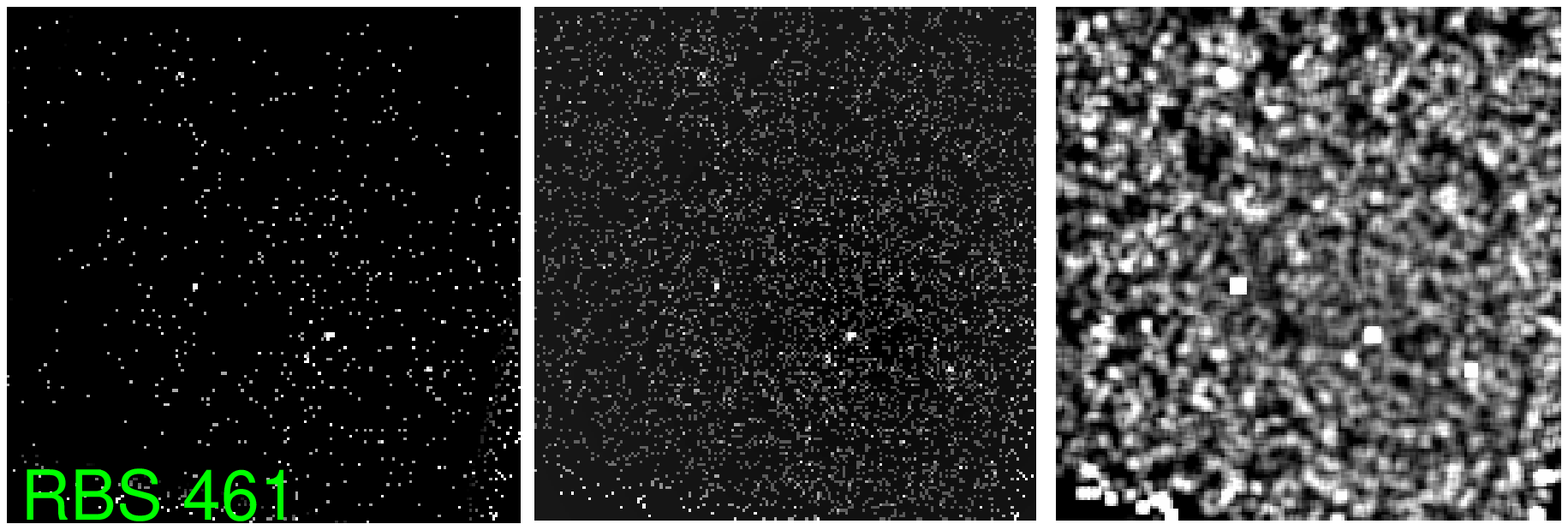}\hfill
\includegraphics[width=0.33\textwidth]{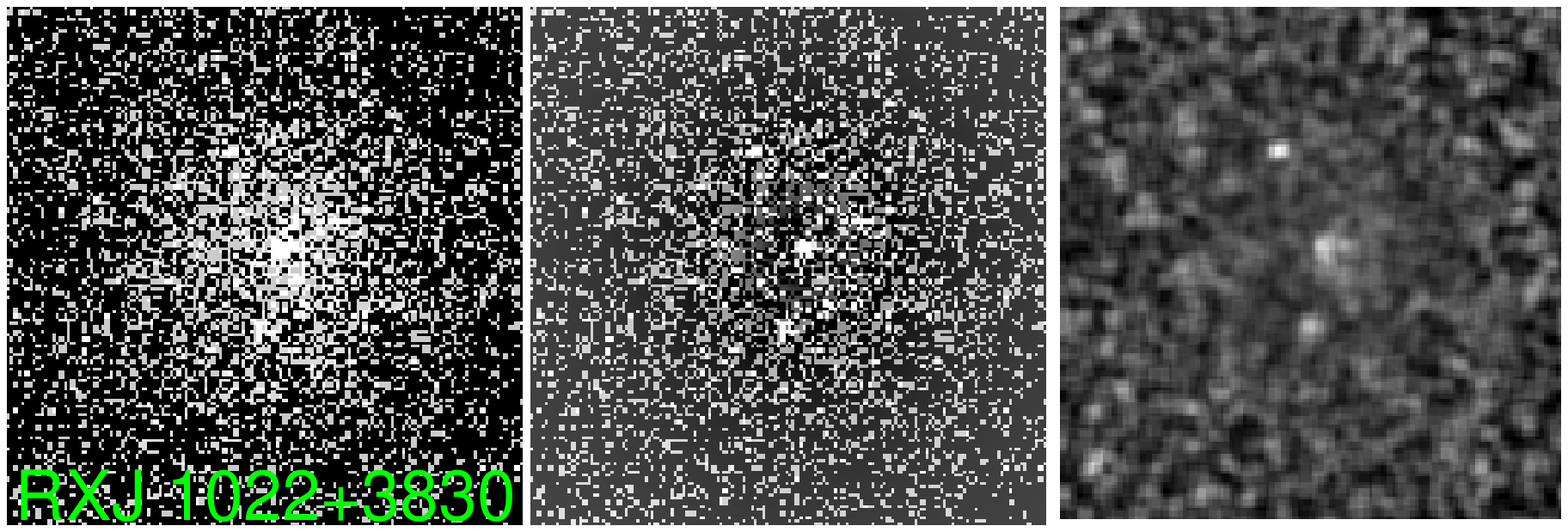}\hfill
\includegraphics[width=0.33\textwidth]{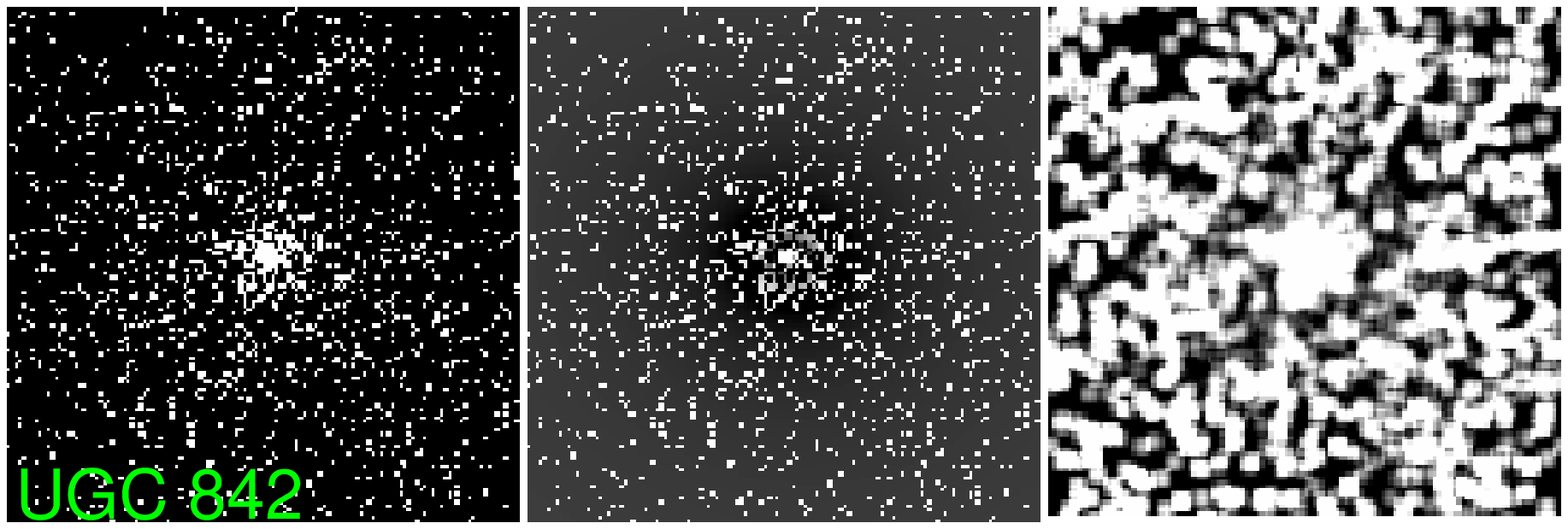}\\
\includegraphics[width=0.33\textwidth]{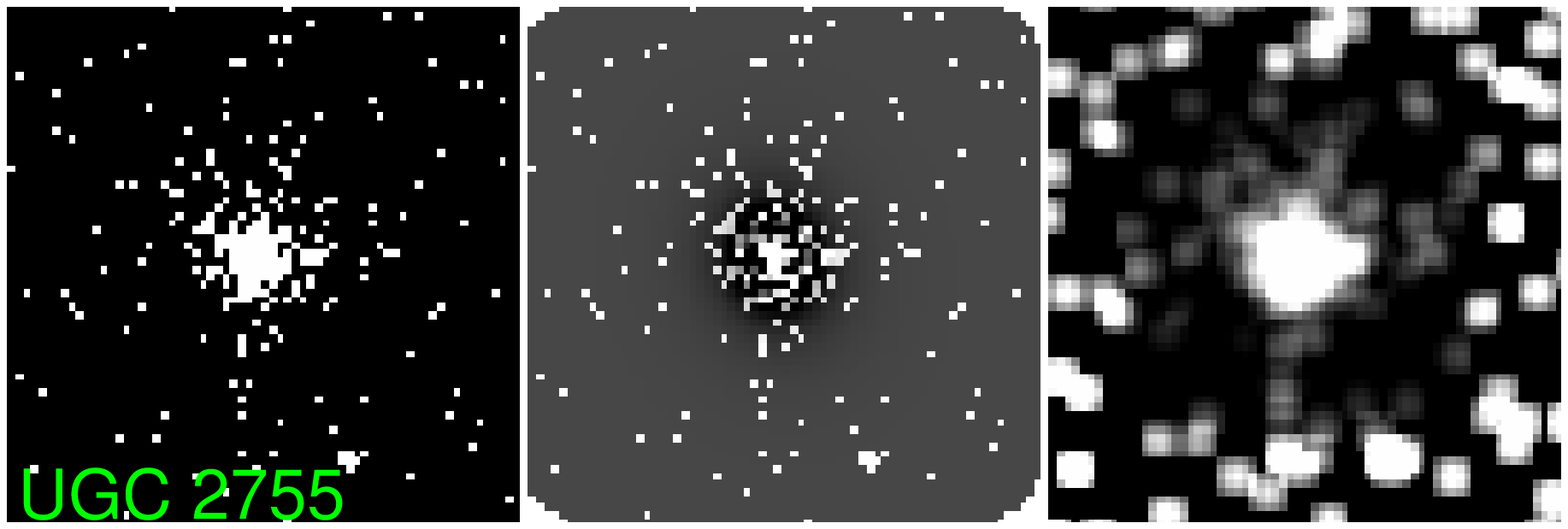}
\figcaption{As Figure~\ref{certain_cavity}, but for the 25 groups in
   the N--sample without clearly identifiable cavities. The residual
   images in this figure have not been smoothed.
   \label{no_cavity}}
\end{figure*}

\section{RESULTS}\label{sec:result}

Based on the above criteria, 26 out of the 51 groups in our sample
were identified as harboring certain or possible X-ray cavities. These
systems are evenly split among the C-- and P--samples. In
Figure~\ref{certain_cavity} we show the input image, the residual
image from model fitting, and the quotient image from unsharp masking
for all groups in the C--sample. Figure~\ref{possible_cavity} shows
the corresponding results for the P--sample. For further comparison,
we also show the images for the N-- sample in Figure~\ref{no_cavity}.

Among the 26 groups in the combined C-- and P--samples, about half (8
in the C-- sample and 6 in the P-- sample) show evidence of containing
two cavities, and these are always at symmetric positions relative to
the group center. We did not identify any groups showing evidence for
more than two cavities as otherwise reported for a number of galaxy
clusters (e.g., \citealt{san07,wis07}). At least nine groups in our
sample have been identified in previous work as hosting cavities:
IC\,1262 \citep{tri07}, A262 \citep{bir04}, HCG\,62 \citep{mor06},
NGC\,507 \citep{kra04,all06}, NGC\,5846 \citep{all06}, NGC\,741
\citep{jet07}, NGC\,4325 \citep{rus07}, NGC\,5044 \citep{gas09}, and
NGC\,5098 \citep{ran09}. Eight of these were classified as belonging
to the C-- or P--sample in our analysis. The only exception is
NGC\,4325, which shows weak hints of the presence of cavities but
without satisfying our criteria for inclusion in the P--sample. The
identification of cavities in this system by \citet{rus07} was based
on a detailed Bayesian model-fitting approach which clearly exceeds
the level of sophistication employed in our analysis. Furthermore, two
of the above nine groups (NGC\,507 and NGC\,741) were here classified
as belonging to the P-- rather than C--sample. These comparisons
confirm our suspicion that we have been fairly conservative in our
classification. This might be anticipated given that we are not
incorporating, e.g., radio data to aid in cavity identification.

\begin{figure}
\begin{center}
\epsscale{1.3}
\mbox{\hspace{-5mm}
\plotone{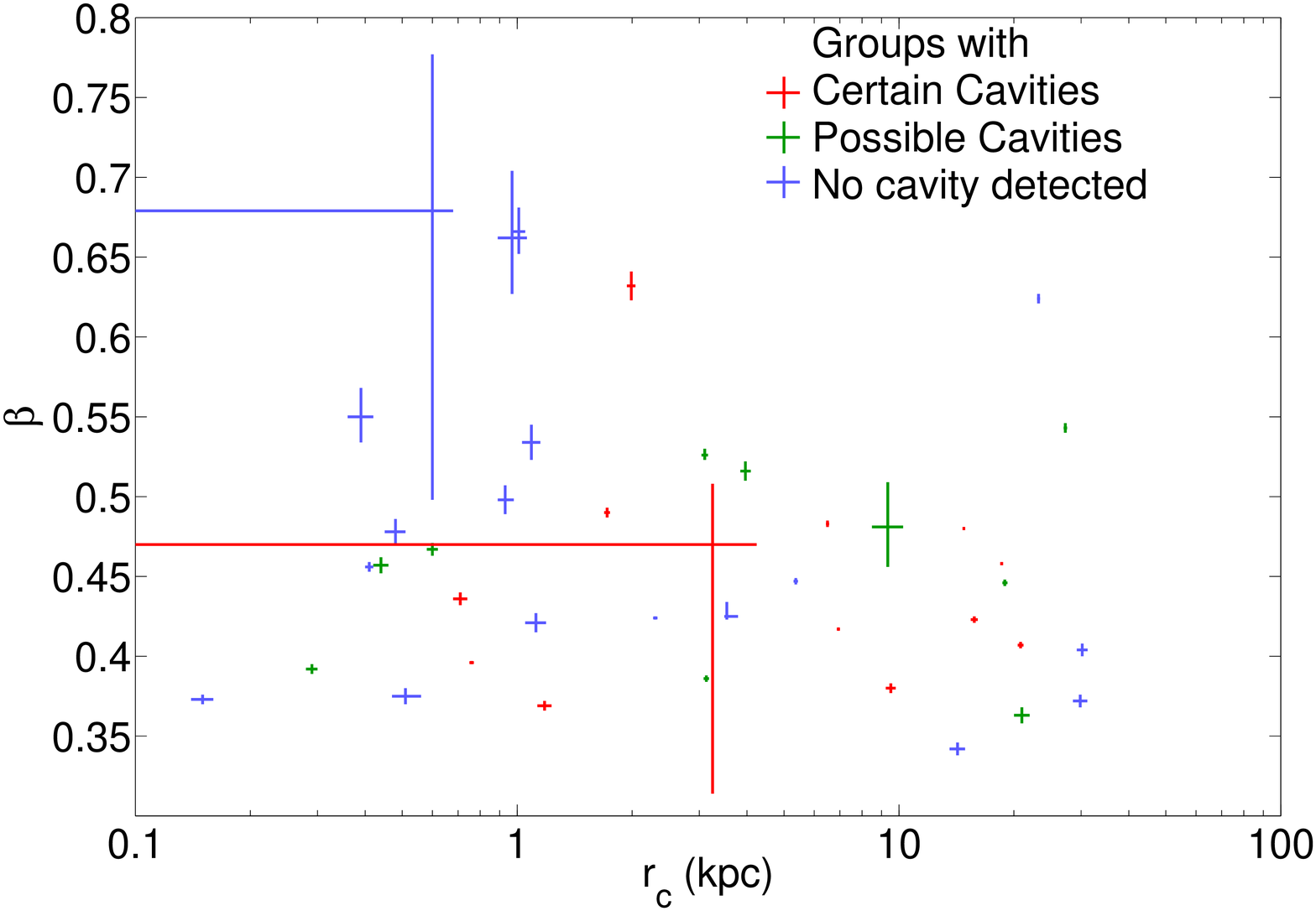}}
\end{center}
\figcaption{Fitted $\beta$--values and core radii with $1\sigma$ error
   bars for the C--sample (red), P--sample (green), and N--sample
   (blue).
  \label{beta_rc}}
\end{figure}

We list the $\beta$--model fit results and cavity measurements (where
applicable) for all the groups in Table~\ref{table:cavity}, along with
the total number of exposure-corrected 0.3--2~keV photons from diffuse
emission within the model fitting region considered for each group
(typically the central $\sim 5'\times 5'$). In agreement with many
previous studies (e.g., \citealt{osm04}), we note that the mean
$\beta$--value derived for the sample is $\beta=0.47$, considerably
lower than the typical value of $\beta \approx 2/3$ seen in clusters
\citep{arn99,moh99}. We also tested for a relationship between $\beta$
and core radius $r_c$ for all three subsamples of groups, as
illustrated in Figure~\ref{beta_rc}. No clear correlation between
these two parameters is evident, neither for the individual subsamples
nor for all groups combined. In addition, mean values of the two
parameters do not show any clear variation among the subsamples, being
($r_c=7.9$~kpc, $\beta=0.45$) for the C--sample, ($r_c=8.8$~kpc,
$\beta=0.46$) for the P--sample, and ($r_c=6.5$ kpc, $\beta=0.49$) for
the N--sample.

\begin{figure}
\begin{center}
\epsscale{0.75}
\plotone{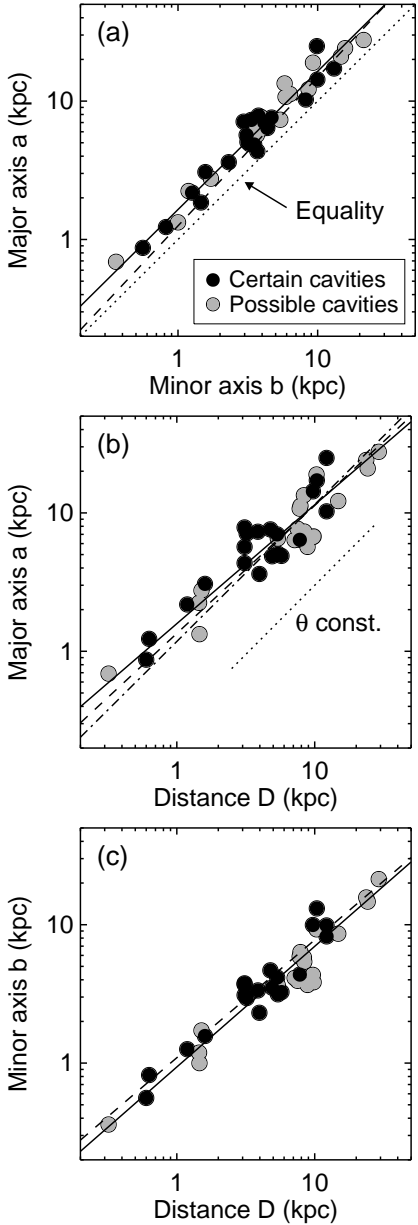}
\end{center}
\figcaption{(a) Cavity major axis $a$ as a function of minor axis $b$
  for the C-- and P--samples. Dotted line represents equality. (b)
  Major axis as a function of projected cavity groupcentric distance
  $D$. Dotted line illustrates the (arbitrarily normalized)
  expectation for cavities enclosed by a cone of constant opening
  angle $\theta$, and dot-dashed line shows the fitted relation of
  \citet{die08}. (c) Minor axis as a function of $D$. Solid lines in
  all plots show the best-fit linear relations for our group sample,
  equations~(\ref{eq:ab})--(\ref{eq:bd}), while dashed lines represent
  our fits to the cluster results of
  \citet{bir04}. \label{size_distance}}
\end{figure}

From the results in Table~\ref{table:cavity}, we can explore any
correlations among the sizes and groupcentric distances of the
identified cavities. Figure~\ref{size_distance}(a) shows the
relationship between the cavity size in the tangential direction $a$
and radial direction $b$ for the C-- and P--samples.  Remarkably,
these two quantities appear strongly coupled, despite the potentially
considerable systematic uncertainties associated with their
estimation. Motivated by the appearance of the data in
Figure~\ref{size_distance}, we fitted linear relations to the data in
log--log space to quantify the observed cavity behavior. For this, we
used the orthogonal regression approach of \citet{iso90}, since there
are (unquantifiable) errors on both parameters, but we note that
results are consistent with those obtained using various ordinary
least squares methods (e.g., $Y$ vs.\ $X$, $X$ vs.\ $Y$, bisector).
For the data in Figure~\ref{size_distance}(a), we then find
\begin{equation}
\mbox{log\,}a = (1.00\pm 0.03) \mbox{\,log\,}b + (0.22\pm 0.02), 
\label{eq:ab}
\end{equation}
for $a$ and $b$ in kpc and with a high correlation coefficient of
0.95. This shows that cavities maintain broadly similar shapes, with a
mean ratio of $a/b\approx 1.7$ that is roughly constant for all cavity
sizes. Cavity sizes are also strongly correlated with their projected
distance $D$ from the group center.  Figure~\ref{size_distance}(b) and
(c) show $a$ and $b$ as functions of $D$ (also in kpc), with best-fit
linear relationships of
\begin{equation}
\mbox{log\,}a = (0.86\pm 0.04) \mbox{\,log\,}D + (0.20\pm 0.03), 
\label{eq:ad}
\end{equation}
for a correlation coefficient $r=0.94$, and
\begin{equation}
\mbox{log\,}b = (0.87\pm 0.03) \mbox{\,log\,}D - (0.03\pm 0.02), 
\label{eq:bd}
\end{equation}
for a coefficient $r=0.88$. As illustrated by the dotted line in
Figure~\ref{size_distance}(b), equation~(\ref{eq:ad}) implies that
cavities are enclosed by a cone of roughly constant opening angle
$\theta$ $\sim 60^\circ$ as they rise and expand within the
surrounding medium.

In Figure~\ref{size_distance}, the fits of
equations~(\ref{eq:ab})--(\ref{eq:bd}) are represented by solid
lines. For comparison, we also show by dashed lines our corresponding
fits to the results for the cluster sample of B\^{\i}rzan et~al.\
(2004; their table 3), namely
\begin{eqnarray*}
\mbox{log\,}a & =  (1.07\pm 0.07)  \mbox{\,log\,}b & +  (0.10\pm 0.07),\\ 
\mbox{log\,}a & =  (0.92\pm 0.07)  \mbox{\,log\,}D & +  (0.13\pm 0.07),\\ 
\mbox{log\,}b & =  (0.85\pm 0.06)  \mbox{\,log\,}D & +  (0.04\pm 0.06).
\end{eqnarray*}
In addition, Figure~\ref{size_distance}(b) further shows the relation
between $a$ and $D$ found by \citet{die08} for an even larger cluster
sample. Despite the somewhat subjective way cavity sizes are estimated
by different authors, there is generally remarkable agreement between
these three studies and hence between results for our groups and more
massive clusters. In particular, the slopes of
equations~(\ref{eq:ab})--(\ref{eq:bd}), along with two out of three
intercepts, are statistically consistent at the 1-$\sigma$ level with
the results of \citet{bir04}.

We next tested for a correlation between the 1.4~GHz radio luminosity
$L_{\rm 1.4\,GHz}$ of the central brightest group galaxy (i.e., that
of any central radio source associated with the galaxy itself rather
than that of any radio lobes coinciding with the X-ray cavities) and
the detection of cavities. Values of $L_{\rm 1.4\,GHz}$ were extracted
from the NRAO VLA Sky Survey (NVSS), and are listed in
Table~\ref{table:cavity}. If data were missing at 1.4~GHz but
available at a nearby frequency, we extrapolated to 1.4~GHz assuming a
power-law spectrum of index unity, $S\propto \nu^{-1}$.  For groups
with NVSS coverage but with no radio flux listed in NED, we used the
99$\%$ completeness flux limit in the NVSS survey (3.4~mJy;
\citealt{con98}) to estimate an upper limit to $L_{\rm
1.4\,GHz}$. Finally, if a source was not covered in NVSS or if the
1.4~GHz emission could not be unambiguously associated with the
central group galaxy in NVSS or FIRST \citep{bec95} data, we left
$L_{\rm 1.4\,GHz}$ as undetermined.

\begin{figure}
\begin{center}
\epsscale{1.15}
\plotone{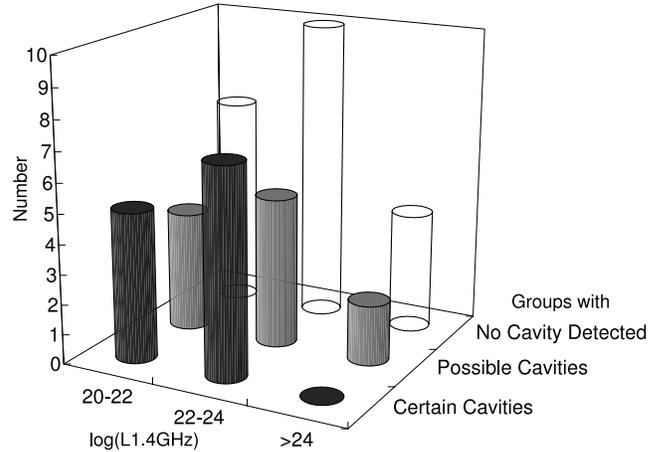}
\end{center}
\figcaption{Distribution of central 1.4~GHz radio luminosities for the
  different subsamples. \label{radio_stat}}
\end{figure}

Figure~\ref{radio_stat} shows the resulting distribution of $L_{\rm
  1.4\,GHz}$ for the different group subsamples. For groups with upper
limits to $L_{\rm 1.4\,GHz}$, we have assumed this upper limit when
displaying these results, but we note that the results would not be
substantially affected by excluding those groups. There is no strong
link between the presence of detectable cavities and the current radio
power of the central group galaxy.  For example, the geometric mean of
$L_{\rm 1.4\,GHz}$ is very similar for the three subsamples,
$10^{22.44}$~W~Hz$^{-1}$ (C--sample), $10^{22.58}$~W~Hz$^{-1}$ (P),
and $10^{22.49}$~W~Hz$^{-1}$ (N). Again, excluding groups with upper
limits to $L_{\rm 1.4\,GHz}$ does not change this conclusion. We also
note that no strong trend is seen between projected cavity distance
$D$ and $L_{\rm 1.4\,GHz}$, with the data showing a small correlation
coefficient of 0.4.

\begin{figure*}
\begin{center}
\epsscale{1.0}
\plotone{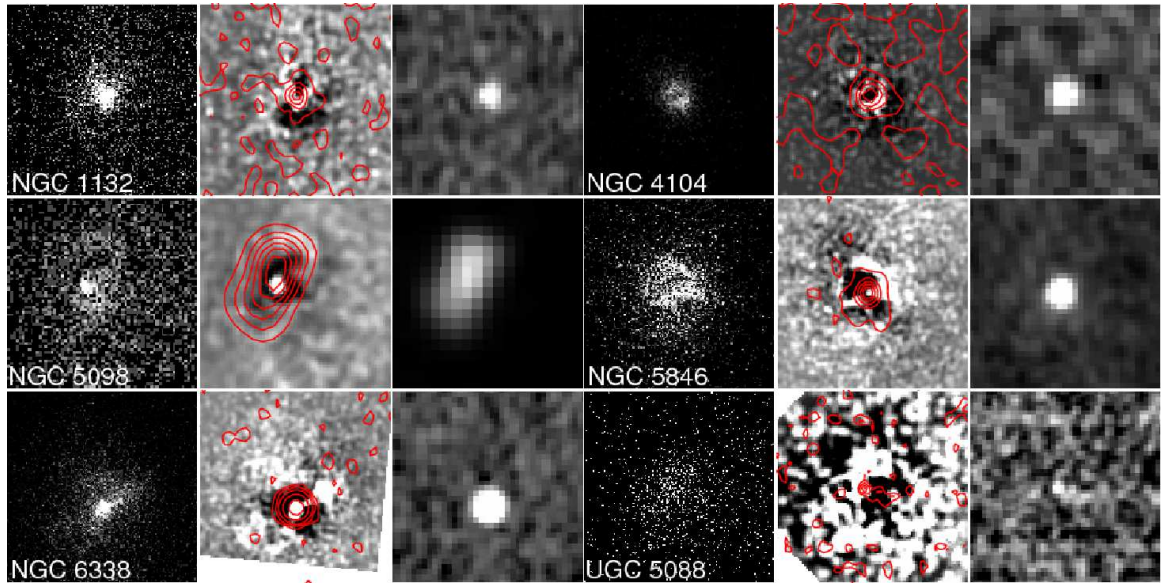}
\end{center}
\figcaption{Comparison of X-ray and radio morphology for the 6 groups
  in the C-- and P--samples with available FIRST radio data. For each
  group, left panel shows the exposure-corrected image, center panel
  the residual image, and right panel the FIRST 1.4~GHz image.
  Contours extracted from the radio data have been overlayed on the
  residual images. The first five groups belong to the C--sample,
  while UGC\,5088 belongs to the P--sample.
  \label{radio_images}}
\end{figure*}

Finally, to search for evidence of a 1.4~GHz radio plasma at the
location of the cavities, we have compared the X-ray and radio
morphology of the groups in the C-- and P--samples, using radio data
extracted from NVSS and FIRST. Since the cavities are generally
present on spatial scales well below the resolution of NVSS data,
comparison between {\em Chandra} and NVSS data generally provides no
clear evidence of any morphological similarities. However, 12 of our
groups have higher-resolution FIRST data in NED, and of those 12
systems, we show the six groups belonging to the C-- or P--samples in
Figure~\ref{radio_images}. No clear association between the X-ray
cavities and any radio lobes is seen, however. It remains a
possibility that the cavities coincide with radio emission at
frequencies well below 1.4~GHz (e.g., \citealt{gia09}).

\section{TESTING CAVITY DETECTION WITH MOCK DATA}\label{sec:mock}

To test the performance of the adopted methods for cavity detection
and better understand the biases inherent in either method, we next
generated a set of mock images designed to match typical {\em Chandra}
exposures within our sample. These images were then analyzed in a
manner similar to that of the real data. This Section describes the
generation and analysis of these mock data.

\subsection{Model Setup}\label{sec:mock2}

The mock images were generated by first modeling the group emission as
an elliptical $\beta$--model. Cavities were superposed as circular
depressions in local surface brightness, and a spatially uniform
background was added.  The following sets of model parameters were
considered:
\begin{itemize}
\item $\beta =  $\, \{0.35, 0.50, 0.65\}.
\item Core radius $r_c = 10$~pixels.
\item Total number of source counts $N_{\rm tot} =$\, \{5,000, 15,000,
30,000\}.
\item Cavity \lq\lq strength\rq\rq \ (the local surface brightness
depression factor): \{1.5, 2.0, 2.5\}.
\item Projected cavity distance $D$ from X-ray peak in pixels: \{15,
35\}.
\end{itemize}
This choice of parameters was generally motivated by the results for
our real sample.  The adopted $\beta$--values span the range of fitted
values covered by all but four of our groups. To keep the number of
output images at reasonable levels, $r_c$ was fixed for all mock
data. This should not introduce any substantial systematic bias in the
results, since Figure~\ref{beta_rc} shows that $r_c$ does not vary
systematically with $\beta$ for our groups. Values of $N_{\rm tot}$
were chosen to range from the median of the groups with no detected
cavities ($\sim 5000$) to a value below that of HCG\,62, one of the
groups with the most prominent and obvious cavities.  For the cavity
radii, we assumed $R = 0.6 D$, in rough agreement with
equation~(\ref{eq:ad}).

A total of 54 mock images were generated to span all combinations of
the above parameters. For each image, a uniform background level was
added, drawn from a Gaussian distribution with a mean equal to the
approximate 0.3--2~keV value for a typical 50-ks ACIS-S exposure
($\approx 0.1$~counts~arcsec$^{-2}$), and with $\sigma$ =
0.2$\times$mean.  The minor-to-major axis ratio of the $\beta$--model
was drawn from a Gaussian with mean 0.8 and $\sigma = 0.12$
\citep{moh95}, but was restricted to $\geq 0.65$. Position angles of
the $\beta$--model and of the cavities themselves were independently
drawn randomly from a uniform distribution. The resulting surface
brightness model was then convolved with the {\em Chandra} point
spread function, derived using the \lq\lq mkpsf\rq\rq \ tool in {\sc
ciao} at a photon energy of 1~keV and at the detector position of the
ACIS-S3 aimpoint. Finally, Poisson noise was added to the mock
images. Other instrumental effects (or point sources) were not
included, since the model fitting to the real data was performed on
exposure-corrected and point-source excised images.

Admittedly, the model setup is a rather simplistic one; we assume the
number of cavities in each system is always two, they represent
similar (local) brightness depressions, appear circular, and are at
the same distances from the center of the X-ray emission. As such,
orientation effects were not taken into account (e.g., the cavities
were assumed to be in the plane of the sky). However, since the
purpose of this exercise was simply to test and understand our ability
to recover cavities in the real data under reasonably favorable
conditions, a detailed exploration of the full cavity parameter space
would be beyond the scope of this work.

\subsection{Cavity Detectability}\label{sec:mockresult}

\begin{figure}
\begin{center}
\epsscale{1.17}
\plotone{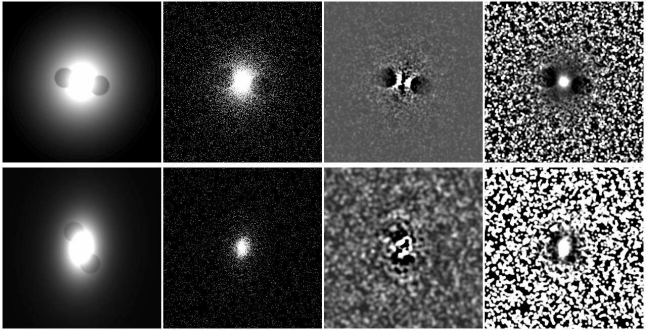}
\end{center}
\figcaption{Two examples of raw and processed mock images: Top panel
  shows a model with total source counts $N_{\rm tot}=$\,30,000,
  $\beta=0.65$, cavity strength 2.5 (prominent cavities) and cavity
  distance $D=35$~pixels. Bottom panel: $N_{\rm tot}=$\,5,000,
  $\beta=0.65$, cavity strength 1.5 (weak cavities) and $D= 35$
  pixels. From left to right: Input source model, raw mock image,
  residual image from the $\beta$-fitting method, and the quotient
  image from unsharp masking (both slightly smoothed).
 \label{mock_image}}
\end{figure}

\begin{figure}
\begin{center}
\epsscale{1.1}
\plotone{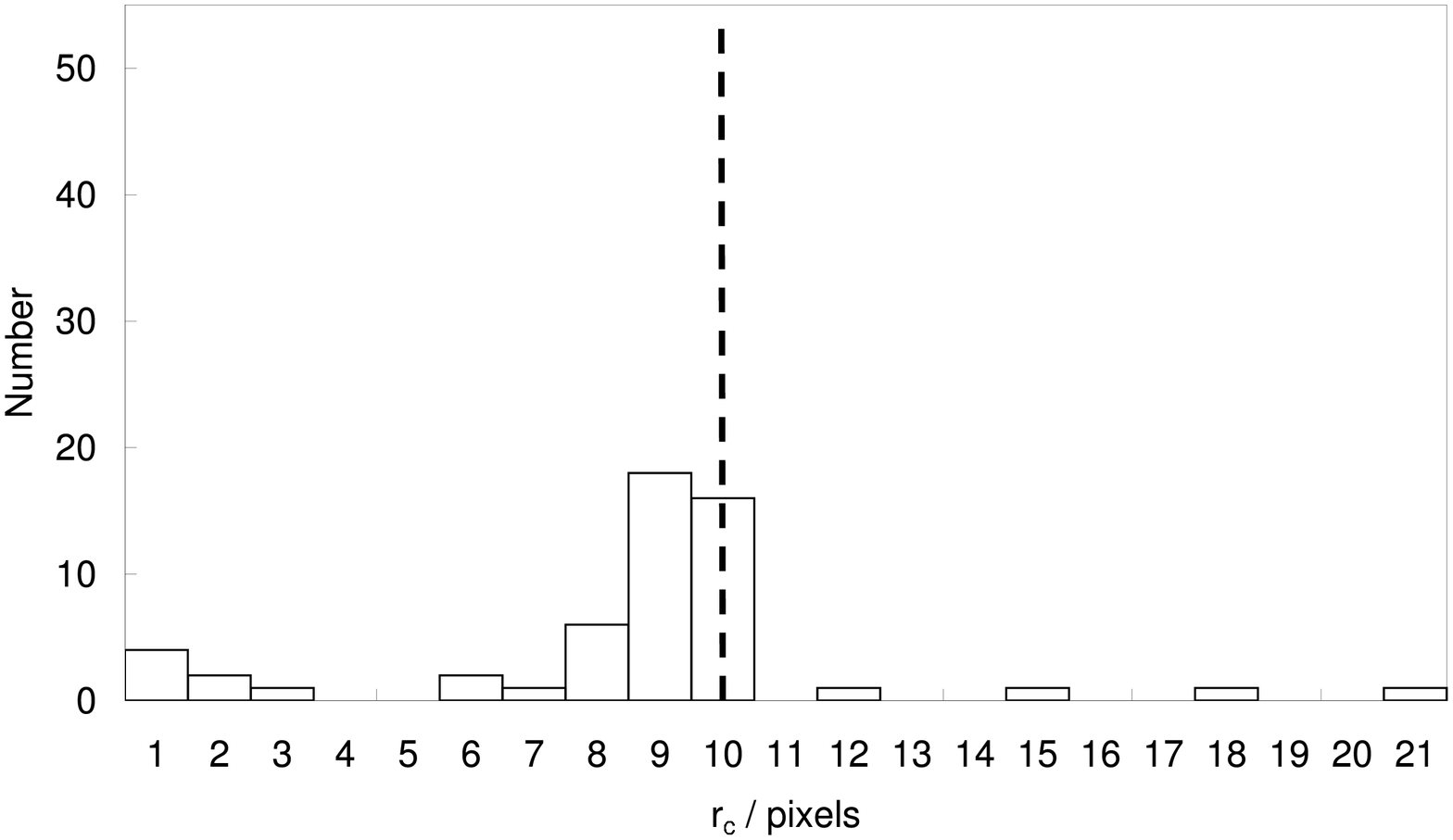}\\
\plotone{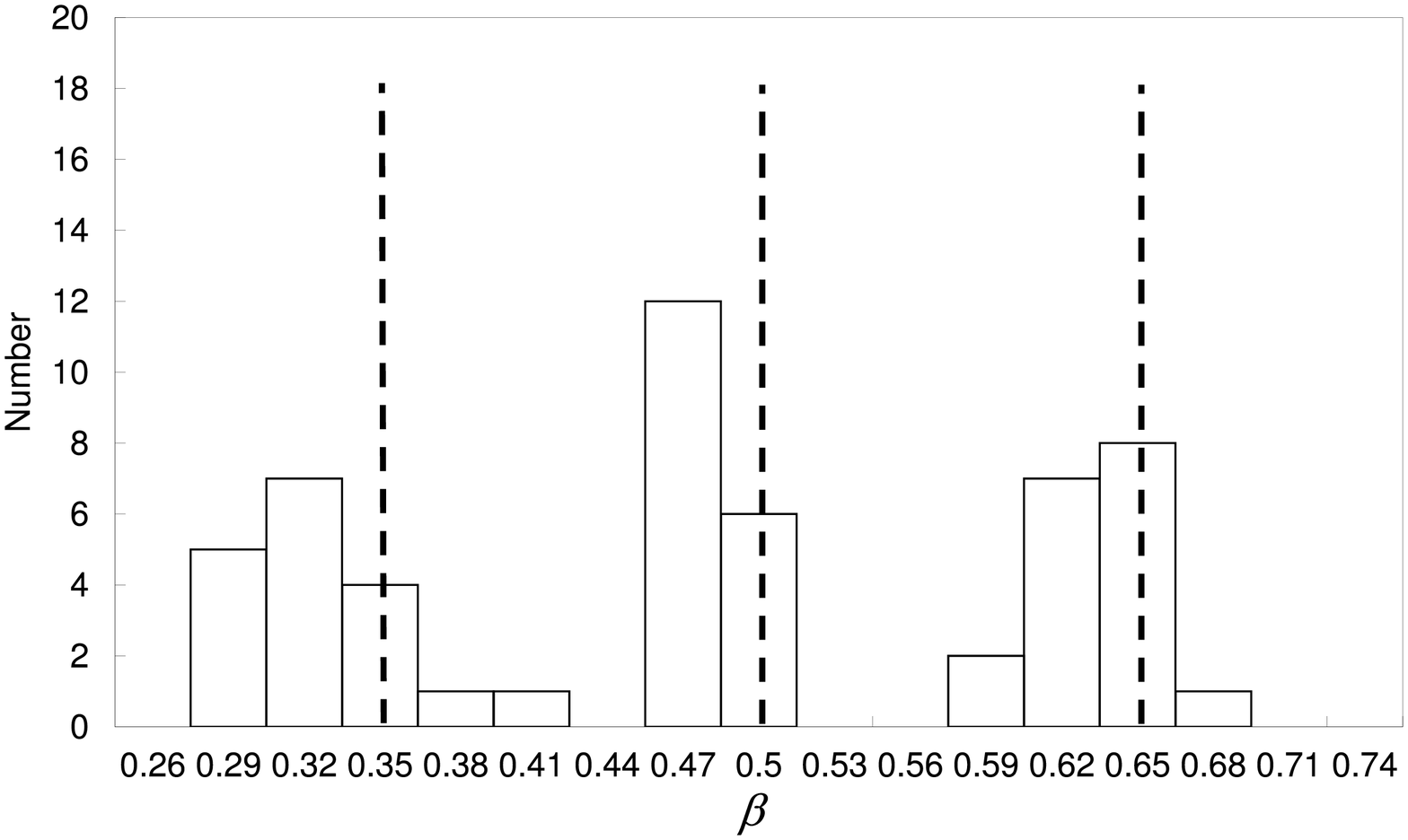}
\end{center}
\figcaption{Distribution of fit results for (top) core radius $r_c$
  and (bottom) $\beta$ for the mock images. Vertical dashed lines
  represent the mock input values.  \label{mock_rcbeta}}
\end{figure}

\begin{figure}
\begin{center}
\epsscale{1.15}
\plotone{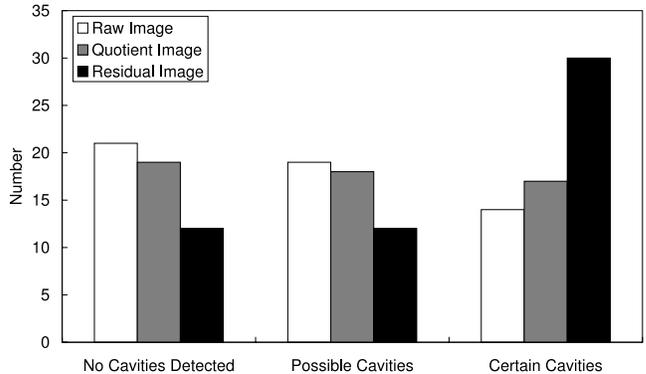}
\end{center}
\figcaption{Number of mock data sets identified as harboring certain,
  possible, or no cavities based on the unprocessed mock images (empty
  bars), quotient images (grey), and on the residual images from model
  fitting (black).\label{mock_eff}}
\end{figure}

We applied both the methods described in Section~\ref{sec:analysis} to
the mock images in a test of the veracity of either approach.
Figure~\ref{mock_image} shows two example mock images, with $N_{\rm
tot}=$\,30,000 and 5,000, respectively, along with the results of our
two methods. For the brighter source, the input cavities are easily
recognizable in both output images. For the fainter case, however,
there is some ambiguity as to the apparent location of cavities in the
two output images.  We will return to this issue below.
Figure~\ref{mock_rcbeta} shows the distributions of best-fit values of
core radius and $\beta$ obtained from surface brightness modeling of
all the mock images. Encouragingly, all distributions are peaked at
the input values, lending credibility to the use of this method for
describing group emission on large scales.  For example, even though
the mock sources are, on average, fainter than our real ones, there is
no overlap between the resulting distributions for the three input
values of $\beta$. Figure~\ref{mock_eff} summarizes the results of
searching for cavities in both the output mock images, the quotient
images, and the residual images. While the rate of ``certain'' cavity
recovery is comparable for the unprocessed and quotient images, using
the residual images is clearly more efficient, with the success rate
increasing by a factor of two (from 14--17 to 30 out of the 54 data
sets).

\begin{figure}
\begin{center}
\epsscale{1.25}
\mbox{\hspace{-7mm}
\plotone{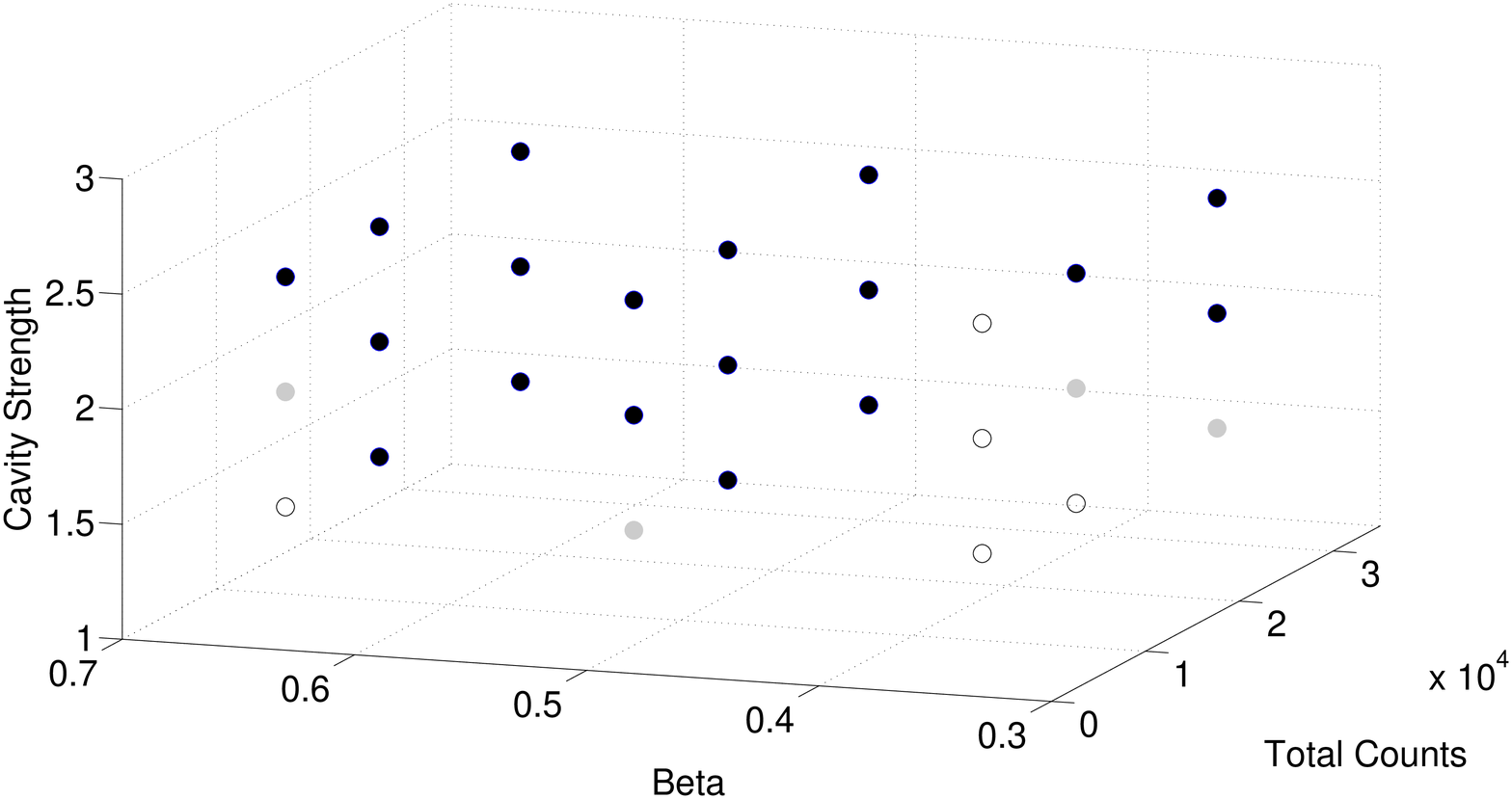}}
\mbox{\hspace{-7mm}
\plotone{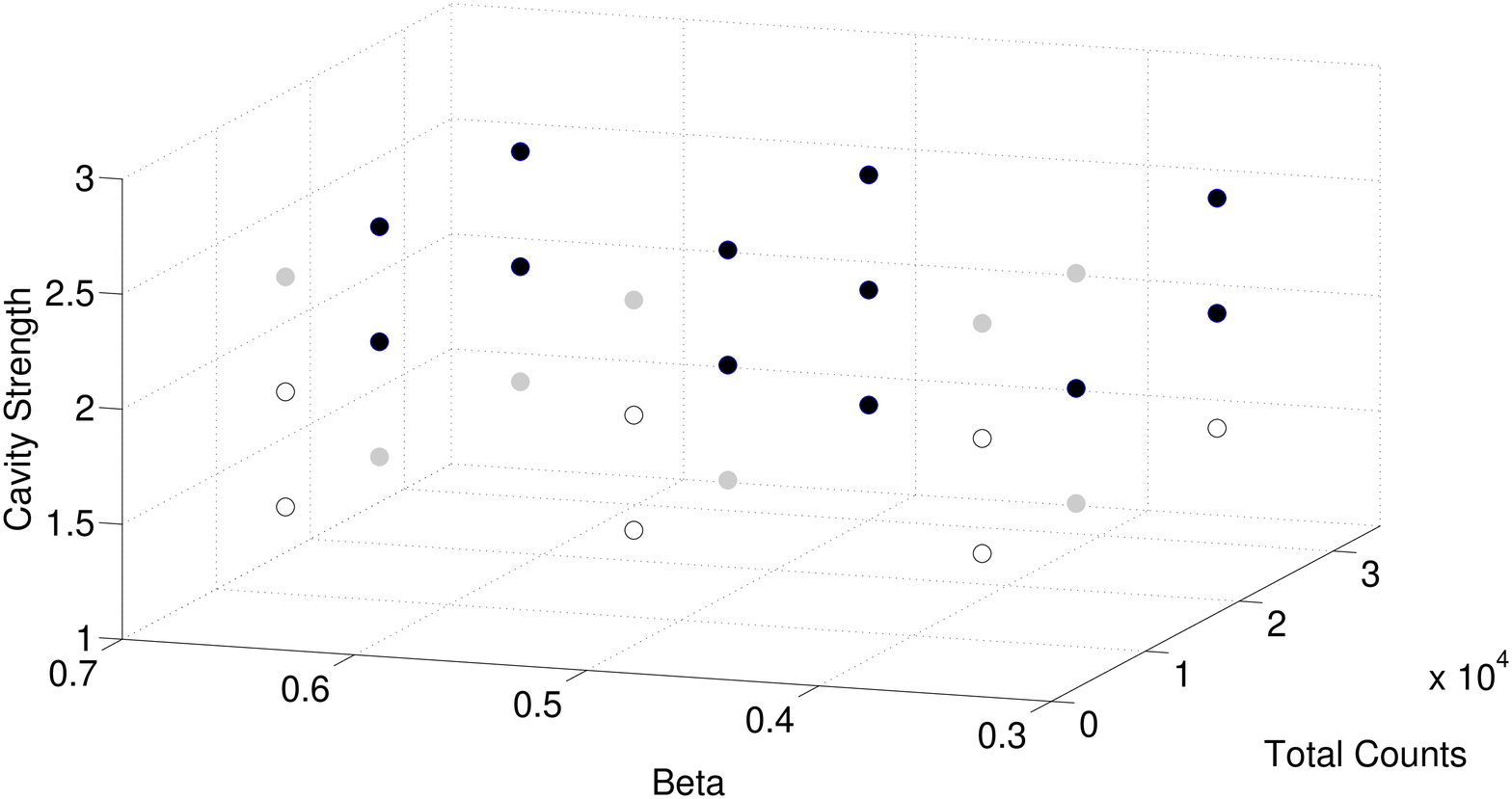}}
\end{center}
\figcaption{Influence of total source counts, $\beta$, and cavity
  strength on cavity recovery in the residual images from
  $\beta$--model fitting. Top panel shows the detection result for the
  27 mock simulations with close cavities ($D=15$~pixels from the
  group center) and bottom panel the corresponding results for the
  ones with distant cavities ($D=35$~pixels). Dark circles represent
  certain cavity detections, gray circles possible ones, and empty
  circles data sets with no detectable cavities.\label{mock_3d}}
\end{figure}

We found that all four controllable input parameters ($\beta$, $N_{\rm
tot}$, cavity strength and distance) influence the identification of
cavities, even when applying the relatively successful model fitting
approach.  This is illustrated in Figure~\ref{mock_3d}, which outlines
the impact of changing the various model input parameters, again using
the classification described in Section~\ref{sec:detection}. In
particular, the value of $N_{\rm tot}$ emerges as a crucial factor for
cavity recovery.  Among the 18 mock images with $N_{\rm tot}
=$\,5,000, only eight were identified as hosting possible cavities in
the residual images, while for $N_{\rm tot}=$\,30,000, the
corresponding value is 17 out of 18. Remaining input parameters also
play a role. For example, at fixed $N_{\rm tot}$, cavity recovery is
more efficient for $\beta\geq0.5$ compared to $\beta=0.35$, and for
higher cavity strength and smaller distance.  Another lesson from the
mock data was that Poisson noise can easily mask one of the two
cavities in X-ray faint sources. In other words, visual identification
of only one cavity in unprocessed data does not necessarily exclude
the presence of a second cavity.  Finally, we note that we found no
clear correlation between the identification or input properties of
cavities and various statistical moments of the photon number
distribution in the residual images.

Naturally, we cannot with confidence exclude the possibility that a
few cavities have been falsely identified in the {\em Chandra
data}. This problem necessarily also plagues all other similar studies
to some extent, but could be particularly acute for the present work
since we have not incorporated radio data or other means to aid in
cavity detection. However, we have reasons to believe that this issue
is not important within our present analysis. First, as already
discussed in Section~\ref{sec:result}, our comparison to existing
results suggests that we have been rather conservative in our cavity
identification. Second, none of the identified cavities appears
severely discrepant in Figure~\ref{size_distance} (note in particular
the tightness of the observed correlation between estimated major and
minor axis), which argues against a significant population of falsely
identified cavities. Third, as mentioned above, within the lowest-S/N
mock subsample (featuring 5,000 net counts), we identified cavities in
eight out of 18 cases. As can be gleaned from
Table~\ref{table:cavity}, the corresponding fraction for all the real
data sets with $<$\,5,000 counts is comparable but slightly smaller,
seven out of 19. This further supports the notion that the incidence
of falsely identified cavities in our {\em Chandra} sample must be
small.

\subsection{Surface Brightness Modeling versus Unsharp Masking}

The two methods considered here for revealing small-scale structure in
extended emission each have their strengths and weaknesses. The
model-fitting approach clearly requires data of fairly high
signal-to-noise ratio to ensure convergence in the fitting process. In
addition, it assumes a model for the underlying large-scale emission
which may not always provide a good description. With unsharp masking,
no assumptions are needed for the large-scale morphology, but the
method requires a highly non-trivial choice of smoothing
scales. Furthermore, the method normally assumes a circular smoothing
kernel, so the smoothing on large scales could potentially generate
spurious cavities in systems with a highly elliptical X-ray
morphology.

A general conclusion based on all 54~mock images is that the
model-fitting method seems superior to unsharp masking in recovering
clearly identifiable cavity structures, cf.\ Figure~\ref{mock_eff}.
This is perhaps not surprising, given that the underlying group
emission in these data is perfectly described by a $\beta$--model
(modulo Poisson uncertainties on source and background counts).
However, this finding is not limited to our mock data; despite careful
consideration of the smoothing scales employed,
Figures~\ref{certain_cavity} and Figure~\ref{possible_cavity} show
that unsharp masking often produces a lower contrast between the
cavities and their surroundings than the output from the model fitting
approach.  The top panel of Figure~\ref{mock_image} further
illustrates this point for the mock data, showing another example in
which cavities in the residual image are more prominent than those in
the quotient image. Another interesting conclusion can be reached from
the bottom panel of Figure~\ref{mock_image}, which emulates a low-S/N
data set with only weak cavities. In this case, the two input cavities
are reasonably well recovered in the residual image, whereas the
quotient image from unsharp masking would lead one to conclude,
incorrectly, that cavities are present to the east and west of the
center of the group emission. Hence, even in an idealized case such as
this, unsharp masking applied to low-S/N data can produce misleading
results.  This serves as a cautionary note if using this method alone
to search for cavities.  Consistency tests using other methods, and/or
the inclusion of radio data to support the existence of cavities, may
be crucial for reliable identification of cavities in X-ray faint
systems.

In addition, the model fitting approach is occasionally able to
recover structure that is not seen in the quotient images. For
example, in the residual image of A1991 (Figure~\ref{certain_cavity}),
two cavities are clearly visible, while unsharp masking only hints at
the presence of the northern cavity. Nevertheless, unsharp masking can
still provide a very useful supplement to visual inspection and
surface brightness fitting. This is particularly true in the cases
where model fitting clearly fails to converge on sensible parameters
(typically in relatively low--S/N data), and for groups in which there
are more complicated structures which cannot be described by an
elliptical $\beta$--model. As seen in Figures~\ref{certain_cavity} and
\ref{possible_cavity}, a number of our groups belong to the latter
category. For example, IC\,1262 presents a sharp surface brightness
discontinuity in the central regions, possibly related to a cold front
\citep{tri07}.  NGC\,4104 shows a clear depression in central surface
brightness, possibly due to cavities being oriented along the line of
sight. For RXJ\,1159+5531, the fitting approach fails to clearly
uncover the two cavities hinted at north-east and south-west of the
group center in the unprocessed image. These are located at either end
of an X-ray jet, itself situated at the center of the diffuse emission
and visible in the raw data (although not clearly so in
Figure~\ref{possible_cavity}). In the residual image, the cavities are
overshadowed by some structure generated in the fitting process as a
consequence of the complex core morphology in this system.  In all
these cases, unsharp masking provides valuable aid in cavity
identification.

\section{DISCUSSION}\label{sec:discuss}

\subsection{Cavity and Central Radio AGN Fractions}
\begin{figure}
\begin{center}
\epsscale{1.17}
\plotone{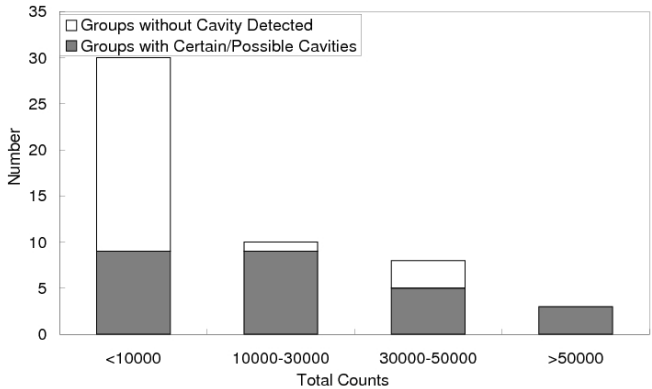}
\end{center}
\figcaption{Cavity detectability and the distribution of total source
   counts $N_{\rm tot}$ for all 51 observed groups. \label{cts_stat}}
\end{figure}

In total, we find that 26 out of our 51 groups ($51\pm 12$\%) show at
least tentative evidence for the presence of cavities. While this is
statistically inconsistent at the $\sim 4\sigma$ level with {\em all}
our groups showing cavities, our experimentation with mock images
suggests that the real fraction might be higher. In particular, since
the total number of source counts plays such an important role for the
detectability of cavities in our mock data, it is relevant to examine
the distribution of total source counts for our real sample.
Figure~\ref{cts_stat} shows the connection between $N_{\rm tot}$ for
the observed groups, as defined in Section~\ref{sec:result}, and our
cavity detection results. Groups with $N_{\rm tot} \geq$\,10,000 are
very likely to be identified as harboring cavities, while more than
two-thirds of the systems with lower $N_{\rm tot}$ show no clear
evidence of these. Successful cavity detection is thus clearly biased
towards groups with high-S/N data.  Despite this, there {\em are} real
(and mock) groups with a large number of counts ($N_{\rm tot}\geq
$\,30,000) for which cavities are not detected.  In real data, this
could certainly be because cavities are absent or present a low
intrinsic X-ray contrast, as suggested by our mock results.
Alternatively, it may be due to orientation effects, the impact of
which we have not tested with the mock data. Also, as discussed above,
some of our groups show features which are clearly not well described
by a $\beta$--model, a fact which could also complicate cavity
recovery when relying on surface brightness modeling in the process.

Nevertheless, given the lower mean S/N of the mock systems with
non-detected cavities, our inability to detect cavities in many of the
low-S/N {\em Chandra} data sets may still be consistent with the
presence of cavities in these systems. As mentioned, for the
lowest-S/N mock subsample, cavities are identified in 8/18 systems,
consistent with the 7/19 found in real data. Including brighter
systems ($\leq\,$15,000 net counts), corresponding values are $19/36 =
53\pm 15$\% (mock) and $13/35= 37\pm 12$\% (real), just consistent at
1-$\sigma$. The total rate of systems showing evidence of identifiable
cavities in the mock data of 42/54 ($78\pm 16$\%) is higher than that
of the {\em Chandra} sample however, but the fractions are still
consistent at the 90\% confidence level.  In that sense, our results
are broadly consistent with the possibility that most, if not all, of
our real groups do harbor cavities. However, we strongly caution
against over-interpreting these numbers. Our simplified mock setup was
not designed to reproduce the full complexity of the real sample, and
the above comparison should be considered indicative at best. Hence,
we do not claim our results to suggest that {\em all} our groups
contain cavities, only that the actual fraction is likely considerably
higher than the $\sim 50$\% suggested by our analysis.

A somewhat related question is that of the incidence of central radio
AGN within our sample. Since six of our groups are excluded from such
considerations due to source confusion or lack of 1.4-GHz coverage, we
cannot make statistically robust statements for the full
sample. However, among the remaining 45 systems, 34 ($76\pm 17$\%)
host a detectable central radio source. This fraction is higher than
the corresponding incidence of detectable cavities (23/45, $51 \pm
13$\%), but only marginally so, and it rises to 21/23 (consistent with
100\%) if only considering the groups in the combined C-- and
P--sample. As such, the fraction of detectable central radio AGN is
comparable to that of identifiable cavities, especially considering
that the latter is likely a lower limit to the true fraction. We do
note that while the N--sample is statistically underrepresented among
groups with detectable radio AGN, this is not necessarily the case if
adopting a fixed cut in $L_{\rm 1.4\,GHz}$ so as to focus on
radio-loud AGN alone. This is already hinted at in
Figure~\ref{radio_stat}, which indicates no connection between the
presence of detectable cavities and the current radio power of the
central group galaxy. Furthermore, given that NVSS sensitivity is
nearly uniform \citep{con98} and that the weakest detected radio
source within our sample is also the nearest and has a flux just below
the NVSS 99\% completeness limit, it is clear that all our groups
could potentially harbor a central AGN with a 1.4-GHz luminosity
brighter than this source ($L_{\rm 1.4\,GHz} \approx 1\times
10^{20}$~W~Hz$^{-1}$).

\subsection{Comparison to Results for Clusters}\label{sec:clustergroup}

With half of our sample showing some evidence of cavities, our groups
are lying between results for individual giant elliptical galaxies
($\sim1/4$; \citealt{nul09}) and the X-ray brightest cool-core
clusters ($\sim2/3$; \citealt{dun06}).  Our results in
Section~\ref{sec:mock} suggests that these differences can be at least
partly explained by observational biases that favor cavity detection
in high--flux systems. In addition, as discussed in
Section~\ref{sec:result}, the surface brightness distribution of
groups is generally flatter than in clusters across the relevant
radial ranges.  The dependence on $\beta$ of the cavity recovery rate
(cf.\ Figure~\ref{mock_3d}) may therefore also play a role for the
above results.

Alternatively, the observed differences in detectable cavity fraction
may reflect a more fundamental connection between the presence of
high-contrast X-ray cavities and global system properties such as
X-ray luminosity or total mass. Under the plausible assumption that
X-ray cavities are generated by a radio outburst in the central AGN,
the lower cavity fraction in smaller systems could be explained by a
systematic variation in central AGN duty cycle with system mass.  If
X-ray fainter systems generally have shorter central radio outbursts,
clearly detectable cavities are perhaps inflated by the AGN in a
smaller fraction of those systems. The observed lack of a link between
the 1.4~GHz luminosity of any radio source within the central galaxy
and the presence of detectable cavities in our sample seems consistent
with this possibility. If the outburst phase is relatively short,
bright 1.4~GHz emission within the central galaxy may already have
subsided when detectable cavities emerge in the intragroup
medium. This scenario should be testable with lower-frequency radio
data which trace the older, less energetic electron populations from
the outburst. Such studies are currently underway \citep{gia09}.

With regard to the properties of the detected cavities themselves,
another finding of our study is that cavity radii in the radial and
tangential directions are highly correlated, with the ratio of the two
remaining roughly constant regardless of cavity size. The implication
of equations~(\ref{eq:ab})--(\ref{eq:bd}) is that bubbles retain a
mean axis ratio of $a/b\sim 1.7$ as they rise and expand in the
surrounding medium, until eventually shredded by hydrodynamic
instabilities. This finding is in excellent quantitative agreement
with results for massive galaxy clusters \citep{bir04}, as illustrated
in Figure~\ref{size_distance}, suggesting that it is a generic feature
of cavity evolution regardless of the depth of the local gravitational
potential and the entropy gradient of the ambient medium (which set
the characteristic buoyancy rise time of the bubbles). Moreover, the
tight correlation between the size of the cavity and the projected
cavity distance from the group center (as shown in
Figure~\ref{size_distance}) indicates that the apparent angle opened
by the cavity to the center of the group remains nearly constant at
$\theta \sim 60^\circ$ during the evolution of the cavity. This is
also in remarkable agreement with the case of clusters
\citep{bir04,die08}, indicating that similar physical processes are
responsible for bubble evolution, and their disruption, in systems
covering a large range in total mass.

\subsection{Implications for the Formation and Evolution of 
Cavities}\label{sec:obsersimula}

It is interesting to note that the gas temperature profiles for our
sample \citep{ras07,sun09} indicate that 43 of our groups host a cool
core, with the eight non-cool core systems all belonging to the
N--sample. Hence, the (C-- and P--sample) cavity fraction is $60\pm
15$\% for cool-core groups, in contrast to the complete absence of
identifiable cavities among our non--cool core systems.  However, this
need not reflect any intrinsic differences between cool-core and
non-cool core systems, since cavity detectability is conceivably
biased toward cool-core systems with bright central regions (as
suggested by Figure~\ref{cts_stat} and by the increase in cavity
recovery rate with both $N_{\rm tot}$ and $\beta$ for our mock data
sets).
On the other hand, barring differences in normalization, the
surface brightness profiles of our different subsamples show
very similar behavior, with comparable mean values of $\beta$ 
and $r_c$. As such, there is no clear evidence that the surface
brightness distribution is more centrally peaked for the C-- and
P--samples, making it worthwhile to briefly discuss whether the
different cavity detection rates for cool-core and non-cool core
systems reflect a real effect.

While a detailed discussion of the cool core/non--cool core
distinction and its possible origins is well beyond the scope of this
work, we do note that an increased fraction of detectable cavities
among cool-core groups would fit the picture of a self-sustained
cooling-driven feedback loop: Strong central cooling of ICM material
feeds the supermassive black hole in the central group galaxy, which
responds with a radio outburst that creates cavities in the
surrounding medium and temporarily re-heats it, until radiative
cooling can re-ignite the process. Some support for this comes from
the work of \citet{die08b}, based on a {\em Chandra} study of 36
ellipticals, ten of which are central group galaxies in our sample
(and of which eight show evidence of cavities in the present
work). These authors found a relation between the inner radial gas
temperature gradient and the central radio AGN power, in the sense
that positive (and steeper) inner gradients are preferentially found
in optically brighter ellipticals with stronger central radio sources
(but see also \citealt{fuk06} and \citealt{hum06}). In addition,
practically all our groups with identifiable cavities host a
detectable central radio source. Although these results suggest a
connection between strong central ICM cooling, a powerful AGN radio
output, and the presence of detectable cavities, we emphasize that
there is still very little {\em direct} evidence for central gas
cooling to low temperatures in groups or clusters. In particular, deep
X-ray grating spectroscopy of A262, A3581, and HCG\,62, all belonging
to our C--sample, has failed to reveal material cooling to below
$T\approx 0.5$~keV in their central regions \citep{san10}. It is also
unclear how the central black hole would accomplish the distributed
heating required to completely halt catastrophic cooling at the group
core, even if only temporarily. Thus, while the above picture may seem
appealing, many of its details still await observational and
theoretical verification.

Once formed, the properties of cavities themselves may help to shed
light on the processes responsible for their production. In recent
years several authors have employed numerical techniques to study the
evolution of buoyant bubbles in galaxy groups and clusters, including
purely inviscid hydrodynamical simulations in two
\citep{chu01,bru02,rey02,bru03,ste07} and three dimensions
\citep{qui01,bas03,omm04,pav08,sca08,bru09}.  Results generally show
that although under certain circumstances the bubbles could be
long-lived, purely hydrodynamical evolution usually cannot prevent
bubbles from being rapidly shredded by Rayleigh--Taylor,
Kelvin--Helmholtz, or Richtmyer--Meshkov instabilities (e.g.,
\citealt{die08}). To prevent bubble disruption, additional physics
such as magnetic fields, thermal conductivity, and viscosity are
probably required. With certain geometries, a magnetic field could
help to protect the bubble \citep{bru01,rob04, jon05, rusz07, liu08,
dur08, one09,don09}.

Our observation of cavities at projected distances of $D\ga 20$~kpc
from the group core implies that bubbles are reasonably resilient to
disruption processes, thus confirming the necessity of incorporating
such additional physics in numerical work \citep{jon05}.
Alternatively, it has been demonstrated that if a bubble is
continuously being inflated by a purely hydrodynamical jet
\citep{piz06} or a magnetically dominated jet
\citep{li06,nak06,nak07}, under some circumstances the development of
hydrodynamical instabilities could also be suppressed, thus keeping
bubbles intact for longer \citep{die07}. However, the fact that we
find no clear link between the presence of detectable cavities and the
current luminosity of any central radio source seems to argue against
this mechanism as a general feature in our groups.

To briefly explore the possibility of distinguishing between different
model predictions for cavity evolution, we can adopt the approach of
\citet{die08}.  These authors consider a range of different models for
the formation and evolution of cavities, under the reasonable
assumptions that the bubbles rise subsonically in the ICM in pressure
equilibrium with their surroundings, with the ICM background density
described by a $\beta$--model. The models considered include one of
purely hydrodynamic, adiabatically expanding bubbles containing gas
with an adiabatic index $\Gamma=5/3$ (here denoted ``AD53'' for
consistency with \citealt{die08}), and a model in which bubbles are
generated by a current-dominated jet (``CDJ'', see, e.g.,
\citealt{li06}) and have magnetically dominated pressure. In both
these models, predicted bubble radii $a$ grow with clustercentric
distance $D$ as $a \propto [1+(D/r_c)^2]^\alpha$, with $\alpha =
\beta/(2\Gamma)$ and $\alpha=3\beta/4$ for the AD53 and CDJ models,
respectively.  The asymptotic growth of the bubbles (at $D \gg r_c$)
is thus a powerful discriminator of these different models. Among the
models discussed by \citet{die08}, AD53 represents the slowest and CDJ
the fastest expansion of the bubbles. When plotted against $D/r_c$,
\citet[their figure~2]{die08} find that observed cavity sizes at $D\ga
4r_c$ suggest that bubbles expand relatively fast, with CDJ the
preferred model.

\begin{figure}
\begin{center}
\epsscale{1.1}
\plotone{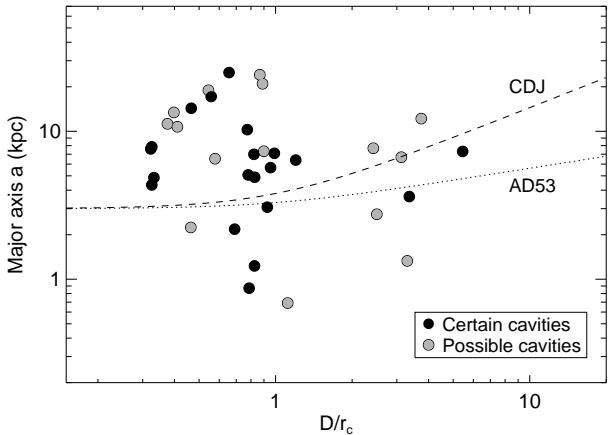}
\end{center}
\figcaption{Cavity major axis $a$ as a function of projected
  groupcentric distance $D$ normalized by $r_c$.  Dotted line
  (``AD53'') shows the predicted evolution of a purely hydrodynamical
  bubble, and dashed line (``CDJ'') that of a magnetically dominated
  bubble inflated by a current-dominated jet (see text for details),
  both for an arbitrarily chosen initial bubble size of $a=3$~kpc.
 \label{models}}
\end{figure}

For comparison, we show a similar plot for our cavity sample in
Figure~\ref{models}. Overplotted are model predictions for the AD53
and CDJ models, both assuming a fixed value of $\beta=0.45$, the mean
value for our groups with detectable cavities.  It is clear that we
cannot easily distinguish between these two simple models. This is
largely due to a dearth of systems with distant cavities in our
sample, since --- unlike \citet{die08} --- we have only a handful of
cavities detected well outside the ICM core radius. A larger group
sample, particularly including systems with distant cavities, would be
required to make robust statements on the nature of cavities in
groups. However, since the cavities within our sample appear to obey
the same scaling relations as seen in more massive systems,
constraints from studies of cavities in clusters \citep{die08} are
likely to apply in our groups as well.

\section{CONCLUSIONS}\label{sec:conclude}

Based on {\em Chandra} archival data of 51 galaxy groups, we have
presented the first systematic search for, and analysis of, X-ray
cavities in a large sample of groups. At least nine of these groups
have previously been reported to show evidence of such cavities. In
the present work, the cavities are identified from {\em Chandra} data
alone, either by subtracting an elliptical $\beta$--model fit from the
X-ray surface brightness distribution, or by performing unsharp
masking using two Gaussian smoothing kernels.  Based on visual
inspection of the resulting images, 13 groups ($\sim1/4$) within our
sample are identified as clearly harboring cavities, and another 13
groups ($\sim1/4$) show tentative evidence for cavities. All of these
26 systems show clear evidence for a cool core. The remaining 25
groups have a lower cool-core fraction ($17/25\approx 70$\%) and do
not show convincing evidence for the presence of cavities. The total
fraction of groups with some evidence of cavities in our sample is
thus 50\%, rising to 60\% if only considering the cool-core
systems. In this respect, our groups are lying between results for
individual giant elliptical galaxies ($\sim1/4$; \citealt{nul09}) and
cool-core clusters ($\sim2/3$; \citealt{dun06}). We found no clear
link between the current 1.4~GHz radio luminosity of the central
brightest group galaxy and the detection of cavities.

To test our ability to identify cavities using the two adopted
approaches, we generated a set of mock images designed to mimic
typical {\em Chandra} images of the observed groups. For these data,
we find that surface brightness modeling is generally superior to
unsharp masking in reliably recovering cavity properties, a conclusion
echoing one already hinted at from our analysis of the real data.  The
results also show that cavity recovery is strongly influenced by a
number of factors, including the projected distance between the cavity
center and the group center, the significance of the surface
brightness depression represented by the cavity, and most importantly,
the signal-to-noise ratio of the diffuse group emission. Cavity
detectability is highly biased toward X-ray bright systems, implying
that the observed cavity fraction in our sample is likely a lower
limit.

We find tight correlations between the radial and tangential radii of
the cavities in our group sample, and between the size and projected
groupcentric distance of cavities. While we are not able to clearly
distinguish between simple model predictions for the evolution of
cavities on the basis of our sample alone, we note that the observed
correlations are in excellent quantitative agreement with results for
more massive clusters. This suggests that very similar physical
processes are responsible for cavity evolution and disruption in
systems covering a large range in mass.

\acknowledgments

We thank the referee for helpful comments which improved the
presentation of our results.  This research has made use of data
obtained from the Chandra Data Archive and the Chandra Source Catalog,
and software provided by the Chandra X-ray Center (CXC) in the
application packages CIAO, ChIPS, and Sherpa.  This research has also
made use of the NASA/IPAC Extragalactic Database (NED) which is
operated by the Jet Propulsion Laboratory, California Institute of
Technology, under contract with the National Aeronautics and Space
Administration.  JR acknowledges support provided by the National
Aeronautics and Space Administration through Chandra Postdoctoral
Fellowship Award Number PF7-80050 issued by the Chandra X-ray
Observatory Center, which is operated by the Smithsonian Astrophysical
Observatory for and on behalf of the National Aeronautics and Space
Administration under contract NAS8-03060. JSM acknowledges partial
support for this work from NASA grant NNG04GC846.

\clearpage

\begin{deluxetable*}{lcccc}
\tabletypesize{\scriptsize} \tablewidth{0pc} \tablecaption{The group
sample} \tablehead{ \colhead{Group}& \colhead{Obs.\
ID\tablenotemark{a}} & \colhead{Obs.\ Date}& \colhead{Exposure
Time\tablenotemark{b}} & \colhead{Distance\tablenotemark{c}} \\ & &
\colhead{(yyyy-mm-dd)} & \colhead{(ks)} & \colhead{(Mpc)}} \startdata
3C\,442A & 6392 & 2006-01-12  & 32.57 & 105 \\
3C\,449 & 4057 & 2003-09-18 & 29.09 & 66.8 \\
A262 & 7921 & 2006-11-20 & 110.63 & 64.3 \\
A744 & 6947 & 2006-10-22 & 39.29 & 321 \\
A1139 & 9387 & 2008-03-28 & 10.05 & 174 \\
A1177 & 6940 & 2006-12-27 & 33.59 & 138 \\
A1238 & 4991 & 2004-03-28 & 21.57 & 324 \\
A1275 & 6945 & 2006-20-05 & 49.32 & 264 \\
A1692 & 4990 & 2004-08-12 & 21.47 & 373 \\
A1991 & 3193 & 2002-12-16 & 38.30 & 255 \\
A2092 & 9384 & 2007-11-13 & 9.99 & 291 \\
A2462 & 4159 & 2002-11-19 & 39.24 & 313 \\
A2550 & 2225 & 2001-09-03 & 59.01 & 546 \\
A2717 & 6974 & 2006-04-10 & 19.79 & 205 \\
A3581 & 1650 & 2001-06-07 & 7.16 & 99.9 \\
A3880 & 5798 & 2004-12-23 & 22.20 & 247 \\
AS1101 & 1668 & 2001-08-13 & 9.95 & 246 \\
ESO\,306-017 & 3189 & 2002-03-09 & 14.05 & 152 \\
ESO\,351-021 & 5784 & 2005-04-24 & 39.46 & 242 \\
ESO\,552-020 & 3206 & 2002-10-14 & 23.62 & 132 \\
HCG\,42 & 3215 & 2006-09-08 & 31.70 & 60.1 \\
HCG\,51 & 4989 & 2004-02-15 & 30.75 & 113 \\
HCG\,62 & 921 & 2001-10-09 & 48.50 & 61.6 \\
IC\,1262 & 2018 & 2001-08-23 & 30.73 & 137 \\
MKW\,4 & 3234 & 2002-11-24 & 29.95 & 88.5 \\
NGC\,383 & 2147 & 2000-11-06 & 44.39 & 66.7 \\
NGC\,507 & 2882 & 2006-08-28 & 43.63 & 64.5 \\
NGC\,533 & 2880 & 2002-07-28 & 37.58 & 72.8 \\
NGC\,741 & 2223 & 2001-01-28 & 30.32 & 73.4 \\
NGC\,1132 & 3576 & 2003-11-16 & 39.58 & 93.8 \\
NGC\,1407 & 791 & 2007-06-05 & 48.45 & 22.8 \\
NGC\,1550 & 5800 & 2005-10-22 & 44.37 & 50.2 \\
NGC\,2300 & 4968 & 2006-05-24 & 45.57 & 25.8 \\
NGC\,3402 & 3243 & 2002-11-05 & 29.51 & 68.7 \\
NGC\,4104 & 6939 & 2006-02-16 & 35.82 & 123 \\
NGC\,4125 & 2071 & 2007-01-04 & 64.22 & 20.2 \\
NGC\,4325 & 3232 & 2003-02-04 & 30.07 & 113 \\
NGC\,5044 & 9399 & 2008-03-09 & 82.65 & 41.7 \\
NGC\,5098 & 6941 & 2005-11-01 & 38.53 & 164 \\
NGC\,5129 & 7325 & 2006-05-14 & 25.78 & 100 \\
NGC\,5846 & 7923 & 2007-06-14 & 90.00 & 26.2 \\
NGC\,6269 & 4972 & 2003-12-29 & 39.58 & 147 \\
NGC\,6338 & 4194 & 2003-09-17 & 47.33 & 115 \\
NGC\,7619 & 3955 & 2006-07-05 & 36.75 & 46.9 \\
RBS\,461 & 4182 & 2003-03-11 & 23.45 & 120 \\
RXJ\,1022+3830 & 6942 & 2006-10-14 & 41.43 & 213 \\
RXJ\,1159+5531 & 4964 & 2004-02-11 & 75.06 & 356 \\
RXJ\,1206-0744 & 9388 & 2007-11-15 & 9.99 & 295 \\
UGC\,842 & 4963 & 2005-02-13 & 39.24 & 188 \\
UGC\,2755 & 2189 & 2001-02-07 & 15.62 & 100 \\
UGC\,5088 & 3227 & 2002-03-10 & 34.32 & 117
\enddata
\tablenotetext{a}{{\em Chandra} Observation ID.}
\tablenotetext{b}{Cleaned {\em Chandra} exposure time.}
\tablenotetext{c}{\,Luminosity distance from the NASA/IPAC
Extragalactic Database (NED), in the reference frame defined by the
cosmic microwave background.} \label{table:sample}
\end{deluxetable*}

\clearpage

\LongTables
\begin{deluxetable*}{lcccccccc}
\tabletypesize{\scriptsize} \tablewidth{0pc} \tablecaption{Fit
results and cavity properties} \tablehead{\colhead{Group}&
\colhead{Cavity\tablenotemark{a}}& \colhead{$r_c$} &
\colhead{$\beta$}& \colhead{$D$\tablenotemark{b}} &
\colhead{$a$\tablenotemark{c}} & \colhead{$b$\tablenotemark{d}}&
\colhead{$N_{\rm tot}$\tablenotemark{e}} &
\colhead{log\,$L_{\rm 1.4\,GHz}$\tablenotemark{f}} \\ \colhead{} &
\colhead{} & \colhead{(kpc)} & & \colhead{(kpc)} & \colhead{(kpc)} &
\colhead{(kpc)} & & \colhead{(W~Hz$^{-1}$)}} \startdata
A262 & C &$ 6.94 ^{+0.06}_{-0.06 }$&$ 0.417 ^{+0.001}_{-0.001 }$& 5.7 & 4.9 & 3.3 & 190240 & 22.51 \\
 & & & & 5.4 & 5.1 & 3.1 & & \\
A1991 & C &$ 18.57 ^{+0.15}_{-0.15 }$&$ 0.458 ^{+0.001}_{-0.001 }$& 10.4 & 17.2 & 13.1 & 59248 & 23.48 \\
 & & & & 12.2 & 24.9 & 9.9 & & \\
A3581 & C &$ 9.52 ^{+0.29}_{-0.28 }$&$ 0.380 ^{+0.003}_{-0.003 }$& 3.1 & 7.9 & 3.8 & 17932 & 23.89 \\
 & & & & 3.1 & 4.3 & 3.7 & & \\
A3880 & C &$ 15.75 ^{+0.33}_{-0.33 }$&$ 0.423 ^{+0.002}_{-0.002 }$& 12.2 & 10.3 & 8.2 & 12693 & \ldots \\
HCG\,62 & C &$ 6.50 ^{+0.06}_{-0.06 }$&$ 0.483 ^{+0.002}_{-0.002 }$& 5.3 & 7.0 & 4.1 & 45614 & 21.35 \\
 & & & & 7.8 & 6.4 & 4.4 & & \\
IC\,1262 & C &$ 20.83 ^{+0.36}_{-0.36 }$&$ 0.407 ^{+0.002}_{-0.002 }$& 9.7 & 14.3 & 10.0 & 35068 & 22.56 \\
NGC\,533 & C &$ 1.72 ^{+0.03}_{-0.03 }$&$ 0.490 ^{+0.003}_{-0.003 }$& 1.2 & 2.2 & 1.3 & 14851 & 22.26 \\
 & & & & 1.6 & 3.1 & 1.6 & & \\
NGC\,1132 & C &$ 0.71 ^{+0.03}_{-0.03 }$&$ 0.436 ^{+0.004}_{-0.004 }$& 3.9 & 7.3 & 3.3 & 12131 & 21.75 \\
NGC\,4104 & C &$ 1.99 ^{+0.05}_{-0.05 }$&$ 0.632 ^{+0.009}_{-0.009 }$& 0.0\tablenotemark{g} & 1.9 & 1.5 & 4916 & 21.96 \\
NGC\,5044 & C &$ 14.79 ^{+0.07}_{-0.07 }$&$ 0.480 ^{+0.001}_{-0.001 }$& 4.8 & 7.6 & 4.7 & 329715 & 21.90 \\
 & & & & 4.9 & 4.9 & 3.5 & & \\
NGC\,5098 & C &$ 1.18 ^{+0.05}_{-0.05 }$&$ 0.369 ^{+0.003}_{-0.003 }$& 4.0 & 3.6 & 2.3 & 11807 & 23.60 \\
NGC\,5846 & C &$ 0.76 ^{+0.01}_{-0.01 }$&$ 0.396 ^{+0.001}_{-0.001 }$& 0.6 & 1.2 & 0.8 & 42072 & 21.23 \\
 & & & & 0.6 & 0.9 & 0.6 & & \\
NGC\,6338 & C & $3.25 ^{+0.99}_{-3.25}$ &$ 0.470 ^{+0.038}_{-0.156 }$& 3.1 & 5.7 & 3.1 & 15596 & 22.84 \\
 & & & & 3.2 & 7.1 & 2.9 & & \\
\hline
3C\,442A & P & & & 9.8 & 6.8 & 3.8 & 2456 & 24.64 \\
 & & & & 8.9 & 5.7 & 3.7 & & \\
A2550 & P &$ 18.94 ^{+0.29}_{-0.28 }$&$ 0.446 ^{+0.002}_{-0.002 }$& 10.3 & 18.9 & 9.3 & 14834 & \ldots \\
 & & & & 7.8 & 10.7 & 5.9 & & \\
A2717 & P &$ 20.98 ^{+1}_{-0.97 }$&$ 0.363 ^{+0.005}_{-0.005 }$& 7.9 & 11.2 & 6.3 & 7443 & 24.41 \\
 & & & & 8.4 & 13.4 & 5.8 & & \\
AS1101 & P &$ 60.94 ^{+0.28}_{-0.28 }$&$ 0.570 ^{+0.003}_{-0.003 }$& 24.2 & 21.0 & 14.7 & 40594 & \ldots \\
 & & & & 23.6 & 24.1 & 15.7 & & \\
ESO\,351-021 & P &$ 3.96 ^{+0.13}_{-0.12 }$&$ 0.516 ^{+0.006}_{-0.006 }$& 14.8 & 12.2 & 8.6 & 4756 & 22.72 \\
NGC\,507 & P & & & 7.2 & 6.4 & 4.1 & 19787 & 22.69 \\
NGC\,741 & P &$ 0.60 ^{+0.02}_{-0.02 }$&$ 0.467 ^{+0.004}_{-0.004 }$& 1.5 & 2.8 & 1.7 & 4388 & 23.84 \\
NGC\,1407 & P &$ 0.29 ^{+0.01}_{-0.01 }$&$ 0.392 ^{+0.003}_{-0.003 }$& 0.3 & 0.7 & 0.4 & 16126 & 21.74 \\
NGC\,1550 & P &$ 3.13 ^{+0.05}_{-0.05 }$&$ 0.386 ^{+0.002}_{-0.002 }$& 1.5 & 2.2 & 1.2 & 49021 & 21.70 \\
NGC\,2300 & P &$ 0.44 ^{+0.02}_{-0.02 }$&$ 0.457 ^{+0.005}_{-0.005 }$& 1.5 & 1.3 & 1.0 & 4234 & 20.36 \\
RXJ\,1159+5531 & P &$ 3.10 ^{+0.06}_{-0.06 }$&$ 0.526 ^{+0.004}_{-0.003 }$& 7.5 & 7.7 & 3.9 & 8128 & $<$ 22.71 \\
 & & & & 9.7 & 6.7 & 4.3 & & \\
RXJ\,1206-0744 & P & & & 29.1 & 27.6 & 21.4 & 949 & $<$ 22.55 \\
UGC\,5088 & P &$ 9.35 ^{+0.9}_{-0.85 }$&$ 0.481 ^{+0.028}_{-0.025 }$& 8.4 & 7.3 & 5.4 & 2097 & 21.08 \\
 & & & & 5.4 & 6.5 & 3.6 & & \\
\hline
3C\,449 & N &$ 0.39 ^{+0.03}_{-0.03 }$&$ 0.550 ^{+0.018}_{-0.016 }$& & & & 1764 & 24.29 \\
A744 & N &$ 29.83 ^{+1.3}_{-1.27 }$&$ 0.372 ^{+0.004}_{-0.004 }$& & & & 5260 & $<$ 22.62 \\
A1139 & N & & & & & & 1948 & $<$ 22.09 \\
A1177 & N &$ 0.97 ^{+0.09}_{-0.08 }$&$ 0.662 ^{+0.042}_{-0.035 }$& & & & 3819 & $<$ 21.89 \\
A1238 & N & & & & & & 1395 & \ldots \\
A1275 & N &$ 30.20 ^{+0.99}_{-0.97 }$&$ 0.404 ^{+0.004}_{-0.004 }$& & & & 7811 & $<$ 22.45 \\
A1692 & N & & & & & & 1543 & \ldots \\
A2092 & N & & & & & & 789 & $<$ 22.54 \\
A2462 & N & & & & & & 10261 & 25.24 \\
ESO\.306-017 & N &$ 1.12 ^{+0.07}_{-0.07 }$&$ 0.421 ^{+0.006}_{-0.006 }$& & & & 5718 & \ldots \\
ESO\.552-020 & N &$ 0.51 ^{+0.05}_{-0.04 }$&$ 0.375 ^{+0.005}_{-0.005 }$& & & & 5414 & $<$ 21.85 \\
HCG\.42 & N &$ 3.54 ^{+0.25}_{-0.05 }$&$ 0.425 ^{+0.009}_{-0.002 }$& & & & 7297 & 22.02 \\
HCG\,51 & N &$ 0.48 ^{+0.03}_{-0.03 }$&$ 0.478 ^{+0.008}_{-0.008 }$& & & & 3335 & $<$ 21.73 \\
MKW\,4 & N &$ 5.37 ^{+0.07}_{-0.07 }$&$ 0.447 ^{+0.002}_{-0.002 }$& & & & 32765 & 22.21 \\
NGC\,383 & N &$ 1.01 ^{+0.04}_{-0.04 }$&$ 0.666 ^{+0.015}_{-0.014 }$& & & & 2888 & 24.41 \\
NGC\,3402 & N &$ 2.30 ^{+0.03}_{-0.03 }$&$ 0.424 ^{+0.001}_{-0.001 }$& & & & 39972 & 21.67 \\
NGC\,4125 & N &$ 0.15 ^{+0.01}_{-0.01 }$&$ 0.373 ^{+0.003}_{-0.003 }$& & & & 8878 & 19.95 \\
NGC\,4325 & N &$ 23.21 ^{+0.21}_{-0.21 }$&$ 0.624 ^{+0.003}_{-0.003 }$& & & & 39165 & $<$ 21.71 \\
NGC\,5129 & N &$ 0.93 ^{+0.05}_{-0.04 }$&$ 0.498 ^{+0.009}_{-0.009 }$& & & & 2592 & 21.93 \\
NGC\,6269 & N &$ 1.09 ^{+0.06}_{-0.06 }$&$ 0.534 ^{+0.011}_{-0.011 }$& & & & 2843 & 23.11 \\
NGC\,7619 & N &$ 0.41 ^{+0.01}_{-0.01 }$&$ 0.456 ^{+0.003}_{-0.003 }$& & & & 6666 & 21.73 \\
RBS\,461 & N & & & & & & 8079 & 22.54 \\
RXJ\,1022+3830 & N &$ 14.23 ^{+0.67}_{-0.66 }$&$ 0.342 ^{+0.004}_{-0.004 }$& & & & 6896 & $<$ 22.26 \\
UGC\,842 & N & & & & & & 3592 & 22.23 \\
UGC\,2755 & N & $0.60^{+0.08}_{-0.60}$ & $0.679^{+0.098}_{-0.181 }$& & & & 408 & 24.24 
\enddata
\tablenotetext{a}{Classification according to whether the group
contains certain cavities (C), possible cavities (P), or no
detectable cavities (N).} \tablenotetext{b}{Projected distance from cavity
center to the X-ray center of the group.} \tablenotetext{c}{Cavity
size in tangential direction (projected major axis $a$ defined in
Section~\ref{sec:detection}).} \tablenotetext{d}{Cavity size in
radial direction (projected minor axis $b$ defined in
Section~\ref{sec:detection}).} \tablenotetext{e}{Number of
0.3--2~keV photons from diffuse emission on the central CCD.}
\tablenotetext{f}{Radio luminosity at 1.4~GHz of any radio source
within the central brightest group galaxy, extracted from the NRAO
VLA Sky Survey (NVSS; \citealt{con98}). See text for details.}
\tablenotetext{g}{The cavity in NGC\,4104 is at the very center of
the X-ray emission.} 
\label{table:cavity}
\end{deluxetable*}

\end{document}